\documentclass[twocolumn,tighten]{aastex61}
\usepackage{latexsym}
\usepackage{natbib}
\usepackage{amssymb}
\usepackage{amsmath}

\usepackage{graphicx}
\usepackage{graphics}
\usepackage{fancyhdr}
\received{}
\revised{}
\accepted{}
\submitjournal{ApJ}

\shorttitle{Cosmological interpretation of the CMDs of galaxy clusters}
\shortauthors{Sciarratta et al.}

\begin{document}
 
 \title{Cosmological interpretation of the color-magnitude diagrams of galaxy clusters}

\author{Mauro Sciarratta}
\affil{Department of Physics and  Astronomy, Vicolo dell'Osservatorio, 3 I-35100 Padova, Italy}

\author{Cesare Chiosi}
\affil{Department of Physics and Astronomy, Vicolo dell'Osservatorio, 3 I-35100 Padova, Italy}

\author{Mauro D'Onofrio}
\affil{Department of Physics and Astronomy, Vicolo dell'Osservatorio, 3 I-35100 Padova, Italy}
\affil{INAF Observatory of Padova, Vicolo Osservatorio 5, I-35122 Padova, Italy}

\author{Stefano Cariddi}
\affil{Department of Physics and Astronomy, Vicolo dell'Osservatorio, 3 I-35100 Padova, Italy}

\correspondingauthor{Mauro Sciarratta}
\email{mauro.sciarratta@unipd.it}

\begin{abstract}

We investigate the color-magnitude diagram (CMD) of cluster galaxies in the hierarchical $\Lambda$-CDM cosmological
scenario using both single stellar populations and simple galaxy models. 
First, we analyze the effect of bursts and mergers and companion chemical pollution and 
rejuvenation of the stellar content
on the integrated light emitted by galaxies. The dispersion of the galaxy magnitudes and colors on 
the $M_V-(B-V)$ plane is mainly due to
mixing of ages and metallicities of the stellar populations, with mergers weighting more than
bursts of similar mass fractions. The analysis is made using the Monte-Carlo technique applied to ideal model galaxies 
reduced to single stellar populations with galaxy-size mass to evaluate mass, age and 
metallicity of each object. We show that separately determining  the  contributions by bursts and mergers 
leads to a better understanding of observed  properties of CMD of cluster galaxies.
Then we repeat the analysis using suitable  chemo-photometric models of galaxies whose mass is derived from 
the cosmological predictions of the galaxy content of typical clusters. Using the halo mass function 
and the Monte-Carlo technique, we derive the formation redshift of each galaxy and its photometric history. These are
used  to simulate the  CMD of the cluster galaxies. The main conclusion is that most massive galaxies have acquired 
the red color they show today in very early epochs and remained the same ever since. The simulations nicely reproduce 
the Red Sequence, the Green Valley and the Blue Cloud, the three main regions of the CMD in which galaxies crowd.   

\end{abstract}


\section{Introduction} \label{sec:intro}

The color-magnitude diagram (CMD) is the reference diagnostic tool for
understanding the physical properties of stars and stellar populations of any age, chemical composition and size
going from simple star clusters to galaxies and on recent times the galaxy content of clusters of galaxies.
Leaving aside the classical case of stars and star clusters, for which starting from the original studies by 
 \citet[][]{Hertzsprung1909} and \citet{Russell1914} nowadays there is an immense body of literature, and also those for
our own Galaxy and  galaxies in the Local Group and nearby Universe 
\citep[e.g. Andromeda][]{Baade1944,Sandage1957,Blaauw1959}, 
the  CMD diagnostic has been  extended to the integrated light of galaxies in clusters, e.g.  
like  Virgo and Coma, to understand the physical properties of their galaxies
\citep[e.g.][]{ChesterRoberts64,Chiosi67, Visvanathan_Sandage_1977, Sandage_Visvanathan_1978a, 
Sandage_Visvanathan_1978b}. The advent of modern instrumentation provided much richer CMDs 
\citep[][to mention a few]{BowerLuceyEllis1992,Kodama_etal_1998,Kodama_etal_1999,Terlevich_etal_01, 
Bell_et_al_2004} and allowed to  infer information 
on the past history of the clusters themselves.
Furthermore, since colors are independent on distance,  for any cluster this latter is nearly the same for all member
galaxies, and  the fact that many (if not all) clusters
do share many  common features, their CMDs  have been considered
as key cosmological probes \citep{Tully_etal_1982,BowerLuceyEllis1992}.

In this context, galaxy clusters play a key role in our understanding of not only galaxy formation and 
evolution but also the same cosmogony of the Universe. In brief, studies of nearby clusters have provided 
important informations about the gas content \citep{DaviesLewis73,Kennicutt83},
the very high mass-to-light ratios (several hundreds  solar units) indicating
large amounts of dark matter \citep[DM; e.g.][]{FaberGallagher1979,Adami_etal_1998}, 
the gravitational lensing of distant objects in the Universe 
\citep[][and references therein]{Bartelmann2010}, the 
 strong X-ray emission due to hot gas filling the intra-cluster medium 
\citep[ICM; see e.g.][]{GottGunn1971,SunyaevZeldovich1972}, the
fractions of blue  and red galaxies and their morphological ratios 
\citep[both differ from the ones in the field][]{ButcherOemler1978,Dressler1980},
the so-called galaxy color bi-modality \citep{Baldry_etal_2004} and other issues not mentioned here. 
 These features, together with phenomena like
dynamical friction, tidal disruption, cooling flows and chemical
enrichment of ICM from galactic winds, make clusters very similar objects:
i.e. excellent tools  for exploring the Universe  \citep{BowerLuceyEllis1992}. 
Furthermore, several questions have been asked and answered in the past three
decades thanks to galaxy clusters, e.g. cold DM as the concordance model to describe
dynamical matter in the Universe  
and the non-hierarchical nature of baryonic physics, together with the inherently famous problem
of ``downsizing" of galaxies \citep[see][for a detailed review, and all references therein for 
in-depth description of galaxy clusters]{KravtsovBorgani2012}.
For all these reasons, the study of galaxy clusters and their galaxy content is of paramount importance.

As far as the CMDs are concerned, three main loci were soon identified:
the first was the Red Sequence \citep[firstly noted by][]{devaucouleurs1961},
an almost perfectly linear band throughout a
broad range of luminosities mainly occupied by evolved early-type galaxies. Since then thanks to many large scale surveys, 
magnitudes and colors in different photometric pass-bands, morphological types, redshifts, for hundreds  of thousands of 
galaxies became available. To mention a few we recall the Galaxy Zoo from the Sloan Digital Sky Survey of 
\citet{Blanton_eta_2003}, \citet{Lintott_etal_2008}, \citet{Wong_etal_2012} so that the existence  and evolution of the 
Red Sequence was 
widely confirmed \citep{Stott_etal_2009,Head_etal_2014}, and more recently, the faint-end of this was also investigated
(see \citealt{Boselli_Gavazzi_2014} for the faint-end  of the Red Sequence in high density environments,
\citealt{Roediger_etal_2017} for the faint-end of the Red Sequence in the Virgo clusters at the faintest magnitudes,
and \citealt{Head_etal_2014} for dissecting the Red Sequence in bulge and disk of early-type galaxies in the Coma cluster). Finally, 
\citet{Baldry_etal_2004} quantified the bimodal CMDs of galaxies. Very soon it became evident that, 
in addition to the 
Red Sequence, two more regions in the CMDs are crowded by galaxies: 
the Blue Cloud, where
gas-rich galaxies forming stars at high rates are found; the Green Valley,
in between the first two, where a complicated interplay
between gas conversion into new stars and the redward evolution of
old stars is at work \citep{Menci_etal_2005}. All this is the analog of the bimodal color distribution of star 
clusters  in the LMC  
\citep{Chiosi_Bertelli_Bressan_1988}, and globular clusters in our own and external galaxies 
\citep[see][and references]{Cantiello_Blakeslee_2007}. 
In this context many attempts have been made to firmly define the three regions on the CMD. 
Unfortunately the task is difficult because of the well known age-metallicity degeneracy: stars and 
stellar populations become redder at increasing age and/or metallicity,  to which in the case of galaxies
at least one more cause  must be added, i.e. the past star formation history of the system 
(see \citealt{Tinsley1980} for a classical review of the basic concepts and the recent review by  
\citealt{Silk_Mamon_2012}).

Understanding the origin of the Red Sequence (otherwise named Color-Magnitude-Relation, CMR), 
its slope and width for the galaxies in the Local Universe has been the subject of many studies, 
among which we recall \citet{Baum1959}, \citet{Faber1977}, \citet{Dressler1984}, 
\citet{BowerLuceyEllis1992}, \citet{Burstein_etal_1995}, \citet{Burstein_etal_1997}, 
\citet{Gallazzi_etal_2006}, 
\citet{Mei_etal_2008}, \citet{Valentinuzzi_etal_2011}  to mention a few. In parallel, the galaxy 
CMR and its properties have the subject 
of intense observational and theoretical investigations among many others  we recall 
\citet{Gladders_etal_1998}, \citet{Tran_etal_2007}, \citet{Mei_etal_2009}.  
Basing on the classical scenario of galaxy formation and evolution in presence of  supernova-driven galactic 
winds \citep{Larson1974,ArimotoYoshii1987,Tantaloetal1996,Kodama_Arimoto_1997,Tantaloetal1998,Chiosi_etal_1998}, 
on semi-analytical 
models in 
the hierarchical scheme \citep{WhiteFrenk1991,Kauffmann_1996,KauffmannCharlot1998}, and on NB-TSPH 
simulations of galaxy formation and evolution \citep[e.g.][]{Chiosi_Carraro_2002}, the commonly accepted view emerged
that the  Red Sequence de facto is a metallicity
sequence and not an age sequence, however with an age dispersion that seems to increase at decreasing galaxy mass. 
This conclusion requires an adequate treatment of the chemical evolution in the model galaxies to be taken 
into account, otherwise the slope of the CMR is not reproduced \citep[][]{Kauffmann_1996}. This problem is 
persistent even in very recent studies \citep[][]{Nelson_etal_2018} .
Conversely, understanding the global appearance  of the Red Sequence in the CMDs of cluster galaxies 
is still at its infancy
\citep[e.g.]{Mei_etal_2008,Lee_etal_2017}. This will be the main goal of our study.

In recent times, massive numerical large scale simulations of  hierarchical
galaxy formation in $\Lambda$-CDM cosmogony including both DM and baryonic matter (BM) have been made possible by the new
generation of numerical codes developed by \citet[][]{Hernquist_Springel_2003}, \citet{
Springel_Hernquist_2003a,Springel_Hernquist_2003b}, \citet{Vogelsberger_etal_2012}, \citet{PuchweinSpringel2013},
\citet{Baraietal2013}  and references therein.
In these, much effort is paid to include star formation and chemical enrichment, radiative cooling and heating 
in presence of an UV background
radiation field,  and feedback processes of different nature. With the aid of these simulations, the 
variation of the cosmic star formation rate density, SFRD($z$),  with  redshift 
\citep[see][and references]{MadauDickinson2014,KatsianisTescarietal2017} has been addressed and 
largely explained \citep[][]{RaseraTeyssier2006,Hernquist_Springel_2003,KatsianisTescarietal2017,Pillepich_etal_2017}. 
In brief,  they investigated the effect 
of the energy feedback from supernovae explosions,
stellar winds, and AGN activity on modeling the cosmic star formation. They made use of an improved version
of the code P-GADGET3 by
 \citet{Springel2005} with chemical enrichment \citep{Tornatore-etal2007}, supernova energy and
 momentum-driven galactic winds
\citep{PuchweinSpringel2013}, AGN feedback \citep{SpringelDiMatteo2005,Planellesetal2013}, metal-line cooling
\citep{Wiersma2009a,Wiersma2009b} plus
molecules/metal cooling \citep{Maioetal2007}, supernova-driven galactic winds with feedback
\citep{Baraietal2013}, thermal conduction \citep{Dolagetal2004}.
Some of these simulations take also  the 
photometric evolution of the stellar content of galaxies into account so that
the CMDs are possible, in particular for galaxies belonging to clusters, and consequently are used to prove
the physical assumptions of the large scale cosmological simulations 
\citep[see e.g.][]{Nelson_etal_2015,Nelson_etal_2018}. Furthermore,
theoretical studies \citep[e.g.][]{Bower_etal_2006,Henriques_etal_2015} are devoted to  disentangling 
the effect of DM and BM  on observable quantities: they have proven to be consistent with
observations starting from a hierarchical point of view, even with increasing
complexity of the underlying physics.

In the cosmological view of galaxy formation and evolution,
steps forward have been made thanks to the halo mass functions (HMF),   
\citep[e.g.][]{ShethTormen2002,	Warren_etal_2006,Tinker_etal2008,Angulo_etal_2012,Behroozi_etal_2013}, 
the integral of this over a given mass interval, named the halo growth function (HGF), \citep{Lukicetal2007}, 
and merger trees \citep[e.g.][]{Boylan-kolchin_etal_2009,Guo_etal_2011} 
calculated inside highly-detailed cosmological N-body simulations. They are useful to assemble a sample
of DM halos, hosting galaxies made of BM, in order to follow the evolution of observables through
cosmic times until $z=0$. In this way, several properties of galaxies have been
fully explored, like for instance stellar mass assembly inside DM halos \citep{Moster_etal_2013},
galaxy growth \citep{DeLuciaBlaizot2007} and population in cosmic epochs \citep{Guo_etal_2011}
and the SFRD($z$)  \citep{Chiosi_etal_2017}.

\textsf{Aims of this study}. The great success of the large scale simulations might lead us to conclude that no 
other alternatives are worth being pursued to investigate the formation and evolution of galaxies together with 
their large scale properties.  However,  major   drawbacks of the massive numerical 
simulations are their complexity, high  cost in terms of time and effort,  and lack
 in flexibility and prompt response   to varying key input physics. For these reasons we prefer to investigate the 
properties of the CMDs of cluster galaxies  from a different perspective. The method we are going to propose is 
the same we have used to investigate the SFRD($z$) in \citet{Chiosi_etal_2017}. 
We start from suitable models of 
galaxies made of DM and BM that  are able to catch the essence of NB-TSPH simulations of galaxy formation and evolution
in the classical $\Lambda$-CDM cosmological model of the Universe, of which  we exploit the number density 
of galaxies predicted by
large scale  N-Body simulations.  With the aid of this, we predict the population of galaxies in clusters 
of typical size, and  follow the photometric evolution  of the model galaxies:  we make use of the 
Monte-Carlo technique to simulate bursts of star formation in individual galaxies 
(age and intensity of the bursts are randomly chosen) and to simulate mergers among galaxies of different age, 
mass, chemical enrichment  and stellar content. We derive the CMD and investigate how galaxies 
distribute in the three main groups. In particular we look at the physical status of the galaxies in the groups   
and under which conditions they pass from one group to another.
We also examine and try to quantify the relative  weight of the hierarchical aggregation compared to the intrinsic 
star formation history in galaxies of different mass. In particular we investigate the effect of galactic winds in shaping 
the color evolution of a galaxy.  

The plan of the paper is as follows: 
in Section \ref{obser_data} we briefly present  the CMDs based on the 
WINGS and OMEGA-WINGS surveys for the whole sample and also for a selected group of clusters of which we intend
to interpret the CMDs with the aid of Monte-Carlo 
simulations  based on both SSPs and realistic models of galaxies. 
In Section \ref{galmod} we briefly present
the main physical assumptions  at the base of SSPs and model galaxies. 
In Section \ref{results} we present the Monte-Carlo simulator of bursts of star formation in galaxies and mergers
 among galaxies, 
in which
a generic galaxy is approximated to a SSP of suitable metallicity and age, and for these composite objects we derive 
the CMD.
In Section  \ref{using_gal}  we extend the Monte-Carlo simulator of Section \ref{results} by replacing the SSP 
with realistic models of galaxies. In particular we estimate the population of DM hosting BM galaxies with 
the aid of 
the HMF and HGF  formalism (the definition of the latter will be given therein), derive the galaxy 
demography 
in clusters, generate the path of model galaxies in 
the CMD in presence of   mergers, discuss 
the Red Sequence, Green Valley and Blue Cloud, and finally examine the star formation rate versus galaxy mass 
relationship.
In Section \ref{conclusions} we draw some general conclusions 
and outline the plans for future studies.
Finally, in Appendices \ref{SSP_IMF_COL} and \ref{app_galnum_analysis}
we present a few technical details in form of tables and figures that 
may be of general interest.

\begin{table}
\begin{center}
\caption{ The Color Magnitude Relation  $(B-V)$ = $ (A \pm \Delta A)$  $M_V$ + $(B\pm \Delta B)$ of the Red 
Sequence for a selected sample of galaxy clusters from the WINGS Survey \citep{Valentinuzzi_etal_2011}. }
\label{cmr_slope}
\begin{tabular}{|c c c c c c| }
\hline
  Name  & Number  & $A$      &$(\pm \Delta A)$ &  $B$       &$(\pm \Delta B)$  \\
  \hline
A119    &   116      & -0.042   &      0.007      &    +0.051  &   0.142 \\
A168    &   118      & -0.064   &      0.013      &    -0.458  &   0.266 \\
A754    &   118      & -0.042   &      0.006      &    -0.011  &   0.114 \\
A970    &   117      & -0.044   &      0.010      &    +0.046  &   0.198 \\
A2399   &   116      & -0.051   &      0.008      &    -0.198  &   0.169 \\
A2457   &   114      & -0.045   &      0.009      &    -0.022  &   0.181 \\
A3158   &   122      & -0.042   &      0.006      &     0.033  &   0.128 \\
A3376   &   116      & -0.045   &      0.009      &    -0.176  &   0.180 \\
A3395   &   115      & -0.037   &      0.007      &     0.044  &   0.133 \\
A3809   &   119      & -0.065   &      0.010      &    -0.516  &   0.217 \\

\hline
\end{tabular}
\end{center}
\end{table}

\begin{figure}
	\centering{
		{\includegraphics[width=9.cm, height=9.0cm]{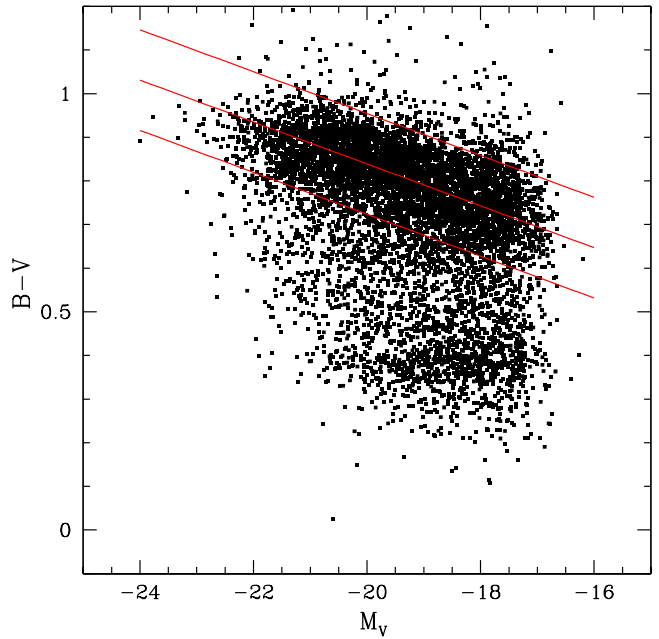}  }    }
	\caption{The total CMD for the 46 galaxy clusters by \citet{Cariddi_etal_2018}. The red lines 
are the 
fiducial CMR derived from the entries for ten clusters of Table \ref{cmr_slope} with 1$\sigma$  uncertainty 
on the zero-point. } 
	\label{total_CMD}
\end{figure}

\section{Observational data and CMR}\label{obser_data}

In recent times, large optical and spectroscopic databases for the galaxy content of nearby clusters
have become available  thanks to the WIde-field Nearby Galaxy-cluster Survey (WINGS) of  
\citet{Fasano_etal_2006} and \citet{Varela_etal_2009} and companion OMEGA-WINGS extension of 
\citet{Gullieuszik_etal_2015} and \citet{Moretti_etal_2017} for a number of clusters in the 
redshift interval ($0.04 \leq z \leq 0.07$).
All this material has been subsequently examined by 
\citet{Cariddi_etal_2018} with particular attention to the problems of the accurate 
determination  of the stellar light  and the 
stellar mass profiles of galaxy clusters.  They measured and examined more than 7,000 galaxies in 46 clusters 
 for which they provide the absolute V and B magnitudes,  the  morphological type according to the  
classification system  RC3  by   
\citet{deVaucouleurs_etal_1991}, \citet{Corvin_etal_1994} and the extension made by 
\citet{Fasano_etal_2012}, and the four different estimates of the star formation rates (SFR) by
\citet{Fritz_etal_2007,Fritz_etal_2011}. 
The issue of membership of the galaxies to the clusters 
under consideration has been addressed 
and examined by \citet{Cava_etal_2009} to whom we refer for all details.
In this study we have considered all the 46 clusters studied 
by \citet{Cariddi_etal_2018}, however with particular attention to a subgroup of ten clusters for which  
more data were available.  The list of ten clusters under 
examination is given in Table \ref{cmr_slope} which contains the 
cluster identification (column 1), the total number of galaxies in each cluster (column 2), and the 
estimated color-magnitude 
relation (CMR) expressed  by $(B-V)$ = $ (A \pm \Delta A)$ $M_V$ + $(B\pm \Delta B)$ (slope and zero point, columns 3 
through 6): data related to the Red Sequence were derived by \citet{Valentinuzzi_etal_2011}. 
 From the CMR of ten clusters we derive 
the fiducial  mean CMR: $(B-V) =  -0.048 M_{V}  - (0.121 \pm 0.173)$ 
whose zero-point is estimated at 1$\sigma$ uncertainty.
This should roughly indicate the slope,  width and location of the  Red Sequence in the CMD. The
$B-V$ color of each galaxy was derived here using the values of the B and V Sextractor 
\verb|mag_auto| given by \citet{Gullieuszik_etal_2015} corrected for galactic extinction and K-correction. 
These are not aperture magnitudes but extrapolated values providing the total galaxy 
luminosity. The colors are therefore average values of the whole visible galactic components.
The cumulative CMD  for all galaxies 
in the full sample of clusters is  shown in Fig. \ref{total_CMD} where the Red Sequence, the Green Valley, 
and the Blue Cloud are well evident. 
In this CMD we also plot the fiducial CMR for the ten best clusters (the red lines).    

\begin{figure*}
	\centering{
		{\includegraphics[width=8.5cm]{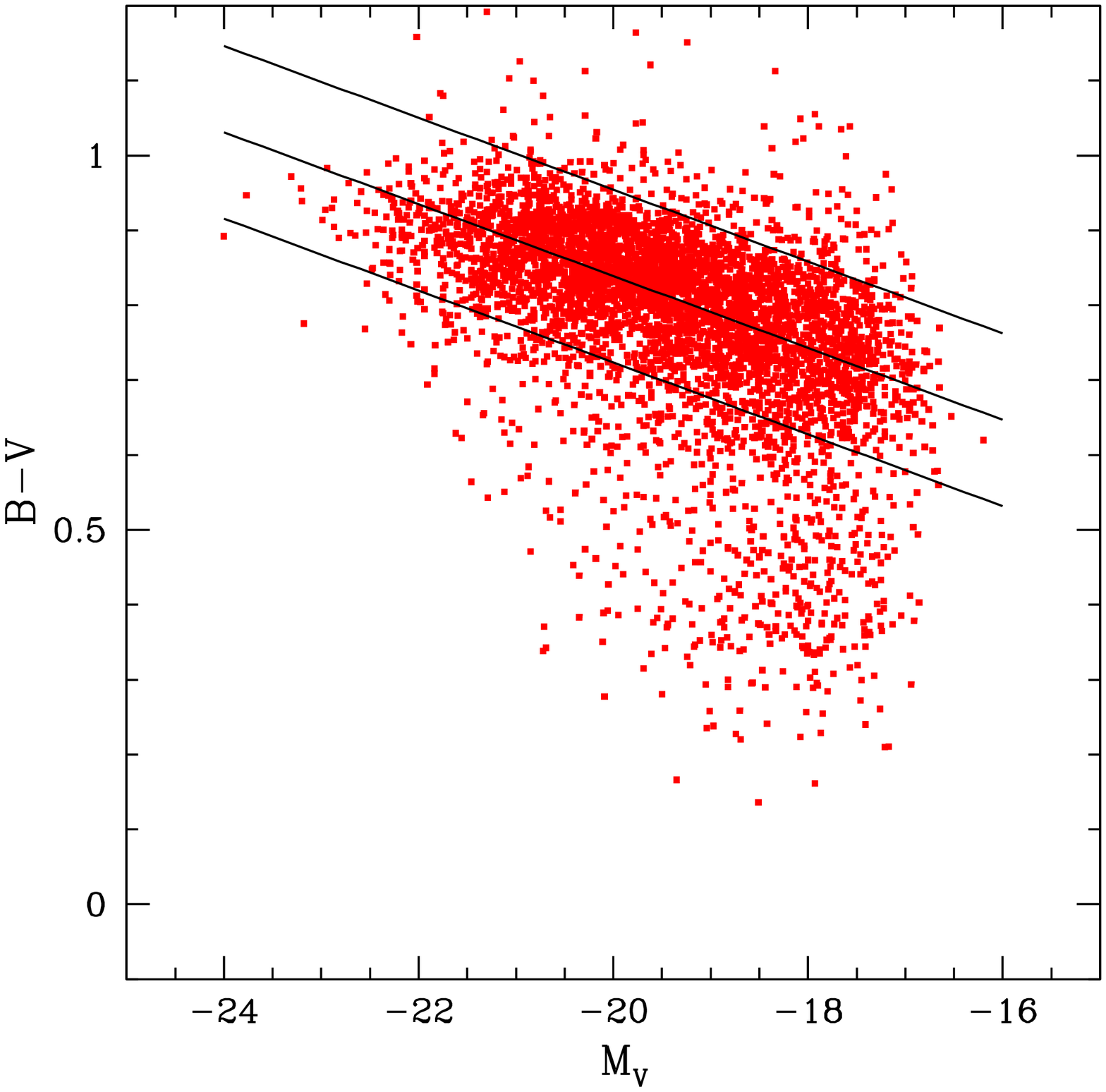}  } 
		{\includegraphics[width=8.5cm]{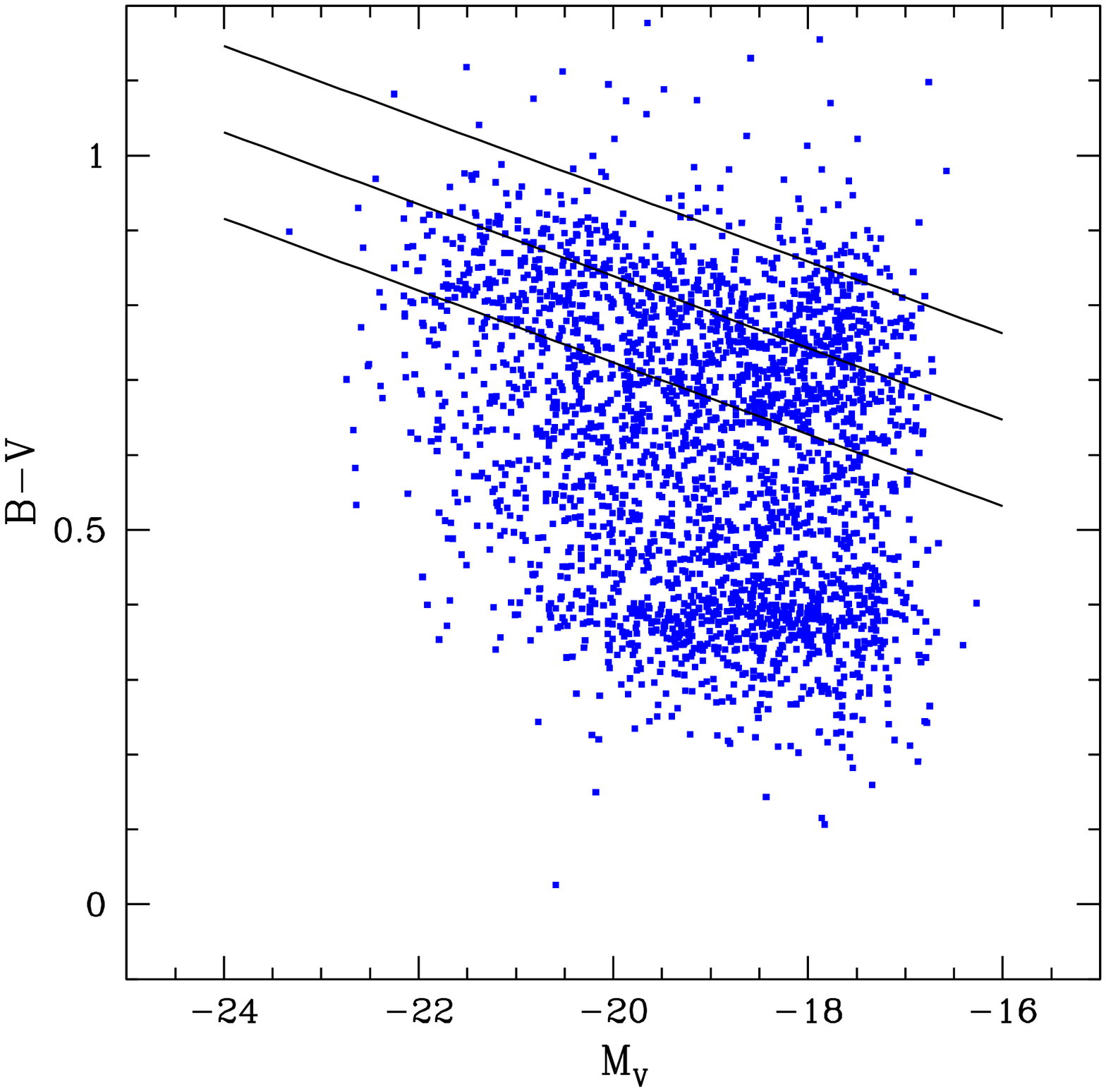} }   }
	\caption{\textsf{Left Panel}: The CMD for early-type galaxies (from cD through S0/a) of all the 46 
		clusters of \citet{Cariddi_etal_2018}.The black lines are the fiducial CMR of the ten 
                clusters listed in Table \ref{cmr_slope} with 1$\sigma$  uncertainty on the zero-point.
		\textsf{Right Panel}: the same as in the left panel but for late-type 
		galaxies from Sa through  cI.}
	\label{all_clu_morph}
\end{figure*}

A few remarks are worth here: (i) The Red Sequence contains a few objects with unusual red  color 
($B-V \geq 1.1$).  Their occurrence could be due to poor photometric data, wrong membership, and/or very high 
reddening. Since they are very few in number compared to the total, we simply drop them off the sample. 
The sample of ten 
reference clusters has been cured for the presence of these very red objects.
(ii) Roughly speaking,  the Green Valley is confined in the color range $0.6 < B-V < 0.8$, has nearly 
the same mean slope of the Red Sequence and  is populated by fewer objects. 
The scarce population of the Green Valley suggests that galaxies in it are in rather quick 
evolutionary phases of their stellar content.  (iii) Finally the Blue Cloud,
customarily associated to galaxies with ongoing star formation (e.g. spirals and robust dwarfs), falls below the Green 
Valley and has a rich population. (iv) The sharp magnitude boundary at $M_V\simeq -17$ is merely due to observational 
detection limits of the survey and therefore is less of a problem. However, one should keep this in mind when comparing 
theoretical predictions to observational data at the low luminosity side of the CMD.

Now we separate galaxies of the full sample in two groups according to their morphological type:
the first group named the \textit{early-type galaxies} contains all the objects with RC3 type in $[-6,0]$, whereas 
the second group named the \textit{late-type galaxies} are objects with RC3 type in $(0, 11]$
\citep[][Fasano et al., in preparation]{Fasano_etal_2012}.
The two cumulative CMDs   are separately shown in the left
and right panels of  Fig. \ref{all_clu_morph} together with the fiducial CMR 
as a reference (the solid black lines). 

\begin{figure}
	\centering{
		{\includegraphics[width=8.5cm]{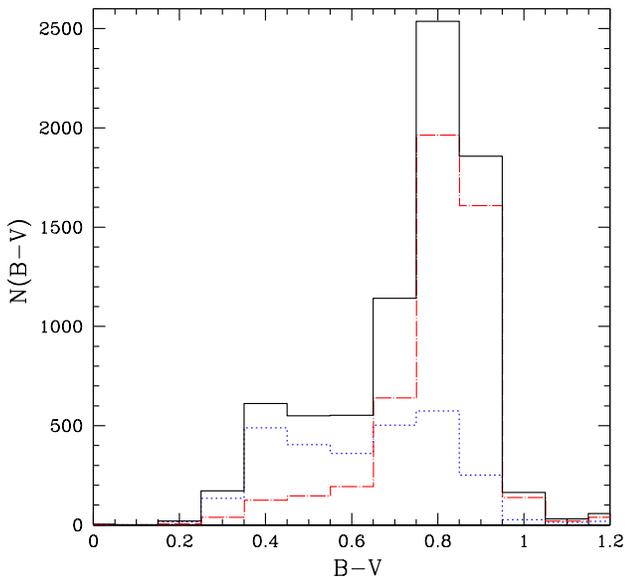}   } }
	\caption{Histograms of the number vs $B-V$ color for all the galaxies of the full sample.  The 
		dashed red line is for early-type galaxies CM (from cD through S0/a); the dotted blue line is for late-type galaxies 
(from Sa through Ic); finally the solid black line is for the two groups of galaxies together.
Rotation of the original CMD has been applied, so that the projection onto the $B-V$ axis is parallel to 
the direction of the the mean CMR.  }
	\label{histo}
\end{figure}

Looking at  CMD for the two subgroups, contrary to what one naively could expect, the two distributions do largely overlap. 
Red spirals have been known for long time, they were called anemic spirals by \citet{van_den_Bergh_1976}, thought to be 
endemic to clusters of galaxies and recently rediscovered by the Galaxy Zoo collaboration \citep{Masters_etal_2010}. 
How  can it be that a spiral galaxy has the same red color of an early-type? Misclassification? Very strong reddening? 
Most likely, in addition to classically referred factors like age and metallicity, both concurring to making a galaxy red, there 
is something else extinguishing  star formation even in spirals and not only in early-type galaxies. The reverse 
side of the coin is that early-type galaxies deeply overlap the region that is typically populated by spirals and 
irregulars, making several early-type galaxies have same colors as late-type galaxies. To cast light on this
issue, we derive the number distribution in color for the whole sample of galaxies. 
In order to better separate the two groups and better put in evidence  the Green Valley,  we apply 
a rotation of  the CMD plane around a central point determined by visual inspection of the data (coordinates are
$B-V =0.70$ and $M_V=-20$) so that the projection on the $B-V$ axis is nearly parallel to the direction of the 
Green Valley, which in turn is nearly parallel to the Red Sequence. The choice of the center of rotation is not an 
severe issue and of course the rotation angle is very small. The results are shown in Fig. \ref{histo}.  
The comparison with the same diagram but without rotation (not shown here) soon 
clarifies that rotation does not affect our conclusion: the sample of early-type galaxies overlaps that of the late-type ones,
and viceversa. It implies that the three loci in the CMD have boundaries not so strictly defined, with 
the Green Valley, in the middle of Red Sequence and Blue cloud, being populated by objects independently of their morphology.

It is worth recalling that while the red peak is quite robust, the height of the blue peak may be affected by a certain 
degree of incompleteness in the number of low luminosity galaxies. In any case, the key point we want to make here is 
that the two distributions greatly overlap, thus lending some support to the idea we would like to advance and explore 
in this study. In the  environment of a cluster, spiral galaxies may 
lose their gas and soon after become red objects, whereas early-type galaxies may occasionally acquire some gas and 
by reviving 
their star formation activity becoming temporarily blue objects. Because of the gravitational confinement of the galaxies 
inside a cluster the above picture should be more probable than in the case of field galaxies, for which 
an evolutionary history in isolation should be more likely. Finally, we would like to note that the CMR for the 
ten best clusters is nearly indistinguishable from the  
one displayed in Fig. \ref{total_CMD}. In the following when comparing theory with observations we will make use of 
the former for the sake of clarity.

As a final remark, we would like to remind the reader that in this study we consider each 
galaxy cluster as a whole leaving aside the possible  radial dependence of the galaxy content (i.e. 
morphological type, galaxy mass, mean star formation rate of each galaxy, etc...) and the presence of diffuse 
intra-cluster medium that could affect the evolution of a generic galaxy via the possible effects of gas 
stripping and harassment. Each CMD is built by plotting the integrated colors and magnitudes of each 
galaxy without considering the geometrical structure of the cluster itself.
   
\section{Baryonic Matter in Dark Matter Halos: Galaxy Models} \label{galmod}

In this section we briefly introduce the models we have used  to describe galaxies in terms
of chemical evolution and photometric.  
For the sake of our investigation, we make use of both single stellar populations (SSPs), with suitable 
chemical composition (the ``metallicity'' $Z$) and age, and of simple model galaxies based on 
the one-zone description,
which is proven to successfully describe the star formation and metal enrichment histories 
and the photometric properties of galaxies in cosmological contexts
\citep[in addition to following Subsections and works therein referred to, we refer the reader to][for full 
descriptions of the models]{Chiosi1980,Tantaloetal1996,Portinarietal1998,Chiosi_etal_2017}.

\subsection{Single Stellar Populations: spectra, magnitudes and colors}
  
The classical tool for this kind of studies is the evolutionary population synthesis (EPS)
introduced long ago by   \citet{Tinsley1968,Tinsley1980} and ever since developed to a high 
degree of sophistication by many authors \citep[][and references]{Bressan_etal_1994,Tantalo_etal_2010,
Conroy_2013,Conroy_etal_2014,Pasha_etal_2018,Dahmer_etal_2018}. 

In the following we adopt the method developed by \citet{Bressan_etal_1994}. The monochromatic 
flux generated by the stellar content of a galaxy of age $T_G$ is defined as:

\begin{equation}
F_{\lambda}(T) = \int_{0}^{T}\Psi[t,Z(t)]sp_{\lambda}[\tau ',Z(\tau ')]\,dt
\end{equation}

\noindent where $\Psi[t,Z(t)]$ is the SFR at the  age $t$ and metallicity $Z(t)$ 
(chemical content in general),
$sp_{\lambda}[\tau ',Z(\tau ')]$ the integrated monochromatic flux of a single stellar population (SSP), 
a coeval chemically homogeneous assembly of stars,  with given age 
$\tau '\equiv T_G-t$
and $Z(\tau ')$. The flux of a SSP is in turn given by

\begin{equation}
sp_{\lambda}[\tau ',Z(\tau ')] = \int_{M_l}^{M_u} \phi(M)f_{\lambda}[M,\tau ',Z(\tau ')]\,dM
\end{equation}

\noindent where $\phi(M)$ is the stellar initial mass function (IMF) and $f_{\lambda}[M,\tau ',Z(\tau ')]$
is the monochromatic flux of a star of mass $M$ inside the SSP. $M_l$ and $M_u$ denote the mass
range within which stars are generated by each event of star formation. The metallicity dependency
of SFR $\Psi(t,Z)$ is customarily neglected and so it is for the time and metallicity of the IMF.
Sources of stellar tracks, isochrones, spectra, and SSPs in different photometric systems are from
\citet{Bertellietal2008,Bertellietal2009,Tantalo05}.

In this paper we adopt the IMF of \citet{Salpeter1955}, which is $\Phi(M)\propto M^{-2.35}$. It is
normalized by choosing the fraction $\zeta$ of stars more massive $M_n$, i.e. the mass interval
most contributing to chemical enrichment over the whole Hubble time:

\begin{equation}
\zeta=\frac{\int_{M_n}^{M_u}\phi(M)\,M\,dM}{\int_{M_l}^{M_u}\phi(M)\,M\,dM}
\end{equation}

\noindent where $M_n$ and $M_u$ are fixed to $M_n\simeq 1\, M_{\odot}$ and $M_u=100 \,M_{\odot}$,
and $M_l$ follows from the normalization.

The database of SSPs  in use has been computed by
\citet{Tantalo05}
adopting the library of stellar spectra assembled by
\citet{Girardi_etal_2002}, for ages between $10^{7}$ yr and $13 \times
10^{9}$ yr, metallicities from $Z = 0.0001$ to $0.1$, and for seven
different prescriptions of the IMF.
Fluxes, and hence magnitudes, are calculated for the total mass of a
SSP:

\begin{equation}
\label{mssp}
\int^{M_{u}}_{M_{l}} M \Phi(M) dM = M_{SSP}\,.
\end{equation}

\noindent
The total mass of the SSP depends on the IMF and the values of $M_{l}$ and $M_{u}$.

The SEDs of SSPs mirror the SED of all component stars along the isochrones and how stars of different 
mass populate an
isochrone. Therefore, the magnitudes and colors of SSPs depend on a complicated game among all those 
players and consequently
differences among the many sources of SSPs in literature are expected \citep[see][]{Tantalo05}.
In this study we adopt the wide tabulations of SSPs
magnitudes as function of age and chemical composition and IMFs 
calculated by \citet{Tantalo05} for a large grid of 
photometric systems going from: classical Johnson, Bessell-Brett, 
Hubble Space Telescope (NICMOS, WFPC2, ACS), to the Sloan Digital Sky Survey (SDSS), 2MASS, and others. 
All details about the photometric pass-bands and definition of the magnitudes 
systems can be found in \citet{Tantalo05,Tantalo_etal_2010}.

It is worth recalling here that in recent times evolutionary tracks with chemical 
composition including the enhancement of $\alpha-$elements (like Carbon, 
Oxygen, Titanium etc.) have been calculated \citep[see e.g.][and references]{Salasnich2000,Fuetal2018}.
Therefore, the parameter $[\alpha/Fe]$ should 
be added to those  characterizing the chemical pattern of elements, and SSPs including its effect
should be considered. However, as the tabulations of SSPs with $\alpha-$enhancement in the Padua
database to our disposal are not large enough for our purposes, we do not include this effect in our analysis. 
In any case, they will  induce second order effects on the dominant ones given by the metal content $Z$.

Finally, we would like to comment that as far the integrated spectro-photometric properties of complex 
stellar populations (stellar clusters and galaxies) are concerned magnitudes and colors do not much depend 
on the underlying IMF, whereas the mass-to-light ratios are sensitive to this parameter. 
This can be easily understood by means of the following simple arguments. Excluding the more intrigued case of very 
young SSPs in which massive stars are present, at increasing age very soon SSP reaches the regime in which most 
of the light come 
from evolved stars in the post main sequences stages (red giants, clump and/or horizontal branch stars, 
asymptotic giant branch stars), the rest from main sequence objects at or just below the turn off. The 
remaining main sequence stars play a minor role.   On the other hand the post main sequence stars of a SSP 
span a rather narrow range of masses. Therefore, to induce a significant effect on the photometric properties 
(magnitudes  and colors), the slope of the IMF should vary a lot over the post main sequence mass interval. 
This is not the case for all IMF in literature. 
Therefore, the use of the 
Salpeter IMF is fully justified and adequate to our purposes. 
However, for the sake of clarity and comparison, in Appendix \ref{SSP_IMF_COL} 
we provide a 
summary of  the SSP magnitudes and colors limited to  the solar composition (metallicity $Z=0.019$), the 
Johnson-Bessell-Brett system,   and three different very popular IMFs 
\citep[][and references]{Salpeter1955,Chabrier_2014,Kroupa2008}.
All other details can be found in \citet{Tantalo05}. The use of the Salpeter IMF is also supported by 
the study of field and cluster ellipticals at intermediate redshifts by \citet{Schade_etal1996}.

\subsection{Simple Galaxy Models}\label{simplegalmod}

In this study, we  adapt the multi-shell spherical models of galaxies developed by  
 \citet{Tantaloetal1998}  
to the one-zone description, which is fully adequate to our purposes.
These galaxy models include  many important physical phenomena, 
for instance gas heating by supernova explosions (both type Ia and type II),
stellar winds, gas cooling by radiative emission, galactic winds, and the presence of 
DM in shaping the gravitational potential. In brief, the galaxy of total mass $M_G$ 
is made by the BM component with mass
 $M_{BM}$ and DM component with mass $M_{DM}$, according to 

\begin{equation}
M_{G}(t) = M_{BM}(t) + M_{DM}
\label{sum_bm_dm}
\end{equation}
The DM mass is supposed to be present from the very beginning with a suitable spherical distribution
so that its contribution to the gravitational potential is fixed. The BM mass is supposed to  be originally in form of 
gas, to flow in at suitable rate, to build up its own gravitational potential and, when physical conditions allow it,
 to transform into stars. This kind of 
model, originally developed by \citet{Chiosi1980}, is named the infall model, the essence of which resides 
in the gas accretion into the central region of the
proto-galaxy at a suitable rate (driven by the timescale $\tau$)  and in the gas consumption by
a Schmidt-like law of star formation. The gas accretion and consumption coupled together give rise to a
time dependence of the SFR closely resembling the one resulting from the N-body
simulations \citep[e.g.][]{Chiosi_Carraro_2002,Merlin06,Merlin07,Merlin2012}.
At this point we warn the reader that, throughout this work, whenever we use the expressions formation time
(redshift) or age of a galaxy we refer to the beginning of baryonic history; the viceversa is also true.

At any time $t$ the baryonic mass $M_{BM}$ is given by the sum 

\begin{equation}
M_{BM}(t) = M_g(t) + M_*(t)
\label{mass_gas_stars}
\end{equation}

\noindent where $M_g(t)$ is the gaseous mass and $M_*(t)$ the star mass. At the beginning, both the gas and the
star mass in the proto-galaxy are zero $M_g(t=0)=M_*(t=0)=0$. The rate of BM (and gas in turn) mass  accretion is  
driven by the accretion timescale $\tau$ according to

\begin{equation}
  \frac{dM_{BM}(t)} { dt} = {{M}}_{BM,\tau}\exp(-t/\tau)
\label{inf}
\end{equation}
where $\tau$ is the accretion timescale and ${M}_{BM,\tau}$ a constant with the dimensions of [Mass/Time] 
to be determined by imposing that 
at the galaxy age ${T_{G}}$ the total baryonic mass of the galaxy ${M_{BM}(T_{G})}$ is reached:

\begin{equation}
  {M}_{BM,\tau}  = \frac{M_{BM}(T_{G})} {\tau [1 - \exp(- T_{G}/\tau)]}\,.
\label{mdot}
\end{equation}

\noindent
Therefore, integrating the accretion law the time dependence of
${M_{BM}(t)}$ is

\begin{equation}
 { M_{BM}(t) =  { \frac{M_{BM}(T_{G})}  {[1-\exp (-T_{G}/\tau)] }    }
                      [1 - \exp(-t/\tau)]  }\,.
\label{mas-t}
\end{equation}
The timescale $\tau$ parameterizes the timescale over which the present-day mass $M_{BM}(T_{G})$
is reached, because it is related to the average rate of gas cooling and it is expected to depend on the mass
of the system. In this scheme the total mass of a galaxy at the present time is
$M_G = M_{DM} + M_{BM}(T_G)$.

While in the earliest stages of a galaxy $M_{BM}$ can be considered equal to $M_g$,
soon after the gas cools down and star formation begins. We model
this throughout the whole life of the galaxy with a Schmidt-like law of star formation 
\citep{Schmidt1959}:

\begin{equation} \label{eqschmidt}
\Psi(t) \equiv \frac{dM_*}{dt} = \nu M_g(t)^k
\end{equation}

\noindent where $k$ regulates the dependency of SFR on gas content (classically
it can be linear or quadratic: we fix it as $k=1$) and $\nu$ is the efficiency parameter
of the star formation process. 

In the infall model, because of interplay between gas accretion  and consumption, the
SFR starts low, reaches a peak after a time approximately equal to $\tau$ and then slowly declines.
The functional form that could mimic this behavior, with systematic variation with mass, is the
delayed exponentially declining law:

\begin{equation}
\Psi(t) \propto \frac{t}{\tau}\exp\left(-\frac{t}{\tau}\right)\;.
\end{equation}

\noindent The Schmidt law in eqn. (\ref{eqschmidt}) is therefore the link between gas accretion by 
infall and gas consumption  by star formation. 

As a whole, this kind of approach stands on a number of observational and theoretical arguments among which we recall
(i) the parameters $\nu$ and $\tau$ can be related to morphology
\citep{Buzzoni2002} and to the presence of ongoing star formation activity inside
observed galaxies \citep{Cassara_etal2016}; (ii) the aforementioned quantities can be easily
tuned in order to fit observational data, and also complex
phenomena that would affect the rate of gas cooling, such as active galactic nuclei (AGN), 
can be empirically taken into account without going into detail \citep[see e.g.][]{Chiosi_etal_2017}.

\subsubsection{Energy feedback, gas heating-cooling, galactic winds }

Long ago \citet{Larson1974} postulated that the present-day CMR
of elliptical galaxies 
\citep[see also ][and references]{Terlevich_etal_01}
could be the result of galactic winds powered by supernova explosions: a long
series of chemo-spectro-photometric models of elliptical galaxies stemmed from
this idea \citep[][and references therein]{Tantaloetal1996, GibsonMatteucci1997,Tantaloetal1998}.
In brief, gas is let escape from the galaxy and
star formation is supposed to halt when the total thermal energy of the gas
equates its gravitational binding energy.

The thermal energy of the gas is mainly due to  four contributions, namely Type Ia and
II supernovae,  stellar winds from massive stars, and AGNs (these latter however are not included here; see above):

\begin{equation}
E_{th}(t) = \sum_{J} E_{th}(t)_J 
\label{Eth_tot}
\end{equation}

\noindent
where $J$ indicates the various sources of heating \citep[see][for all details]{Tantaloetal1998}. 
Suffice to recall here that the injected energies incorporate  the cooling laws of 
supernova remnants and stellar winds by radiative cooling processes. 
The condition for the onset of the wind  is

\begin{equation}
E_{th}(t) \geq \left| \Omega_{g}(t)\right|
\label{eth_omg}
\end{equation}

\noindent
where $\Omega_{g}$ is the gravitational potential energy of the gas. To determine
this latter one needs to know the total gravitational potential of DM and BM together $\Omega_{DM+BM}$. 
We strictly 
follow the assumptions and formalism of \citet{Tantaloetal1998} to whom the reader should refer for all details.  
We only mention here that $\Omega_{DM+BM}$ is a function of the ratios 
${M_{BM}/M_{DM}}$ and  ${R_{BM}/R_{DM}}$, with obvious meaning of the symbols.  
In conditions of mechanical equilibrium and with 
the  current estimates of $M_{DM}$ and $M_{BM}$ of the 
$\Lambda$-CDM cosmogony (see below), both ratios are  equal to 0.16. With these values, the DM gravitational 
potential does not exceed 0.3  of that by the sole BM.

\begin{table}
\begin{center}
\caption{ The key parameters of the model galaxies:  mass, $\nu$ and $\tau$.
 Masses are in solar units.}
\label{Model_Galaxies}
\begin{tabular}{|c| c| r| r| r| r| r| r| r| }
\hline
Type       &$\log M$ & 7   & 8  & 9  & 10 & 11 & 12  & 13  \\
\hline
ET$_{ms}$  &$\tau$   & 6   & 5  & 4  & 3  & 2  & 2   & 1   \\
ET$_{ms}$  &$\nu$    & 10  & 10 & 10 & 10 & 10 & 10  & 10  \\
\hline
ET$_{gw}$   &$\tau$   & 6   & 5  & 4  & 3  & 2  & 2   & 1   \\
ET$_{gw}$   &$\nu$    & 10  & 10 & 10 & 10 & 10 & 10  & 10   \\
\hline
LT$_{ms}$  &$\tau$   & 10  & 8  & 6  & 4  & 3  &  2  & 1    \\
LT$_{ms}$  &$\nu$    & 1   & 1  & 1  & 1  & 1  &  1  & 1    \\        
\hline
\end{tabular}
\end{center}
\end{table}

\begin{figure}
	\centering{
		{\includegraphics[width=9.cm, height=9.0cm]{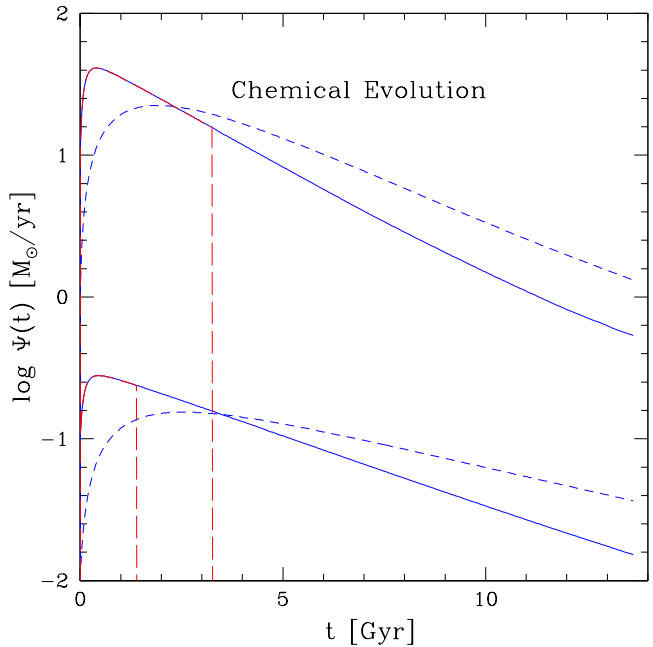}  }    }
\caption{The SFR (on logarithmic scale) versus age of model galaxies. The blue solid lines indicate 
models labeled ET$_{ms}$, the blue short-dashed lines  
those labeled  LT$_{ms}$, and the long-dashed lines  those named ET$_{gw}$.  
Two model galaxies are displayed with different mass, i.e. $10^9$ (bottom) and $10^{11} M_\odot$ (top). }
	\label{sfr_age}
\end{figure}

\subsubsection{SFR and galactic winds}
In the wind model of \citet{Larson1974}, when
condition of eqn. (\ref{eth_omg}) is verified, all the gas freely escapes from the galaxy so that  
further star formation does no longer
occur. The evolution of the remnant galaxy is a passive one and all the gas shed by  stars formed in the
previous epochs, either in form of stellar wind or supernova explosion, does no longer generate new stars.
In addition to this, owing to the different gravitational potential well of massive galaxies with respect
to the low-mass ones, the time at which the threshold energy for galactic winds is reached occurs earlier
in low-mass galaxies than in the massive ones. In contrast, more sophisticated treatments of  winds 
with the aid of NB-TSPH models
of galaxy formation and evolution \citep{Chiosi_Carraro_2002,Merlin06,Merlin07,Merlin2012}
have shown that galactic winds are not instantaneous
but take place on  long timescales. Gas heated up by supernova explosions and stellar winds
 and cooled down by radiative processes  not only gradually reaches the escape velocity but
also affects the efficiency of star formation because   the hot gas  is
continuously subtracted. To take this into account in our simple models, we proceed as follows.

First of all, feeling that the cooling algorithm we are using is not as good as the one currently
adopted in NB-TSPH models, we introduce an efficiency parameter $\eta_{th}$ ranging from 0
(no energy feedback)  to 1 (full
energy feedback) and accordingly change the  (\ref{eth_omg}) to the new one
\begin{equation}
\eta_{th} \times E_{th}(t) \geq |\Omega_{g}(t)|
\label{new_th_cond}
\end{equation}

\noindent Second, we change the star formation law of eqn. (\ref{eqschmidt}) redefining the parameter $\nu$
as an effective efficiency given by

\begin{equation}
 \nu_{eff} = \nu \times \frac{|\eta_{th} \times E_{th} - |\Omega_{g}||}{\eta_{th} \times E_{th} + \Omega_{g}}
\label{nu_eff}
\end{equation}

\noindent where $\nu$ is the usual efficiency. By decreasing the efficiency of star formation at 
increasing $E_{th}$
we intend to mimic the fact that hot gas is likely less prone to generate stars by gravitational collapse.
 As a consequence, the threshold stage for the onset of
galactic winds may occur much later in time or even avoided at all. Less gas is turned into stars as
if part of the gas is continuously escaping from the galaxy. For the aims of this study, we 
tune the parameter $\eta_{th}$ in such a way that the SFR is reduced  but never extinguished during the 
whole galaxy lifetime. We refer to these as models with ``modulated star formation'' (ms). 
In parallel we also calculate 
models with the standard prescription for the galactic wind. These are referred as models with 
``galactic wind'' (gw).    

Finally we calculate models with two different combinations of the parameters $\tau$ and $\nu$ 
that are chosen in such a way that (i) the peak of star formation occurs early on in the galaxy life and 
then significantly declines; 
these are named ''Early-Type (ET) models with modulated star formation'' because they are meant to simulate
early-type galaxies; (ii) the peak of star formation occurs at early times but  star formation  
continues at rather high levels of efficiency till the present epoch; 
these models are meant to  simulate either spiral of dwarf galaxies and are indicated as Late-Type (LT). 
In total, we have the following cases:  ET$_{ms}$, ET$_{gw}$, and LT$_{ms}$. The adopted values for the 
parameters 
$\tau$ and $\nu$ are summarized in Table \ref{Model_Galaxies}. The time dependence for the above three cases 
is shown in Fig. \ref{sfr_age}  for two values of the galaxy mass. We highlight that (i) ET$_{ms}$ and 
ET$_{gw}$ models are
identical until gas escapes completely at the onset of winds for the ET$_{gw}$ ones (ii) SFR turns off 
with age increasing with the mass of the system.

\subsubsection{Galaxy Photometry}
The EPS technique and the concept of SSPs allow us to derive the SED of a galaxy under 
any history of star formation and chemical enrichment.  Knowing the SED and the photometric system in use, 
magnitudes and colors as 
function of the age and chemical parameters are straightforward; the procedure  is the same as in  
\citet{Bressan_etal_1994}. Here we limit ourselves to quickly address the issue of the dust.
Magnitudes and colors of both SSPs and galaxies are affected by the presence of dust. In the case
of SSPs, dust is always present
in the  envelopes of  O-type and AGB stars (i.e. at young and oldish ages), so that attenation of
radiation occurs in the stellar nearby. 
in addition, additional reddening affects radiation before it escapes completely from a galaxy
due to interstellar interstellar dust
\citep[see][for exhaustive discussions and technical details]{Piovan_etal_2006a,Piovan_etal_2006b,
Piovan_etal_2011a,Piovan_etal_2011b,Piovan_etal_2011c,Cassara_etal_2013,Salaris_etal_2014,Cassara_etal_2015}. 
In this paper, the self-attenuation of the spectrum of young and oldish SSPs is already included 
in the dataset in use, while for galaxies we proceed as \citet{Bressan_etal_1994} and adopt the simple 
method of \citet{Guiderdoni_RoccaVolmerange_1987}. In brief, they  introduce the effective optical 
thickness of the gaseous component at the wavelength $\lambda$:

\begin{equation}
	\tau_{\lambda} = 3.25(1-\omega_{\lambda})^{0.5}(A_{\lambda}/A_V)_{\odot}[Z(t)/Z_{\odot}]^sG(t)
\end{equation}

\noindent
where $\omega_{\lambda}$ is the albedo of dust grains, 
for which we take the mean value 0.4 
\citep{DraineLee1984}, 
$A_{\lambda}/A_V$  the
extinction law \citep[here we adopt the relation by][]{CardelliClaytonMathis1989}, $s=1.3-1.4$ for
$\lambda\leq2000 \text{ \AA}$ and $s=1.6$ for $\lambda\geq2000 \text{ \AA}$, and $G(t)$  the
gas fraction. The optical depth  $\tau_{\lambda}$ is used to derive  the transmission function for an 
angle of inclination $i$. The transmission function is defined as

\begin{equation}
\frac{1-\exp(-\tau_{\lambda}\sec i)}{\tau_{\lambda}\sec i}
\end{equation}
\noindent
by which we multiply the monochromatic flux of the rest-frame SED of the 
model galaxies. We adopt here the constant inclination angle $i=0^{\text{o}}$.

\begin{table*}
\begin{center}
\caption{ Empirical mass-metallicity relation for SSP-galaxies. Logarithmic masses are in solar units.}
\label{mass_metals} 
\begin{tabular}{|c| c| c| c| c| c| c| }
\hline 
$\log M$   & 7 - 8  & 8 - 9 & 9 - 10 & 10 - 11 & 11 -12  & 12 - 13 \\
$Z_{min}$ & 0.0004 & 0.001 & 0.004  & 0.008   & 0.019   & 0.040   \\
$Z_{max}$ & 0.0010 & 0.004 & 0.008  & 0.019   & 0.040   & 0.070   \\
\hline
\end{tabular}
\end{center}
\end{table*}

\subsection{Passing from  rest-frame to cosmological evolution of SSPs and galaxies}

The SSPs and  galaxies are let evolve from the redshift of formation $z_{f}$ to the present 
$z=0$,  
i.e. from the rest-frame
age $T=0$ Gyr to the maximum age  $T_{G}$ Gyr where  $T_G=T_U(z=0) - T_U(z=z_f)$
with $T_U(z)$ being  the age of the Universe for the
adopted cosmological model.
If for any reason,
we need to change the redshift of galaxy formation from $z_f$ to $z_{f}^{*}$
(keeping unchanged all other input parameters) 
the same models can be used  provided their rest-frame age is simply
limited to the interval from $T=0$ at
$z_{f}^{*}$ to
$T_{G}^{*}= T_{G,(z_{f})} - T_U(z_{f}^{*})$, where $T_U(z_{f}^{*})$ is the age of the universe
at $z=z_{f}^{*}$.
The cosmological model of the Universe is 
the $\Lambda$CDM concordance cosmology, with the parameters
inferred from WMAP-7 data \citep{Komatsu_etal_2011}: $H_0=70.4\,km\,s^{-1} Mpc^{-1}$,
$\Omega_{\Lambda}=0.73$, $\Omega_m=0.27$, $\Omega_b=0.05$, $\sigma_8=0.81$,
and $n=0.97$.

\section{Playing with SSPs: bursts and mergers} \label{results}

In this section we present a preliminary analysis of CMDs of cluster galaxies using only SSPs of 
different ages and metallicities: masses are set equal to those of typical galaxies in the mass interval 
$10^7$ to $10^{13} \, M_\odot$. We call this type of galaxy model {\it SSP-galaxies}. 
To this aim we set up Monte-Carlo simulations of bursts of star formation 
in already existing seed objects of mass, age and metallicity, and of mergers among galaxies of
different age, 
metallicity and mass. We compare the results of both simulations are compared both with observational 
data related to the ten best WINGS clusters described in Section \ref{obser_data}.

\begin{figure}
	\centering{
		\includegraphics[width=8.5cm]{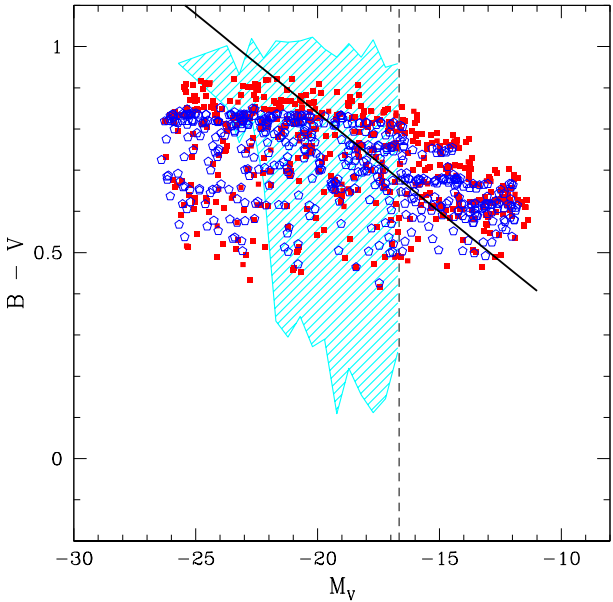}}
	\caption{Simulation of 500 SSP-galaxies with $T_{G,max}=13$ Gyr (red solid squares) and
		$T_{G,max}=6$ Gyr (blue open pentagons), which
		underwent a single burst with $T_{B,max}=5$ Gyr and mass fraction of $5\%$, 
                observed at $z=0$.
		WINGS data related to the ten best clusters (cyan shaded area) are shown for comparison;
		the dashed vertical line represents the faintest object in the WINGS dataset. 
                The black solid line is the fiducial 
                mean CMR already discussed in Sect.  \ref{obser_data}. It is worth noting the discrepancy 
                between the fiducial CMR and the distribution of the red galaxies on the CMD. First for 
                galaxies brighter than $M_V \simeq -20$ the fiducial CMR is steeper than the models,  second 
                for fainter galaxies the slope is nearly the same but the models are redder than the 
                observed galaxies. On one hand the zero point of the fiducial  CMR is uncertain, on the other 
                hand the burst models may not be fully suited to represent real galaxies.  } 
	\label{ccmjburst}
\end{figure}

\subsection{Bursts and mergers}\label{burst_merger}

As a first step, we approximate the complex mix of stellar populations inside a galaxy of mass 
$M_G$ with a SSP 
of suitable age $T_G$,  metallicity $Z_G$ and the same mass.  $T_G$, $Z_G$  are the age and metallicity 
reached at the peak 
of star formation, which according to current galaxy models occur shortly after the formation of the 
galaxy itself
\citep[][]{Chiosi_etal_2017} so that $T_G$ ($z_G$) roughly corresponds to the formation time (redshift).   
With $T_U$  the present age of the Universe 
for the adopted cosmological scenario, then 
$T_{G,z}=T_U - T_G$ is the age (redshift) of the Universe when the galaxy was born.

In the following, we will explore two paradigmatic cases: (i) An already in place galaxy via the  
initial major 
episode of star formation, which later undergoes an additional episode of star formation of minor 
intensity 
(thereafter referred  as \textit{burst} case). With simulations of this kind, we explore the 
consequences of 
adding young stellar components to already evolved stellar assemblies,
in other words we can estimate the effect of a rejuvenation event on an otherwise passively 
evolving stellar system.
This is the analog of simulating either completely wet mergers or a minor stellar activity 
for any internal
reason (eg. re-use of the gas shed by RGB and AGB stars).
(ii)  The other interesting case to consider 
is  the case of a \textit{merger} of two galaxies of different mass, age, and metallicity. This
 would simply tell us how 
the photometric properties of each of the two subsystems added together would give rise to a new 
photometric
appearance of the composed system even in absence of companion star formation. 
This is the analog of a random combination of wet and dry mergers.

\textsf{Bursts}.
The  age $T_G$  of the initial star forming episode is supposed 
to fall in the age range $T_{G,max} > T_G > T_{G,min}$.
Subsequently,  a burst of star formation engaging a certain percentage of the mass (typically up to about
$10\%$) is supposed to occur at any age $T_B$ comprised between $T_{B,max} = T_{G,min}$ and  the present 
time (more precisely
$T_{B,min}=0.1$ Gyr,  the minimum age in the SSP grids).

The rest of the procedure is quite
simple: first, we  take  the fluxes from SSPs of different metallicities,
normalize them to unit of mass (with the Salpeter IMF and $M_{l}=0.1\, M_\odot$ 
and $M_{u}=100\, M_\odot$, $M_{SSP}=5.826\, M_\odot$), and then multiply them
by the mass of the galaxy. Next, we randomize ages and masses of the seed SSP-galaxies together
with the age and mass percentages of the burst episode. To this aim, it is more convenient to express 
the age and masses in 
terms of their logarithms, in order to avoid non-uniform distribution in the randomly chosen values.
The ages (written with lower case symbols to remind the reader that they are expressed as logarithms) 
of the seed galaxies are given by

\begin{equation}
t_{G} = t_{G,max} - r\,(t_{G,max} - t_{G,min})
\label{eqageseed}
\end{equation}

\noindent and those of the bursts by
\begin{equation}
t_{B} = t_{B,max} - r\,(t_{B,max} - t_{B,min})\,;
\label{eqageburst}
\end{equation}

\noindent $r\in(0,1)$ is a random, always different number. Similar procedure is made for the mass 
of the seed SSP-galaxy,
which spans the range $10^7$ to $10^{13}$ $M_\odot$, and the mass percentage $p_B$ of the burst mass with 
respect to the mass of the host galaxy. The percentage $p_B$ goes from 0 to 0.05. Therefore,  
the relative contribution of the two components to the total flux (magnitudes in any pass-band) is 
given by

$$f  = (1 - p_B)\,f_G   + p_B \,f_B$$ 

\noindent with obvious meaning of the symbols.
Finally, since the SSP fluxes (magnitudes and colors) depend on both age and metallicity 
and we know that this latter in turn increases with the galaxy mass, we have taken this into
 account by adopting an empirical 
mass-metallicity relation that is based on chemical models of galaxies and observational data and is 
presented here in Table \ref{mass_metals}. Shortly speaking, metallicity is for simplicity binned 
in terms of logarithmic mass.
  
\textsf{Mergers}. The mass, 
metallicity and age of each galaxy  are derived using the same procedure as above, the only 
difference being that 
the permitted age interval extends now  over nearly the whole Hubble time i.e.  
$T_{G,max} - T_{G,min} \simeq T_{G,max}$.
Denoting with $T_{G,j}$ the  age of the $j-$th component of the merger 
(a single event for simplicity) and using the 
logarithmic notation, the ages  $t_{G,j}$ are  

\begin{equation}
t_{G,j} = t_{G,max,j} - r\,t_{G,max,j} \label{eqageseedmerg} 
\end{equation}

\noindent where $r$ is the random number. 
When  two galaxies merge together, in the resulting 
bigger object 
there are stars from both initial components. At each time, their contribution to the total flux 
is first suitably 
shifted by the age difference between the two components to set up a common clock 
and then weighed by the mass of each component.

To keep our argumentations simple, we will treat mergers as single pairs of galaxies. 
In addition to this, we do
not consider the case of multiple mergers first for the sake of simplicity and second because 
they seem to become inefficient as the clusters reach  virialization 
\citep{Merritt_1988,Richstone_1990}.
In the following we present
some simulations of mergers and their effect on the CMD.

\textsf{Remarks}. In the simulations below, we will use only SSP with solar partition of 
$\alpha-$elements, i.e $[\alpha/Fe]=0$. For each burst, metallicity is for simplicity chosen 
to be equal to that of the seed galaxies: this means that, at variance with
mergers, metallicities will not mix together.

\begin{figure*}
	\centering
	\includegraphics[width=18cm]{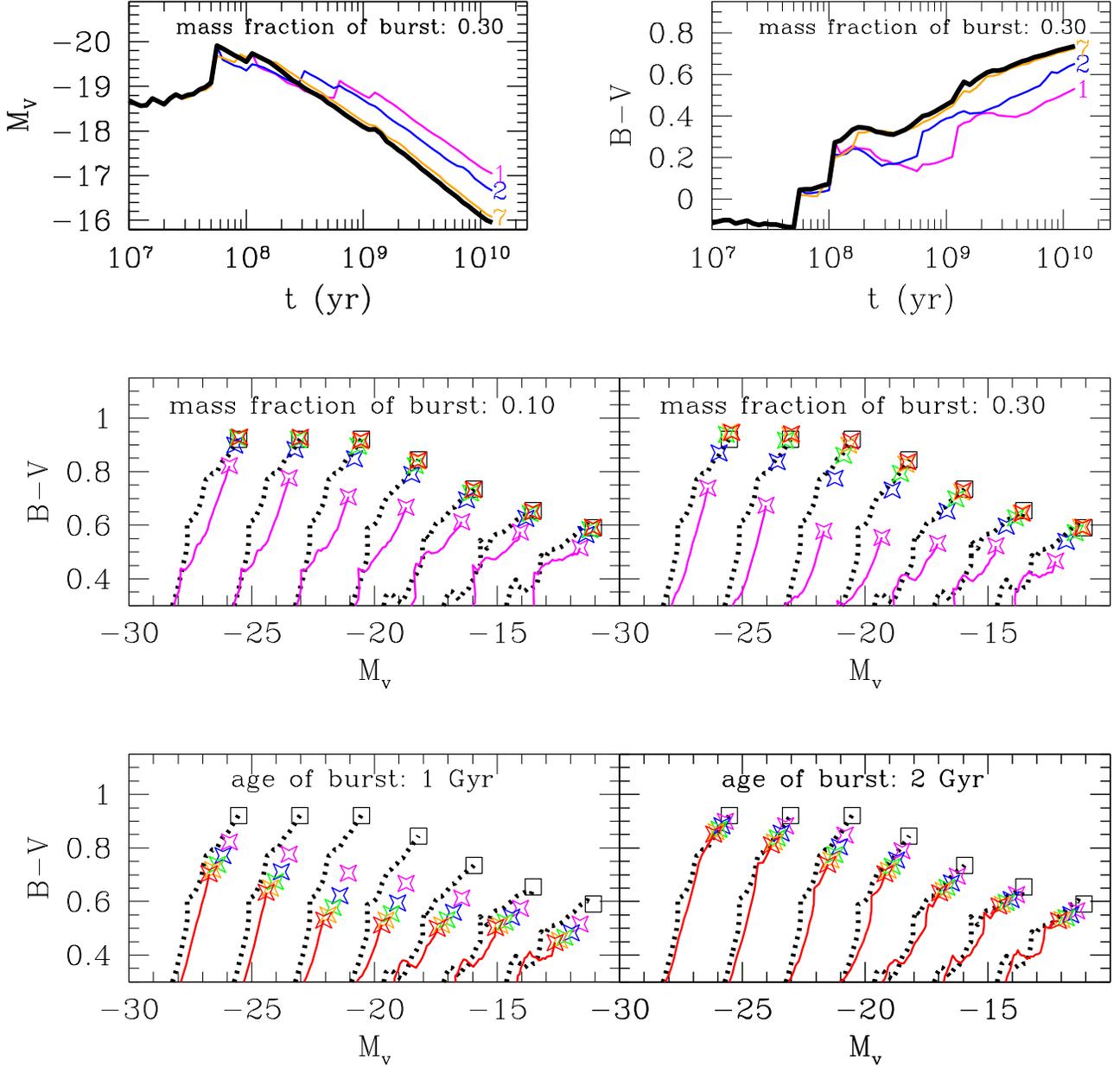}
        \caption{Results of dissection of single bursts inside 13 Gyr-old SSP-galaxies.
	\textsf{Upper Panels}:
	evolution with time of $M_V$ (right) and $B-V$ (left) for the SSP-galaxy 
        with $\log M=9$ ($Z=0.004$) perturbed by
	a single burst with mass fraction of $30\%$; solid thin curves are, in terms of increasing age, 
        magenta (1 Gyr), blue (2 Gyr)
	and orange (7 Gyr); the thick black line is the unperturbed case shown for comparison.
	\textsf{Middle Panels}: CMDs of SSP-galaxies with fixed mass fraction of burst for a set 
        of different ages of bursts, observed at $z=0$:
	open stars indicate the final values reached by each perturbed case; colors are, in terms of 
        increasing age, magenta (1 Gyr), blue (2 Gyr), green (4 Gyr),
	orange (7 Gyr) and red (8.5 Gyr); the path followed by the most extreme case is shown 
        with a thin magenta line; the unperturbed case
	is shown as well (dotted black line ending with an open square).
	\textsf{Lower Panels}: same as Middle Panels, but with fixed age of burst for a set of different
	per cent masses of bursts:	colors are, in terms of increasing percentage,
	magenta (10$\%$), blue (20$\%$), green (30$\%$), orange (40$\%$) and red (50$\%$);
	the path followed by the most extreme case is shown with a thin red line; again
	the unperturbed case is shown as well (dotted black line ending with an open square)}
	\label{evo_burst}
\end{figure*}

\subsubsection{Results}
\textsf{Bursts.} 
In Fig. \ref{ccmjburst} we give a first insight on the impact by
young stars polluting the light emitted by evolved stellar populations:
we simulate single bursts, with $T_{B,max}=5$ Gyr and $T_{B,min}=0.1$ Gyr and a mass
fraction of $5\%$, inside already existing SSP-galaxies with formation interval comprised between
(i) $T_{G,max}=13$ Gyr and $T_{G,min}=5$ Gyr  (red solid squares)  
and (ii) $T_{G,max}=6$ Gyr and $T_{G,min}=5$ Gyr 
(blue open pentagons).  It it worth noting first the broad 
interval for the formation of bulk galaxy in the case (i), and the extreme, largely unrealistic, 
recent age for the bulk activity of case (ii).

After mass and metallicity are fixed,
a complex interplay between ages of galaxy and burst is at work. 
As it will be better shown by means of detailed  chemo-photometric models of galaxies with infall 
of gas,
the generation of  new bright stars strongly enhances the emission in the
blue side of  a galaxy spectrum. Indeed looking at
the two different sets of simulations with SSP-galaxies, it can be already seen that
the main  effect is a shift towards bluer colors and brighter
luminosities in the $V$ band. Even if we are treating single burst events,
Green Valley and Red Sequence are already in place, although the latter appearing by-eye less steep
than the one indicated by the WINGS data (black solid line) in the range of high mass  SSP-galaxies, 
whereas for lower mass
the agreement with observational data is nearly good. Changing $T_{G,max}$ while
keeping fixed $T_{B,max}$ leads to a broadening of the Red
Sequence, with the color $B-V$ becoming redder at increasing ages of the seed galaxies.
Several comments are necessary here. 
First, we note  a large discrepancy 
between the fiducial CMR and the distribution of the red galaxies on the CMD. For galaxies brighter 
than $M_V \simeq -20$ the fiducial CMR is steeper than the models,  whereas for fainter galaxies the slope 
is nearly the same but the models are redder than the observations. Even if the zero point 
of the 
fiducial  CMR is uncertain, this may suggest that the burst models are not  fully suited 
to represent real galaxies. 
Second, very few SSP-galaxies fall into the region  of the Blue Cloud.
 This means that much larger amounts of young stellar populations
should be present inside galaxies of intermediate mass to get the lowest values of $B-V$. 
We could repeat the experiment
by applying more massive bursts, however we prefer not to go further 
because another point of concern is also evident. Third, a large number of SSP-galaxies 
fall into the region of high luminosities ($M_V \lesssim -23 $) and blue colors ($B-V \lesssim 0.8$)
contrary to what shown by the observational data. This latter aspect will be fully clarified while 
dealing  with mergers. Finally the models extend to magnitudes fainter than the observational ones, 
but this is less of a problem because of  the photometric incompleteness of the data.

To cast light on the results presented in Fig.\ref{ccmjburst}  we set  
$r=0$ in eqns. (\ref{eqageseed}) and (\ref{eqageburst}) and fix the burst 
age at some pre-selected values. The results of these experiments are shown   in Fig. \ref{evo_burst}.
In the upper panels, we display  the photometric evolution of a reference SSP-galaxy
 with mass $\log M = 9$, metal content $Z=0.004$ basing on the empirical 
for the mass-metallicity relation,  and age 13 Gyr. We show the unperturbed case (thick black line)
together with three cases of  single bursts with the same fractionary mass ($30\%$) 
and different ages,
namely 1, 2, and 7 Gyr, represented by the magenta, blue and orange
lines respectively. The situation is immediately clear: 
we see that the younger the burst, the greater is the difference in terms both
of magnitudes and colors. Moreover, the 7 Gyr line stands always very close
to the unperturbed case despite the high intensity of the burst: this tells us that an old burst
does not alter significantly the  photometric appearance of an old galaxy. 
At decreasing age of the burst, the effect of this   becomes larger and larger.
Interestingly, a 2 Gyr-old burst
leads to an intermediate state between 1 Gyr and 7 Gyr, confirming that the bulk
of luminosity comes from rapidly evolving, very bright stars. As a final remark,
we warn the reader that slight differences occurring at ages younger than  $10^8$ yr may be
numerical and not strictly physical.

To better understand the consequences of bursts on the magnitudes and colors of a hosting galaxy, 
in the middle
and lower panels of Fig. \ref{evo_burst} we show a set of bursts applied to SSP-galaxies
of different mass (for all the values listed in Table \ref{mass_metals}) at varying the burst intensity 
and age. For each case, the metallicity of the SSP-galaxy and the companion burst vary according to the 
mass-metallicity relation of Table \ref{mass_metals}. For each case, 
the dotted line is the unperturbed SSP-galaxy and the open square shows the
 present-day stage. The solid lines are the most extreme case for the burst either in age (middle panels)
or intensity (bottom panels),  whereas the opens stars are the corresponding end-stages.
Finally, the color code of the open stars  indicates the values taken by the parameter let vary in each 
panel as appropriate and listed in the caption of Fig. \ref{evo_burst}.  

In the middle panels of Fig. \ref{evo_burst} we show what happens for single bursts of different age 
(five values are considered, namely 1, 2, 4, 7 8.5 Gyr) and 
constant fractionary mass (0.1 and 0.3 in the left and right panels respectively). 
 In presence of a burst, the present-day stages (open stars) are bluer and brighter 
 than the unperturbed case. This has important consequences for the  Red Sequence and the Green Valley
that depend on the fractionary mass of the bursts
The Red Sequence  broadens towards bluer colors  in presence either of  massive and relatively young  
bursts 
(fractionary mass from 0.3 and larger and age equal or younger  than 4 Gyr) or less massive but very young 
burst  (mass fraction  lower than 0.20 and age younger than 2 Gyr). 

In the bottom panels we replicate this analysis  but now keeping fixed
the age of bursts. It is quite evident that youngest bursts of any per cent mass  can already 
take into account the
great dispersion discussed in Section \ref{obser_data}.
Bursts with an age of 2 Gyr lead to final values that are not so different with respect
to the unperturbed case: therefore, motivated by what we found in the upper panels for 7 Gyr-old bursts
and in the middle panels for the green open stars,
it is reasonable to say that bursts with age equal or older than, say, 4 Gyr
do not lead to substantial variations on the CMD. 

Finally, recalling  the absence of $\alpha-$enhancement for
these experiments, it can be easily foreseen that taking it into account would likely lead to 
a broader dispersion on the evolutionary path of SSP-galaxies and a larger one in the $B-V$ color. 

\textsf{Mergers.} In the above cases, dealing with bursts implicitly leads to assume
that the seed galaxy is older than the burst component.
From a cosmological point of view, 
 a bound DM halo arises when the density peak height becomes greater
than the local value \citep[e.g.][]{PressSchechter1974,Bond_etal_1991,ShethTormen2002},
therefore, a galaxy can form inside a DM halo when the latter has accreted enough BM
to ignite star formation. There is however a possibility that a galaxy
will not survive for a long time because of cannibalism by other, more massive
galaxies in its nearby surroundings: in this case, gas of the cannibalized galaxy becomes
active fuel for the acceptor. This latter in turn may be  either older or younger
than the former depending on the previous history. In such a case, photometric variations due both to 
stellar activity likely accompanying the merger and the stellar populations already in-situ are  
expected to happen.
This means that together with a mix in ages, a mix in metallicity between different
aged SSP-galaxies is also at work. Therefore, together with bursts mergers are expected to drive 
the dispersion on the CMD, especially inside galaxy clusters.

\begin{figure*}
	\centering{
		{\includegraphics[width=8.5cm]{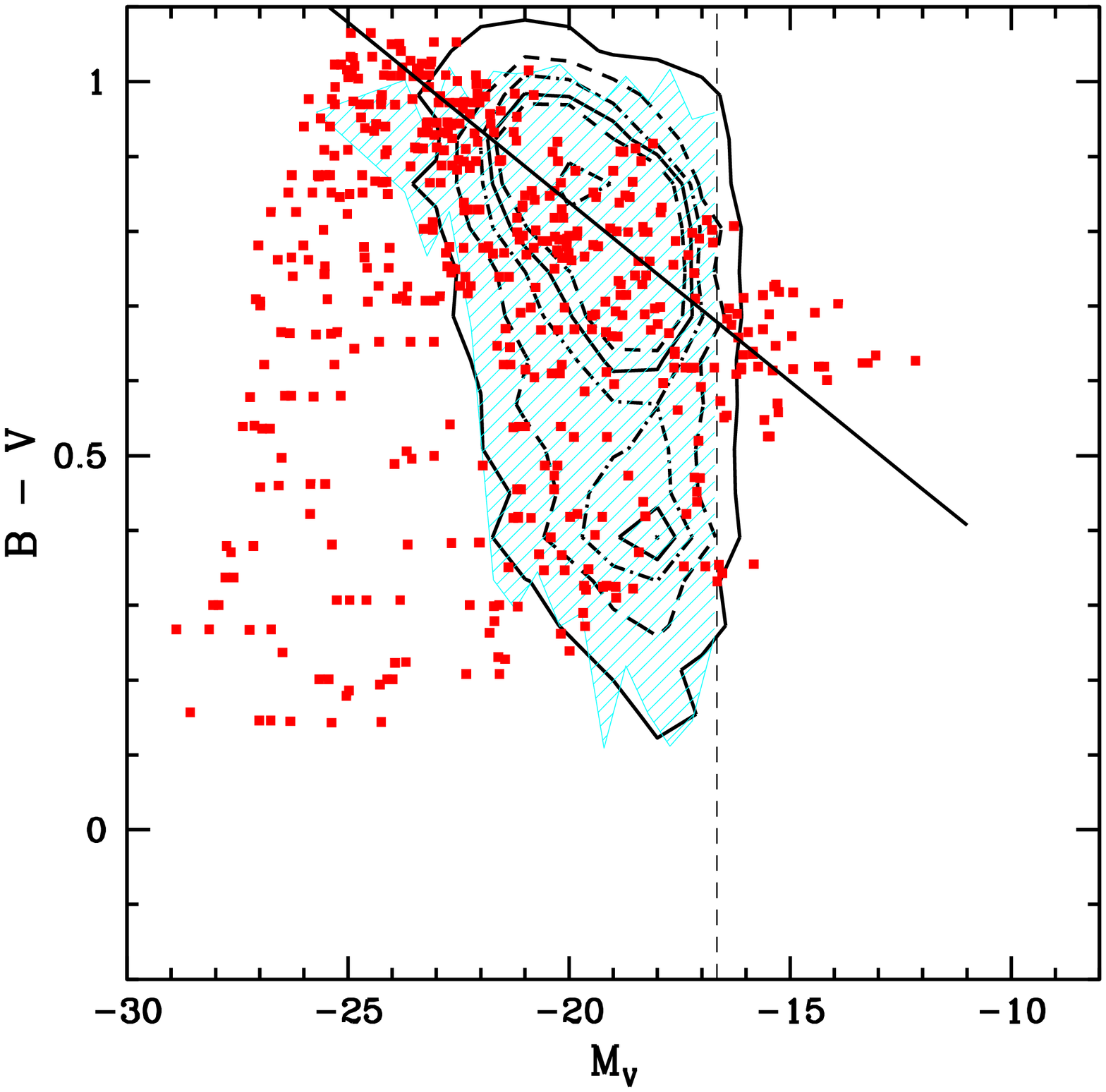}  } 
		{\includegraphics[width=8.5cm]{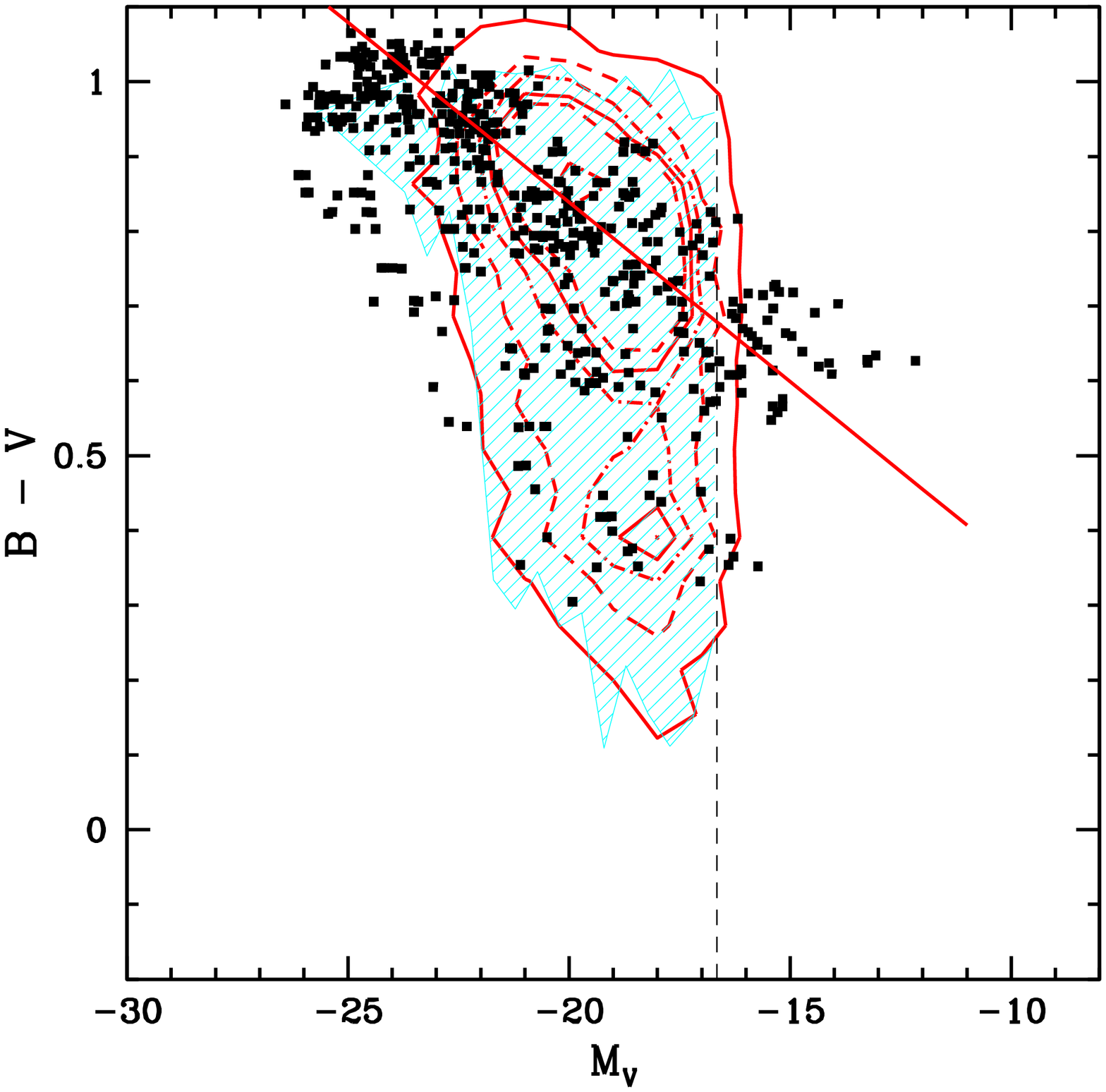} }   }
\caption{\textsf{Left Panel}: Simulation of 500 SSP-galaxies (red squares) with $T_{G,max}=13$ Gyr, which
	underwent a single merger, observed at $z=0$. \textsf{Right Panel}: Black squares represent 
        other 500
	SSP-galaxies simulated with the heuristic filter (the funnel); the black lines delimit
	the range in $B-V$ covered by the theoretical simulations with $\Delta M_V=0.5$. In both panels 
	the WINGS data for the sample of 
	ten clusters (cyan shaded area) are shown for comparison together with the contours of constant 
        relative number of galaxies per elementary area of the CMD, $n(area)/N_{Tot}$. The elementary area is 
        $\Delta M_V = 1.0$ mag $\times$ $\Delta (B-V)= 0.05$ mag. $N_{Tot}=2325$ objects. The contour lines 
        are for $n/N_{Tot}=0.001$, 0.005, 0.01, 0.015, 0.020, 0.050,  moving from the most external one, 
        respectively. In both plots we display the fiducial CMR of Sect. \ref{obser_data} (black line in the 
        left panel, red line in the right panel).  Finally, the dashed vertical line is the limit for the 
        faintest object in this dataset.}
	\label{ccmjmergers}
\end{figure*}

The situation is shown in Fig. \ref{ccmjmergers}: the cyan shaded area 
represents galaxies
inside the WINGS clusters (total number of objects $N_{Tot}=2325$), whereas the red squares are random 
mergers between SSP-galaxies. On the area populated by the observations we also show the contours of constant 
relative number of galaxies per elementary area of the CMD, $n(area)/N_{Tot}$, that is defined by 
        $\Delta M_V = 1.0$ mag $\times$ $\Delta (B-V)= 0.1$ mag. 
For the sake of comparison we plot the fiducial CMR of Sect. \ref{obser_data}.  Finally, the dashed vertical 
line is the limit for the 
        faintest object in this dataset.
 
Using the sole merger hypothesis (left panel of Fig. \ref{ccmjmergers}), the dispersion would be even
much larger than that of the bursts and, in particular, the simulations  would extend also
 to the very blue and bright regions of the CMD. In contrast, these regions  
are avoided by real galaxies (this behavior has also been found  in the case of bursts). 
This implies that a sort of 
selection rule must be at play, evidently breaking randomness and 
forcing real galaxies to avoid the region at the bright side of the Green
Valley and the Blue Cloud. 
This is the photometric counterpart of the long known galaxy formation bias first pointed out
by \citet[][]{Kaiser_1984}. At increasing mass of the collapsing over-density generating a galaxy 
of larger mass or even the cluster itself, clumps of matter that 
already collapsed (and hence formed stars) at previous epochs (higher redshifts) are already there.  
In other words high mass objects are very unlikely to be present in the remote past. Therefore a 
mere random process for mergers cannot be at work. 

In order to take this into account in our simulations, we follow a heuristic simple reasoning: 
we just suppose that the
range of ages admitted for each bin of galaxy masses  gets narrower at increasing mass; this is
equal to say that, instead of simulating encounters between galaxies of purely
random ages, we apply a funnel in order to avoid formation of massive galaxies
at recent times. As a consequence, also  mergers among 
galaxies of comparable high mass are avoided. 
The funnel is represented by the factor $f_M$:

\begin{equation}
f_{M,i} = f_2 - (f_2-f_1)\frac{M_{max}-M_{i}}{M_{max}-M_{min}}
\label{fmfunnel}
\end{equation}

\noindent where $M_i$ is the logarithmic mass of the $i$-th galaxy,
$i=1,\dots,N$ and $N$ is the total number of simulated galaxies,
$M_{min}=6$, $M_{max}=13$ (the logarithm of the minimum and maximum galaxy mass 
in solar units we have adopted), $f_1=1$, $f_2=0.3$. Eqn. (\ref{eqageseedmerg})
then becomes:

\begin{equation}
t_{G,j} = t_{G,max,j} - f_{M}\,r\,t_{G,max,j}\,.
\label{eqageseedmergfunneled}
\end{equation}

\noindent There are a number of reasons motivating this choice. First of all, massive
galaxies are known to have their in-situ star formation already completed at
$z=1$ \citep[e.g.][]{Bower_etal_2006,DeLucia_etal_2006}: as a consequence, 
an evolved massive galaxy cannot likely  merge with a young counterpart of comparable mass, 
simply because
there is a very small number, actually none, of such objects around. Second, at any redshift 
(leaving redshifts above say 10 aside)
the number density per $Mpc^3$ of galaxies with mass
in the interval $10^{10}$ to $10^{12}\, M_\odot$ is down by a factor from $10^5$ to $10^8$ 
or even more with respect
to those in the mass interval  $10^{7}$ to $10^{9}\, M_\odot$ 
\citep[see Table 2 in ][]{Chiosi_etal_2017}. 
This means that high mass galaxies merge only with smaller objects, with a probability increasing
with decreasing mass of the latter ones. 
All this is mimicked by the factor $f_M$ in the eqn. (\ref{eqageseedmergfunneled}) above.

In the right panel of Fig. \ref{ccmjmergers} we show the result of our funnel scheme: the galaxies,
indicated by black squares, occupy the region delimited by black lines, spread in
the CMD over more than 5 orders of magnitude in the V-band luminosity, and seem  to agree
with the observations much more than the red filled squares of the left panel. 
In particular a Red Sequence is clearly present and a 
``zone of exclusion" is in place. The  funnel mechanism is effective in keeping massive galaxies 
out of the left side of the Green 
Valley, i.e. the lower side of the Red Sequence in this range of luminosities,
in fair agreement with observational data. Furthermore, a better bounded Green Valley both 
in magnitude and colors
is also produced by our simulations, and the bluest region of the WINGS dataset begins to be populated, too.
Recalling the motivations for the funnel mechanism 
and the experiments shown in Fig. \ref{evo_burst}, this implies that we do not
observe massive galaxies containing significant percentages of  young stellar populations;
in other words, the great bulk of stars in most luminous galaxies are old and red. The black
dots are therefore demonstrating that, even with our over-simplified description of the merger history 
of a galaxy
(reduced here to a single event), the heuristic funnel mechanism can confine galaxies in regions 
fully compatible with the observational data.

We also remark that, at variance with Fig. \ref{ccmjburst}, a relatively small fraction
of galaxies with $M_V < -22$ may reach $B-V>1$: this result is consistent with the fact that
(i) bursts add bulks of young stars which means bluer colors, whereas mergers always constitute a mix of gas 
(and likely young stars) and
already existing old stars which implies redder colors; (ii) at low redshifts, when the ``cosmic noon"
i.e. the peak of the SFRD($z$) has already occurred \citep[see e.g. Fig. 9 of][]{MadauDickinson2014},
dry mergers are more probable than the wet ones so that an already old ex-situ stellar component is more
probable to be added to a seed galaxy; (iii) simulated bursts of reasonable intensity are never 
strong  enough to parallel the effect of a whole  galaxy;
(iv) the simulations shown in Fig. \ref{evo_burst} tell us that only perturbations 
due to very differently aged stellar populations may sensibly affect the $B$ and $V$ emissions;
(v) in the case of mergers, a more realistic mix of metallicities is also present, so that
a broad and adequately steep Red Sequence can form.

We conclude this section by remarking that single events of rejuvenation of stellar populations
already effectively account for the dispersion in the region of star-forming galaxies, namely 
the whole Green 
Valley and the redder part of the Blue Cloud. One can easily foresee that: 
(i) If galaxies (let us consider intermediate masses only, so that a straight comparison with data 
is possible) containing
significant populations  of young stars are considered, we should find 
many more objects falling in  the Blue Cloud. Adding young stellar populations to  oldish galaxies 
shift them towards bluer
colors and brighter magnitudes than the starting values. (ii) 
An arbitrary mix of bursts and mergers is expected to affect the $B$ and $V$ fluxes more 
efficiently than single events. Starting from this reasoning, the Blue Cloud can be identified  
as the locus of the CMD 
occupied by galaxies containing just newly born 
stars independently of the age of ``their seed'' object; shortly speaking, independently of 
the age (redshift) at which the baryonic history of the galaxy began.

\section{Using model galaxies}\label{using_gal}

The key result of the previous analysis is the suggestion that galaxies may scatter in the CMDs  due 
to several causes such as age, 
chemical composition (metallicity), bursts of stellar activity, and mergers. However, the correct analysis of all these 
issues  should rest  on a suitable cosmological scenario   describing the  formation of galaxies inside 
a given 
volume of space (in other words a prescription for the formation and evolution of galaxies within a cluster)
 and 
models galaxies including their own histories of star formation and chemical enrichment. Two 
different strategies  are possible that have already been successfully applied in cosmological context. 
The first one is the so-called merger branching-tree, in which DM halos and their baryonic content at each 
epoch
are the result of previous mergers. Consequently,  the number of branches increases
with the redshift  \citep[see][for a classical description]{LaceyCole1993}: 
the main advantage of this is that mass growth of DM halos  is easily followed up.
The second one  rests on the notion of the HMF, which is a suitable 
description (possibly analytical)  of the number densities of DM halos as a function of mass and redshift 
in cosmological simulations. Although a detailed prescription of how DM potential wells evolve
with time is missing in this approach, the HGF calculated from the HMF is nicely suited to our 
aims, because it allows us to explore the impact of cosmology in shaping CMDs with very little computational 
effort  as compared to the detailed massive numerical simulations.

We build up our sample of galaxies inside a cluster in two steps:
we first exploit the HMFs to get their relative HGFs,
then using the MonteCarlo technique we randomly  assign the formation redshifts $z_f$ to all  galaxies in the cluster
respecting the mass-radius relation. Also in these simulations we  will assume
the $\Lambda$CDM concordance cosmology we have quoted above 
\citep[namely $H_0=70.4 {\rm \,km\,s^{-1} Mpc^{-1}}$,
$\Omega_{\Lambda}=0.73$, $\Omega_m=0.27$, $\Omega_b=0.05$, ][]{Komatsu_etal_2011}. 

The star formation and chemical enrichment histories of the member galaxies are described by 
suitable simple models that however are evolved in isolation (no mergers are allowed). Despite this major 
limitation,
these galaxy models will prove to be  good enough to provide a first description of the physical evolution of 
a cluster galaxy.

\subsection{DM halos hosting BM galaxies} \label{dmhalos}

To get the population of DM halos hosting BM  galaxies, we need to know how many halos of any given  mass are inside
a volume of arbitrary extension: to this aim, we apply the HGF formalism  \citep{Heitmann_etal_2006, Lukicetal2007} 
to obtain the comoving number density of DM halos as a function of redshift $z$. We briefly report
here the definition of HGF, as the integral of HMF over a suitable range of masses $[M_1,M_2]$:
\begin{equation} \label{eqhgf}
\begin{aligned}
n(M_1,M_2,z) &= \int_{M_1}^{M_2} F(M,z)\;d\,log M \\
&= \int_{M_1}^{M_2} \frac{d\,n}{d\,log M}\;d\,log M\;. \\
\end{aligned}
\end{equation}
The integrand, i.e. the HMF, contains three quantities that depend on the adopted cosmological scenario: 
the background density, $\rho_b(z)$; the variance
of the linear density field, $\sigma$; the fitting function, $f(\sigma)$; see the references above for all details. 
Via \verb|HMFCalc|\footnote{http://hmf.icrar.org/}
\citep{Murray_etal2013} we get the values of HMF and sum them together inside
mass  bins centered at $10^7$, $10^8$, $10^9$, $10^{10}$, $10^{11}$, $10^{12}$
and $10^{13}$  $M_{\odot}h^{-1}$ and wide $\Delta \log M = 0.4$.

The expression for eqn. (\ref{eqhgf}) yields the \textit{average} comoving number density of 
DM halos once fixed $f(\sigma)$, which is known to be
tuned and to be varying from one cosmological simulation to another.
Although it is easily predictable that varying the HMF will not change drastically
the shaping of CMDs, we make use of \verb|HMFCalc|
to explore the effects of implementing different HMFs.
 
\begin{figure} 
	\centering
	\includegraphics[width=8.5cm]{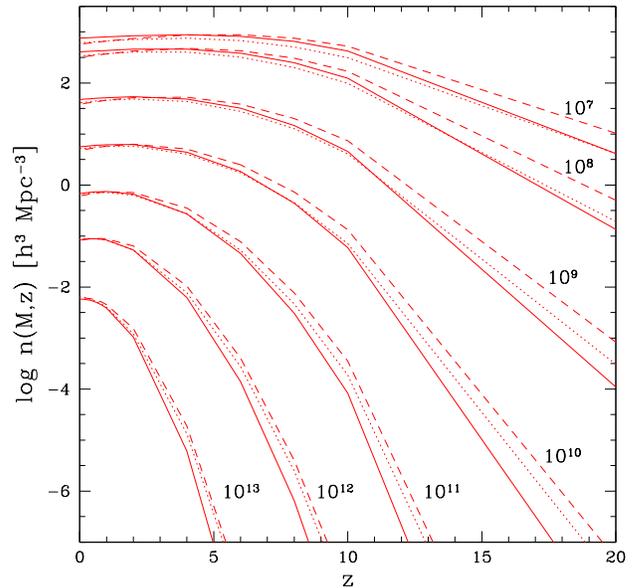}
	\caption{HGFs obtained from the HMFs of \citet{Warren_etal_2006} (dashed), 
                \citet{Angulo_etal_2012} (solid)
		and \cite{Behroozi_etal_2013} (dotted).	The curves are labeled with the values of 
                the mass of the DM halos used to bin the HMFs. See the Appendix \ref{app_galnum_analysis} 
                 for further details.}
	\label{usedHGF}
\end{figure}

We then make use of this  data to calculate simple 4-th order interpolating polynomials:
\begin{equation}
\log n(M_{DM,i}, z) = \sum_{j=0}^4 A_{i,j}(M_{DM,i}) \times z^j \label{Lukic_interp}.
\end{equation}
as made in \citet{ChiosiMerlinPiovan2012} and \citet{Chiosi_etal_2017}; the $i$ index relates the polynomial
interpolation to each mass bin. In this way, we can easily incorporate
the HGFs into generation of $z_f$ for galaxies inside regions of
Universe with desired values of radii (see Subsection \ref{demgal}).

The results of this analytical fit are  shown in Fig. \ref{usedHGF}: the lines represent HGFs calculated
inside the aforementioned mass bins, with labels indicating these latter, from the HMFs of
\citet{Warren_etal_2006} (dashed lines), \citet{Angulo_etal_2012} (solid lines)
and \citet{Behroozi_etal_2013} (dotted lines).
The reader is also referred to Appendix \ref{app_galnum_analysis} for further details about 
the coefficients of the polynomial fits. There are some differences between the curves of
the same mass bin passing from a HMF to another, especially \citet{Angulo_etal_2012} who
extend their  fit to DM halo masses smaller than those considered by \citet{Warren_etal_2006}  and 
\cite{Behroozi_etal_2013}\footnote{Further details about the mass range of best-fits for the three HMFs can be found 
in the cited papers or at the URL 
http://hmf.icrar.org/static/fitting\_function\_table/fitting\_functions.pdf}.
As already pointed out by \cite{Lukicetal2007},  passing from a galaxy  mass
to another, each curve peaks at a different value of the redshift:
this clearly shows that the HGFs already include the mergers among
DM halos. However, the peak values of different HGFs are remarkably
similar once fixed the mass bin: this just anticipates the fact that, like all average observables related to galaxies,
the CMDs of clusters are statistically independent on the underlying precision cosmology.

Starting from Fig. \ref{usedHGF}, there are a few aspects of the HGFs worth being recalled: (i) 
the high merging efficiency leads all bins with $M\leq10^{11}\,h^{-1}\,M_{\odot}$ to reach
a peak at a value of $z$ increasing with decreasing mass; (ii)  the number densities
of intermediate mass galaxies can  both decrease  by  mergers and also  grow  because 
these galaxies are  relatively deep potential wells; (iii) the number densities of the
most massive halos ($M\geq10^{12}\,h^{-1}\,M_{\odot}$) in the recent past keep on being constant
or even increasing, because massive galaxies take long time to \textit{become massive} by accreting DM and BM.

\begin{figure*}
	\centering{
		{\includegraphics[width=8.5cm]{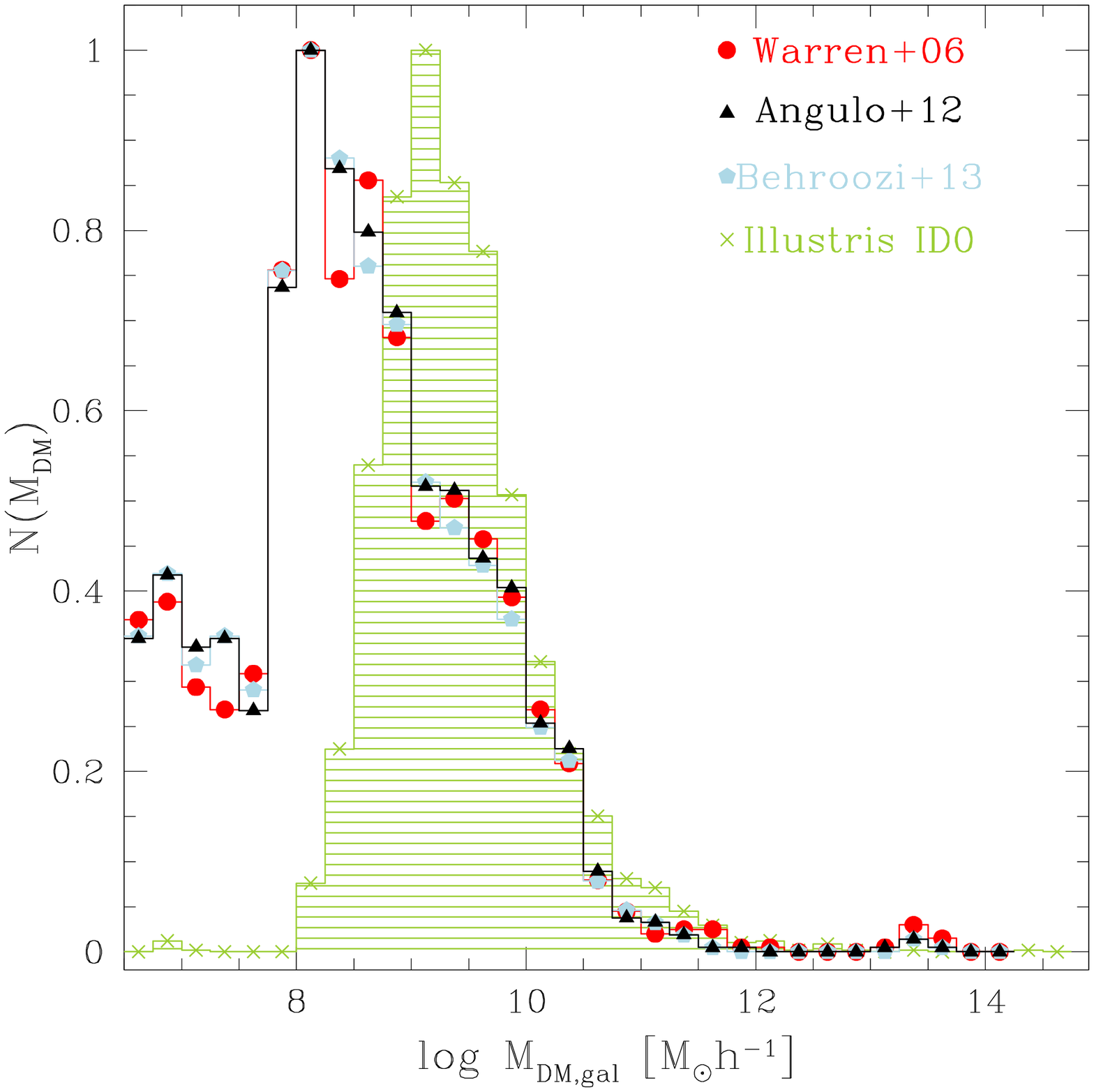}  } 
		{\includegraphics[width=8.5cm]{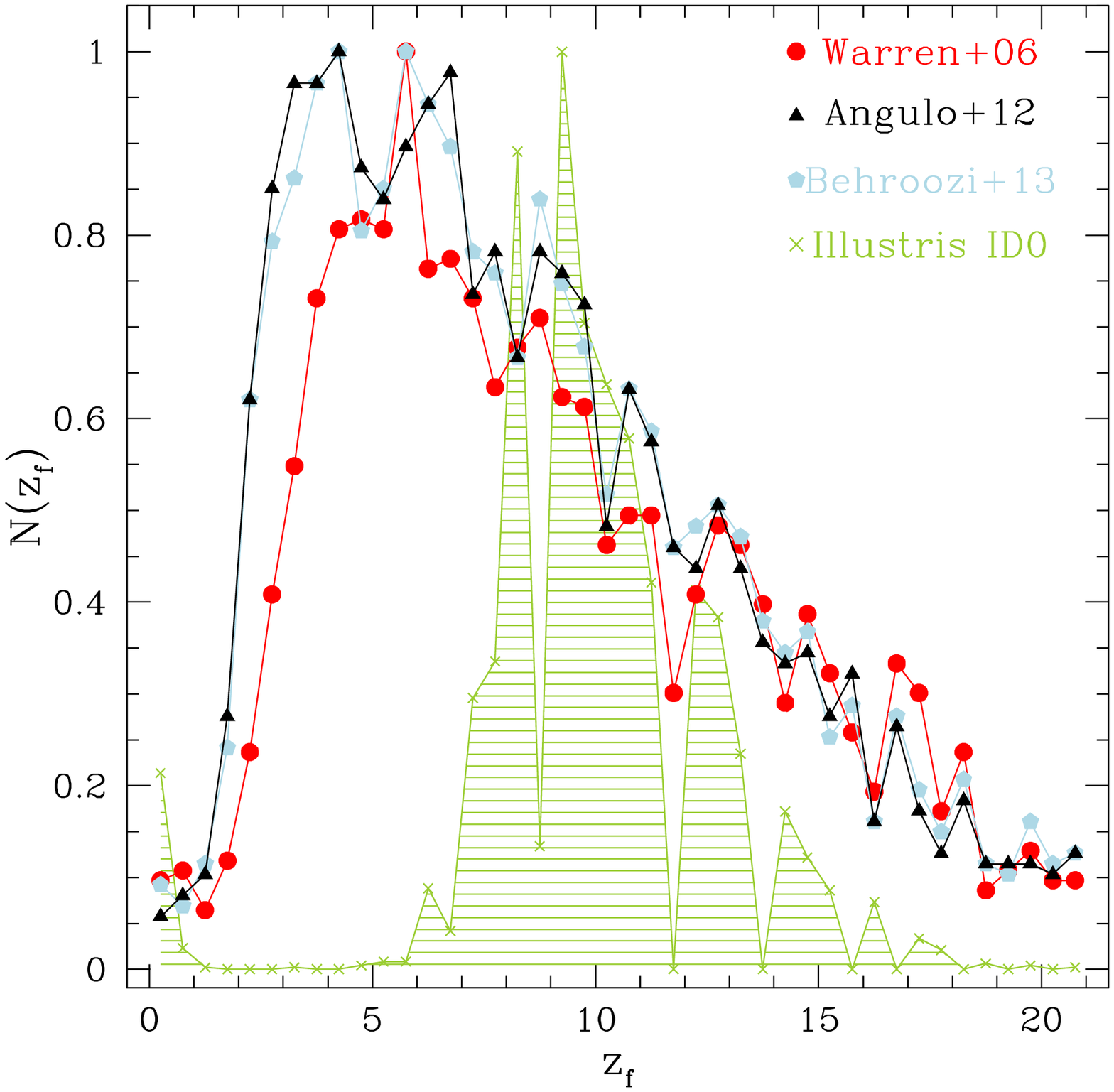} }   }
	\caption{Comparison of demography of DM halos hosting galaxies inside clusters at 
                $z=0$ between ILLUSTRIS simulation
		\citep[][olive green crosses and shaded regions]{Vogelsberger_etal_2014} and
		our random formation redshift generator using the three different HMFs: red dots 
                are \citet{Warren_etal_2006}, black
		triangles are \citet{Angulo_etal_2012} and light blue pentagons are 
                \citet{Behroozi_etal_2013}. In both sides, each histogram is
		normalized to the maximum value (see Table \ref{galdemo_sicofi}  and the text for 
                more details). \textsf{Left Panel}:  histograms of masses.
		\textsf{Right Panel}:  histograms of $z_f$.}
	\label{fig_zfor}
\end{figure*}

\subsection{Galaxy Demography} \label{demgal} 

The HGF can now be used to generate galaxies hosted by DM halos with random masses and formation redshifts $z_f$.
The cosmological limit imposed by the HGF is coupled with the mass-radius relation of
galaxies, as already done by \citet{ChiosiMerlinPiovan2012} whose procedure is shortly summarized here.

The comoving number density $n(M_{DM},z)$ is the result
of two competing effects: the formation of new halos of mass $M_{DM}$ via merger
and/or acquisition of lower mass halos, and the destruction of halos of mass
$M_{DM}$ because  they merge to form higher mass halos. Therefore, the
following equation can be written

\begin{flalign}
n(M_{DM},z) =\,& n(M_{DM},z+\Delta z) + \nonumber \\
& n_{+}(M_{DM},z) - n_{-}(M_{DM},z+\Delta z)
\end{flalign}

\noindent where $n_{+}$ and $n_{-}$ represent the creation and destruction
mechanisms. In particular, the quantity we are interested in is $n_{+}(M_{DM},z)$,
which is the number of new halos of mass $M_{DM}$ which are born at redshift $z$.

The number of halos that merge to form higher mass systems is in turn a
fraction of the number of halos existing at that time, i.e.
$n_{-}(M_{DM},z+\Delta z) = \eta_h \times n(M_{DM},z+\Delta z)$, with $0 < \eta_h < 1$;
so

\begin{flalign}
n(M_{DM},z) =\,& n(M_{DM},z+\Delta z) + \nonumber \\
& n_{+}(M_{DM},z) - \eta_h \times n(M_{DM},z+\Delta z)
\end{flalign}

\noindent and

\begin{flalign}
n_{+}(M_{DM},z)  =\, & n(M_{DM},z)\, - \nonumber \\
&  (1-\eta_h) \times n(M_{DM},z+\Delta z)
\end{flalign}

\noindent The only free parameter here is $\eta_h$, the fraction of halos that merge to form
higher mass systems in the redshift interval $\Delta z=0.145 \times z +0.1$.
In principle, the fraction $\eta_h$   could vary  with
the redshift. However, for the sake of simplicity we assume that $\eta_h$ remains
constant and equal to $0.01$ \citep[see][for details]{ChiosiMerlinPiovan2012}.
For each interval $M_{DM}, M_{DM}+ \Delta M_{DM}$ and $z, z + \Delta z$ we obtain a value $n_{+}(M_{DM},z)$.
This number, renormalized to unit over the whole redshift interval, can be considered as the 
\textit{relative cumulative probability} that a halo of mass $M_{DM}$ is born at redshift $z$.

Finally, for each halo of mass $M_{DM}$ we compare a randomly chosen number
$q \in (0,1)$ with the cumulative probability

\begin{equation}
P_{z_i}=\sum_{z=z_{max}}^{z=z_i} n_{+}(M_{DM},z)
\end{equation}

\noindent until we have $q<P_{z_i}$, and take $z_{f}=z_i$ as its formation
redshift; here $z_{max}$, which is the maximum redshift at which a DM halo hosting a 
galaxy observed at $z=0$
starts to aggregate, is fixed to 21. It is worth recalling that after re-normalization 
the cumulative probability $P_{z_i}$ is also 
confined in the interval $[0,1]$. We then combine the values of $z_f$ of all galaxies
with their values of $M_{DM}$, using the relationship by \citet{Fanetal2010}

\begin{equation}
R_{1/2}=0.9 \frac{S_S(n)}{0.34}\frac{25}{m} \left( \frac{1.5}{f_*} \right)^2
\left( \frac{M_{DM}}{10^{12} M_\odot} \right)^{1/3} \frac{4}{(1+z_{f})}.
\label{mrr}
\end{equation}

\noindent which substantially links the stellar half-mass radius to the mass of its DM halo host.
Here, $m=M_{DM}/M_*$ with $M_*$ the stellar mass, $S_S(n_S)$ is a coefficient related
to the Sersi\'c indexes $n_S$ and to the ans\"atz $R_{1/2}=S_S(n_S)R_g$ relating
gravitational and stellar mass radii, $f_{*}$ the velocity dispersion of the stellar component
with respect to that of DM. For simplicity, we fix these quantities as
$m=10$, $S_S(n_S)=0.34$, $f_{*}=1$ \citep[see][for details]{Fanetal2010}.

\begin{table}
	\begin{center}
		\caption{ Number of simulated galaxies at $z=0$ for two typical values of  cluster 
			radii and each HGF. The acronyms W06, A12, and B13  stand for \citet{Warren_etal_2006}, 
			\citet{Angulo_etal_2012}, and \citet{Behroozi_etal_2013}, respectively.}
		\label{galdemo_sicofi}
		\begin{tabular}{|c|c|c|c|}
			\hline
			$R_c [Mpc]$ & $N_{W06}$      &    $N_{A12}$       &    $N_{B13}$ \\
			\hline
			1.25 & 1656 & 1789 & 1796  \\
			1.50 & 1883 & 2016 & 2033 \\
			\hline
		\end{tabular}
	\end{center}
\end{table}

With this procedure, we derive the populations of galaxies hosted DM halos that would be expected at $z=0$ 
in ideal clusters with typical dimensions, namely with radii
$R_c$=1.25 and 1.5 Mpc. 
The total numbers of galaxies per cluster are listed in Table \ref{galdemo_sicofi}.
We present our samples by examining the histograms of masses and $z_f$ of DM halos
hosting galaxies at $z=0$. Both curves are normalized to the maximum value. 
Our predictions are compared with
the most massive galaxy cluster found inside the ILLUSTRIS-1 simulation 
\citep{Vogelsberger_etal_2014,Nelson_etal_2015},
which is the most detailed, full physics simulation in their 
suite\footnote{http://www.illustris-project.org/data/}. The comparison is shown in Fig. \ref{fig_zfor}.
We briefly recall that in ILLUSTRIS-1 the resolution in DM is $6.26\times10^6\,M_{\odot}$
and the least massive DM friend of friends (FoF) must have 32 particles to be identified as a halo;
the most massive galaxy cluster has a virial mass of $M_{200,mean}=2.89\times10^{14} h^{-1} M_{\odot}$
and a virial radius of $R_{200,mean}=1.66 h^{-1}$ Mpc, with the subscript indicating quantities 
evaluated inside the region having $\rho=200\times\left<\rho\right>$, with $\left<\rho\right>$ the mean 
density of the Universe. The sample analyzed here is made of only those DM halos  hosting a non-zero 
stellar component, i.e. for which photometric data (magnitudes and colors) are given.
The comparison is thus made with our synthetic clusters with physical radius $R_c=1.25$ Mpc.

In the left panel, we see the distributions in mass for DM halos hosting BM galaxies.
It is clear that the two prescriptions lead to distributions peaking at different values: 
we argue that this might
be due to dynamical phenomena acting inside the ILLUSTRIS-1 cosmological box, e.g. 
 mergers, shifting the peak towards higher masses,
and tidal disruption, removing low mass DM halos.
It is however interesting to note
that both distributions resemble by eye some kind of a skewed Gaussian. 
In the right panel, we show the distributions of the same galaxies as a function  of $z_f$.
Here our samples are very different from those of ILLUSTRIS-1. Our galaxies broadly
peak at $3<z<6$, with the bulk of objects already born earlier and a small number  of them
forming later. Inside ILLUSTRIS-1 the birth of galaxies, once reached
a peak in frequency at $z\simeq9$ [$t_{lb}\sim13.2$ Gyr in
terms of look-back time], is afterwards suppressed almost completely, with minor
oscillations around 0 for $1\lesssim z\lesssim6$ and a final uprise at very recent epochs.
Consequently, the ILLUSTRIS-1 sample is apparently  missing galaxies
born at intermediate redshifts [$7.8\lesssim t_{lb}(\text{Gyr})\lesssim12.8$ in
terms of look-back time], whereas in the same time interval our model does predict the formation of galaxies.

The large differences noted in Fig. \ref{fig_zfor} are surely related to the different approximations
underlying  both the complex hydrodynamical simulations of ILLUSTRIS-1 and our simplified 
approach. As recently discussed  by \citet{Chua_etal_2017}, the
distribution of non-collisional matter (i.e. stars and DM) can effectively cool down
in the cores of galaxies formed in the ancient epochs, at variance with
those born more recently: this provides  them much better chance of survival,
or at least stronger resilience against potentially disruptive interactions that
they would surely undergo throughout the cluster environment. A key role is also played by the host
cluster itself because the mass profile of this latter becomes steeper and steeper in the core because of 
dynamical relaxation, consequently  the environment gets denser with time: this trend is found to
be quantifiable in terms of the number of dynamical times
elapsed since the cluster formation \citep[e.g.][]{Jiang_vandenBosch_2016}. Therefore,  each
galaxy has to attain a sufficiently
cold core in order firstly to survive in the cluster environment, and then to grow by accretion of
matter from ICM and by mergers with other substructures. Given these  preliminary considerations,
which explain evaporation of galaxies while falling into more massive halos, the inclusion of the
BM physics in this context is crucial and its effects are still not fully explored. Basing 
on a systematic comparison of the full-physics and dark-matter-only ILLUSTRIS suites,
\citet{Chua_etal_2017} claim that galactic winds and photo-ionization from UV radiation may effectively
inhibits mass aggregation  at least at the low-mass end and in certain regimes of redshifts (see
also \citealt{Despali_Vegetti_2017}
for similar reasoning for DM subhalos of intermediate masses inside ILLUSTRIS and EAGLE
simulations, \citealt{Schaye_etal_2015}). 
In particular, photometric data inside ILLUSTRIS are bound to whether stellar particles,
i.e. stellar populations, are found inside a bound sub-halo: recalling that in ILLUSTRIS-1 the
average gas cell mass is $8.86\times 10^5\,M_{\odot}$, from
which integrated stellar populations might stem, and that a prescription for subgrid physics is always needed to
follow up integrated properties of all kind of particles, these could be a substantial limit of state-of-art
hydrodynamical simulations in describing photometry, especially for galaxies inside more recently
formed DM halos of lower masses. On the other side, our numerical approach, at the basis of
the build-up of samples inside clusters, provides a population of galaxies which should reflect
the observational evidence of the mass-radius relation at $z\sim0$ \citep{ChiosiMerlinPiovan2012}:
if observed galaxies are merely those which survived to all disrupting interactions they 
underwent, then at least a minimum fraction of the sample (not as close to zero as the one of ILLUSTRIS-1) 
should have formed in an evidently large range of look-back time. However, acknowledging that
our approach might not be the most refined one, we feel that the best interpretation might be
a middle ground between the two models.

\begin{figure*}
	\centering{
		\includegraphics[width=8.5cm]{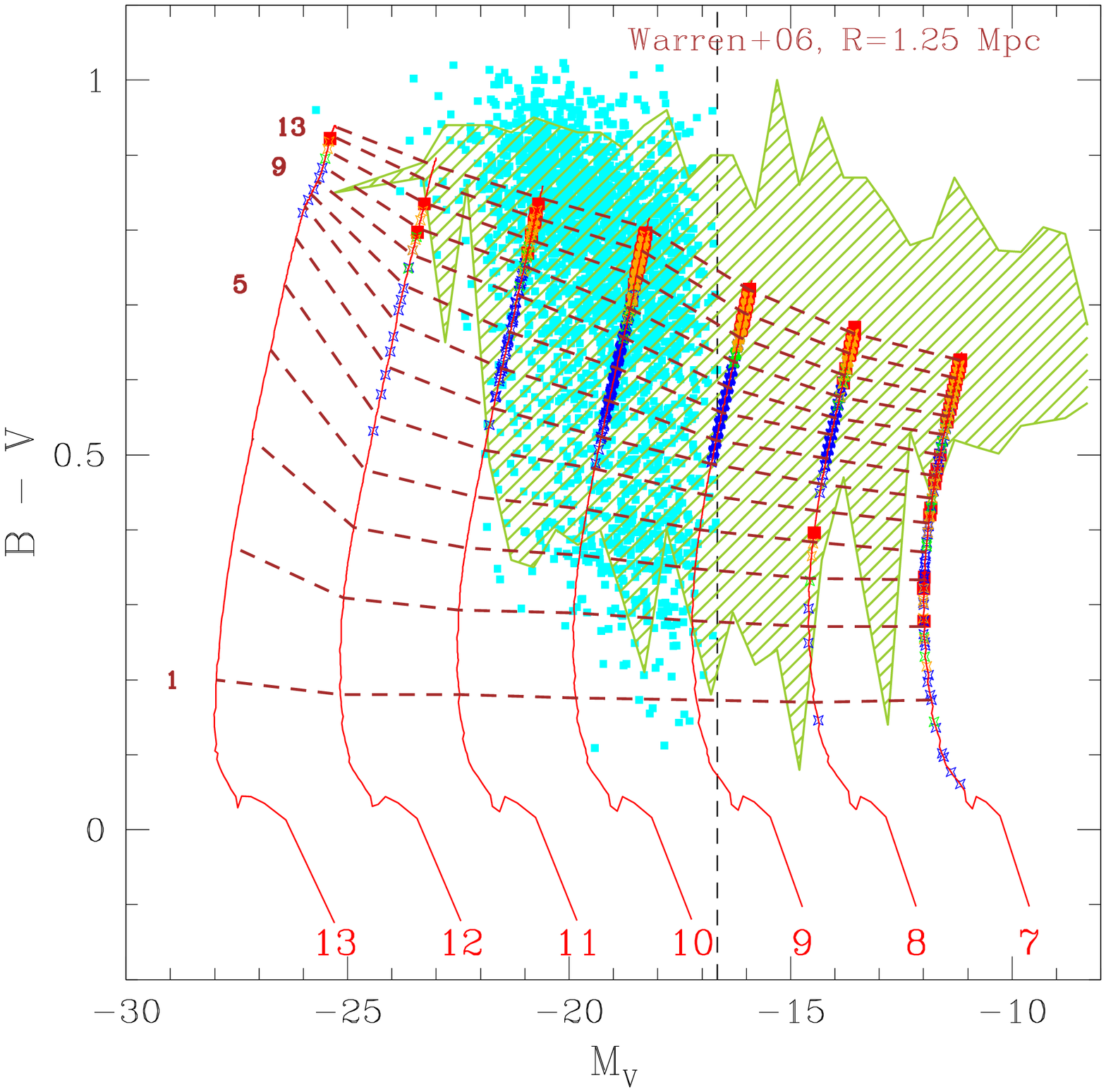}
		\includegraphics[width=8.5cm]{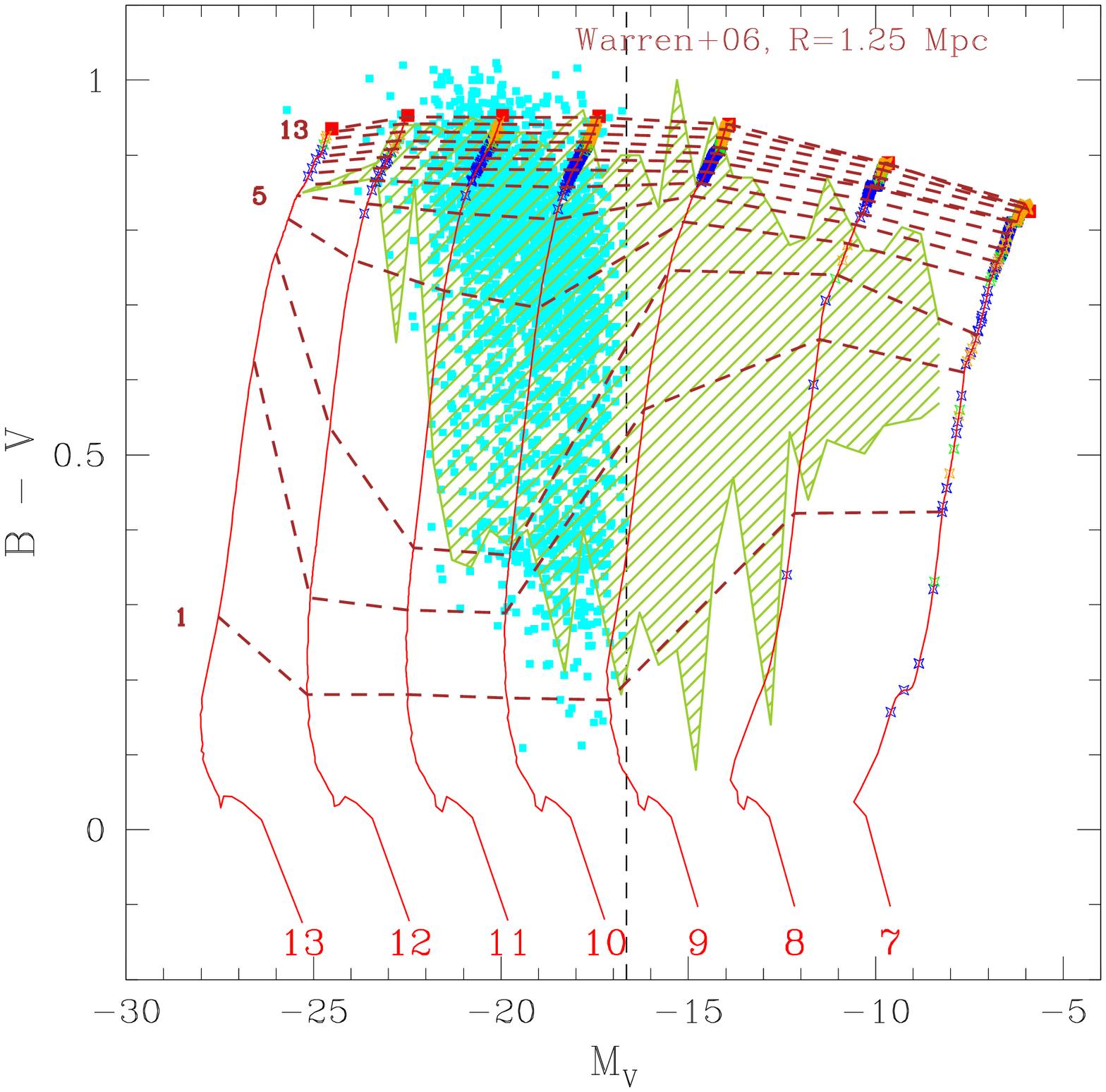}}
	\caption{CMD of galaxies for the reference cluster (see text for details).
		The green shaded area represents galaxies of
		the most massive cluster inside ILLUSTRIS-1 simulation; the cyan points are
		galaxies of the selected sample of the WINGS clusters. The vertical dashed line indicates
                the limit of the faintest object; the red solid lines are the evolutionary paths in the rest frame
                of the model  
                galaxies of different baryonic mass, from $10^7$ (right) to $10^{13} \, M_\odot$ 
                (left) in steps of a factor of ten. The value of the mass (on logarithmic scale and in solar 
                units) is also indicated along each line. Open stars along the lines  are
		selected stages of the model galaxies as they would be observed at various bins of redshifts
		(at decreasing redshift,  blue is for $0.1<z\leq0.5$, 
		green for $0.05<z\leq0.1$, orange for $0.01<z\leq0.05$, red for $z\sim0$). The
                dark red dashed lines connect the stages of equal age for each mass,
		with  time steps of 1 Gyr  from 1 to 13 Gyr from bottom to top.  Ages in Gyr are also 
                indicated along the dashed lines limited to a few cases for the sake of clarity. 
                \textsf{Left Panel}:  models
		with modulated star formation. \textsf{Right Panel}: models with straight galactic wind.} 
	\label{cmd42impro}
\end{figure*}

Summing up, as seen in Section \ref{burst_merger},
especially in Figures \ref{ccmjburst} and
\ref{ccmjmergers}, dispersion inside CMDs is driven by scatter in formation epochs:
we can say that our model, although crude and not closely resembling cosmological simulations,
is still fully adequate in locating the birth of galaxies in 
cosmic epochs.  Moreover, distributions built up from HGFs are evidently
very similar in terms of both masses and $z_f$, with the only slight exception
of the formation redshifts arising from \citet{Warren_etal_2006}  that peak about 
2 Gyr earlier than those of the other two distributions. As it will better proved  in Appendix
\ref{app_galnum_analysis} and anticipating here the results displayed
in Fig. \ref{cmd42impro}, we argue that
when the integrated properties of single galaxies are analyzed, such as light emitted as a whole,
typical informations on precision cosmology like for instance cosmological parameters or mass resolution
are almost completely lost.

\subsection{Path of model galaxies on the CMD: comparison with data}

In Fig. \ref{cmd42impro} we show the CMD of galaxies inside the reference cluster
\citep[radius: 1.50 Mpc; underlying HGF:][]{Warren_etal_2006}. The left panel shows 
galaxy models  with no galactic wind, whereas the right panel those with galactic winds. 
In view of the discussion below it is worth recalling here that the integrated spectra 
of the models galaxies, albeit in a very simple way, take the effects of interstellar dust 
into account. Moreover the simulated galaxy content of clusters does not consider the geometric structure of 
the cluster as far as the star formation,  metal enrichment, and photometric histories of the 
component  galaxies and their position within a given clusters are concerned.

Prior to any other consideration, we need to shortly describe the path of a model galaxy on the CMD. In both 
panels the generic path is indicated by a red solid line. Each path corresponds to a galaxy of different 
total mass as indicated in Table \ref{Model_Galaxies}. Starting from the bottom (the early stages of a 
galaxy's 
evolutionary history) the stellar mass emitting the light is low and the metallicity is low 
as well. As time proceeds, the stellar mass and the
metallicity reached by the gas increase. In other words, the stellar mass is 
a mixture of many generations of stars of different age and metallicity  
each of these contributing to the total light. The contributions are in turn  weighed by the  star 
formation rate at the time of formation.  The integrated light  
whose intensity and color determines the position on the CMD. Since the luminous star mass  
first increases and later decreases  because stars either explode as SNae or leave faint remnants   
 that do not contribute to the light, the path on the CMD first 
goes towards more negative  magnitudes and then bends over to less luminous magnitudes. In the meantime the 
colors become redder and redder  both because of increasing metallicity and  aging of the already existing 
evolving stars. This trend is however  contrasted by newly born stellar generations by the effect of the 
bright, blue massive stars, i.e. in presence of active star formation.  It goes without saying that, as gas 
is used 
up by star formation, this latter effect tends to decrease and eventually disappear when star formation 
ceases completely. Furthermore we like to comment that the mean metallicity reached by any galaxy  affects 
its 
 location on the red sequence both in absence and presence of galactic 
winds. Last point to note is that the dashed line in both panels of Fig.\ref{cmd42impro} connect stages of 
equal age in the rest-frame of the model galaxies.  From  bottom to top the age goes from 1 to  13 Gyr, in 
steps of 1 Gyr.

We plot also the data of the ten best WINGS clusters
(cyan points) and the most massive galaxy cluster inside ILLUSTRIS-1
simulation (green shaded area).
Both the WINGS data and the ILLUSTRIS-1 simulations show the well known drop-off towards the Green Valley and 
the Blue Cloud \citep[see][and references therein]{Nelson_etal_2018}. Roughly speaking,  
the Green Valley is confined in the color range $0.6 < B-V < 0.8$, has nearly 
the same slope of the Red Sequence and  is populated by fewer objects. Finally the Blue Cloud,
customarily associated to galaxies with ongoing star formation (e.g. spirals and robust dwarfs), falls below 
the Green 
Valley and has a numerous population.
The relatively scarce population of the Green Valley may suggest that galaxies  are  
either slowly accelerating  their evolution but still retain some residual stellar activity (left 
panel of Fig. \ref{cmd42impro}) or past the major star forming activity and the sudden interruption of this 
by galactic wind quickly evolve toward the region of red colors (right panel of Fig. \ref{cmd42impro}). The 
alternative depends of the true relative number of galaxies in the three regions that is subject to 
completeness of the observational data. Assessing this issue is beyond the aims of the present study.

Looking at the left panel, in which no galactic winds are considered,  
a number of features are evident. First of all, all evolutionary paths (the red solid lines) end up
at $B-V$'s that increase with the
galaxy mass: thus  defining the locus in the CMD of the present-day stage. 
However, it is worth noting noting that along this line, star formation is still going on albeit at minimal 
levels does
(see the blue solid lines in Fig. \ref{sfr_age} showing the SFR vs age relationship for this type o galaxy 
models). The slope of the constant age loci(dashed lines in Fig\ref{cmd42impro}) becomes more and negative at 
increasing age. Both the location and slope of the present-age stages nearly coincide with the location of 
the bulk galaxies along the observational Red Sequence  (by eye inspection  the theoretical slope is 
half the observed one), thus providing a simple and consistent way  of explaining the CMR. 

However, looking at the slopes of the different datasets
in their most dense regions, 
we see that our end-point line  falls in between those for the WINGS
and the ILLUSTRIS-1 clusters, the slopes of which are steeper and flatter, respectively.

However, we clearly see that a $\Delta(B-V)\sim 0.6$ and $\Delta M_V \sim 6$ -wide region of 
 WINGS and ILLUSTRIS-1 data,  containing the star forming galaxies, requires  
that other mechanisms
must be taken into account to get their much bluer colors:
these are mainly bursts and mergers, the effects of which have already been anticipated with the aid of 
the SSPs-galaxies. 

\begin{figure*}
	\centering{
		\includegraphics[width=8.5cm]{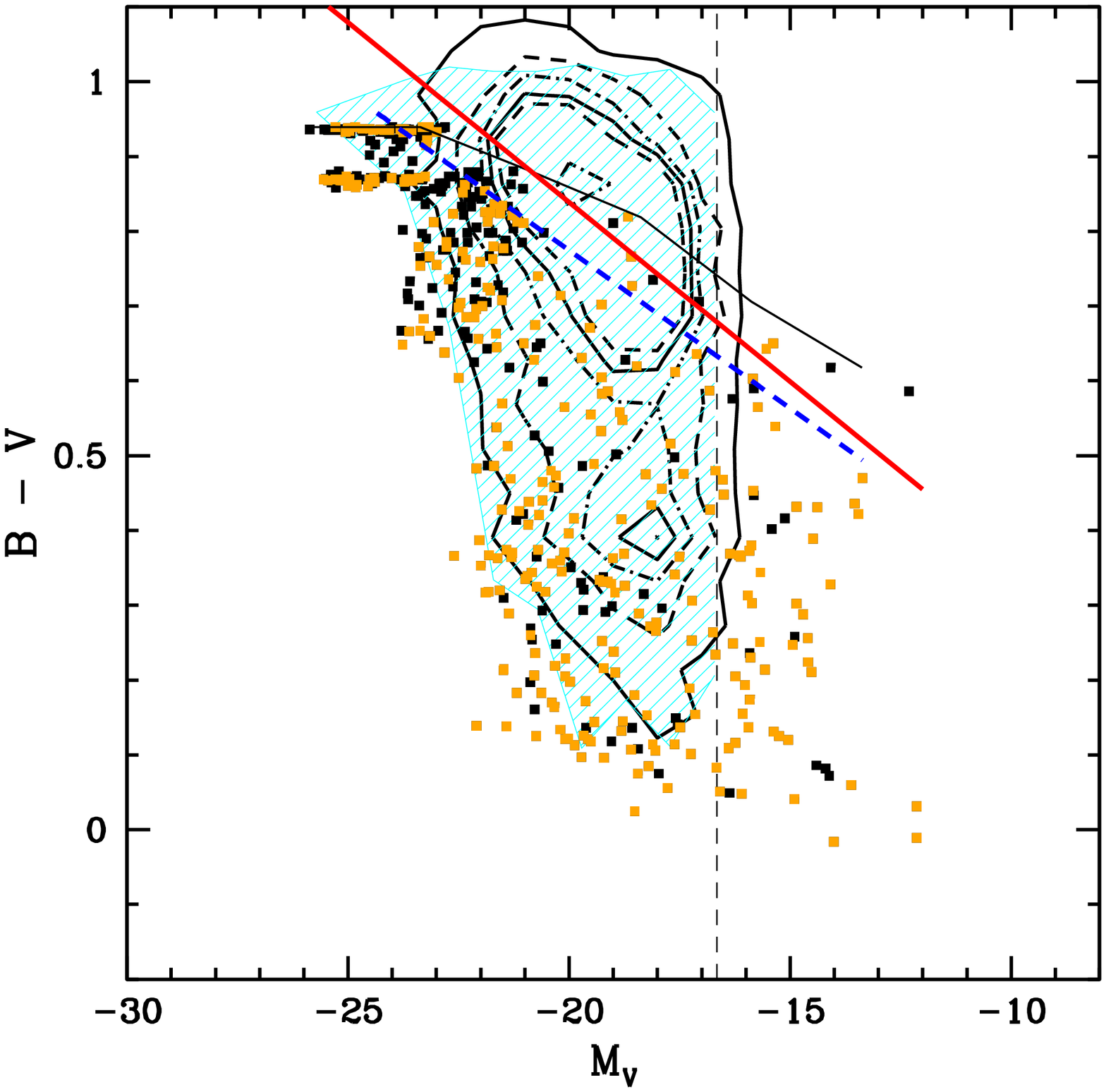}
		\includegraphics[width=8.5cm]{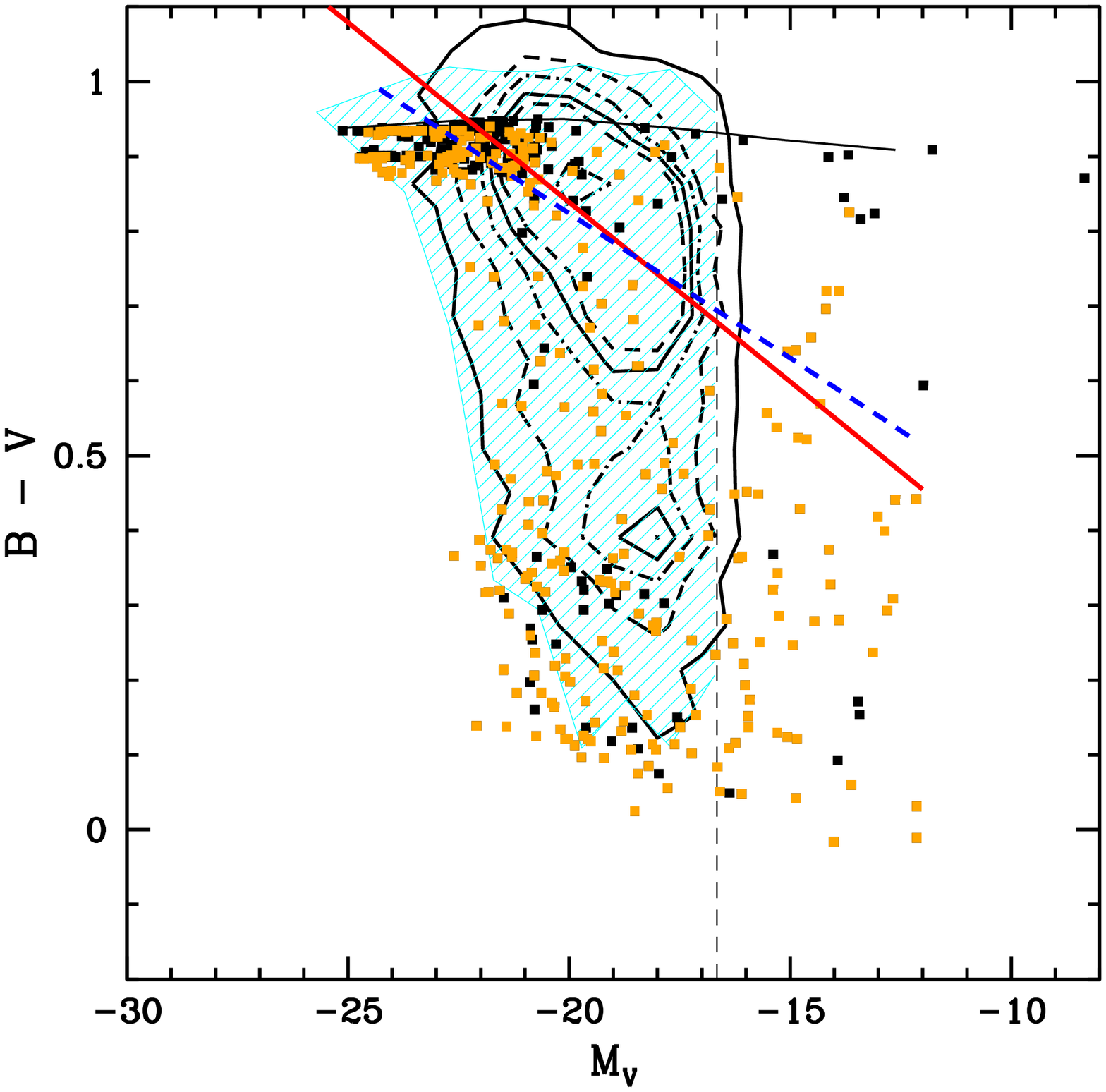}}
	\caption{Galaxy mergers. \textsf{Left Panel}: the case of galaxy models with modulated star formation;
		black squares are the mergers of two early-type galaxies (ET$_{ms}$+ET$_{ms}$). Yellow squares 
		are either the mergers of two late-type objects (LT$_{ms}$+LT$_{ms}$)  or of a late-type 
		plus an early-type galaxy (ET$_{ms}$+LT$_{ms}$). The underlying HGF is the one by \citet{Warren_etal_2006}.
		\textsf{Right Panel}: the same as in left panel but early-type galaxy models now considered have straight
		galactic wind,	ET$_{gw}$. Note the different upper boundary at red $B-V$ colors. 
                In both panels, the shaded area shows 
                the data for the selected sample of WINGS clusters together with the contour lines of 
                constant fractionary number of galaxies per elementary area of the CMD already presented in 
                Fig. \ref{ccmjmergers}, the fiducial CMR of Sect. 
\ref{obser_data} (thick red line in the 
                left panel, and thick black line in the right  panel),  and the best fit of the model 
                galaxies falling within the 1$\sigma$ uncertainty strip of the fiducial CMR (the dashed 
                lines). See the text for more details.} 
	\label{gal_merger}
\end{figure*}

In the right panel we briefly show the consequences of applying galactic winds
to our model galaxies: we proceed as in \citet{Tantaloetal1996,Tantaloetal1998} and \citet{Chiosi_etal_2017}, 
who stop  the star formation process when the energy injection by supernova explosions and stellar winds
overwhelms the efficiency of radiative cooling and the internal energy of the gas exceeds the  gravitational 
energy of this. Therefore, the gas is supposed to leave the galaxy in form of galactic winds and subsequently 
 the  
galaxy to evolve passively.  Since star formation has ceased, galaxies quickly become red and as the 
evolution 
proceeds they get redder and redder but at slower pace. Consequently, after the onset of galactic winds  the lines 
of constant age on the CMD are first squeezed in a narrow color range but also run much flatter than in the previous case.
 In the CMD, all galaxies but for those of the lowest mass  stack along a locus of very red colors 
confined in the interval $0.8<B-V<1$  for magnitudes in the range $-12 > M_V > -25$. In this case the 
theoretical Red Sequence  runs
 flatter than in the previous case and fails to match the bulk red galaxies.

The two ridge lines of the reddest colors in both panels of Fig. \ref{cmd42impro}  can be thought as a sort 
of theoretical boundaries for the
Red Sequence (locus of $z=0$, hence maximum age for galaxies on the CMD). In other words the reddest models 
 in both panels highlight the region of uncertainty or natural thickness of the 
Red Sequence due to the existence of galactic winds, or in a more general way the existence of competing 
and contrasting effects caused by gas heating by energy injection,  gas cooling by radiative processes, and gas 
trapping by gravitation. Since the interplay among all these factors likely varies from a galaxy to another, this
may easily be an additional source of  large dispersion along the Red Sequence towards the reddest colors.
 
Basing on the results of this simple analysis of the effects engendered by galactic winds, we may perhaps 
suggest that
(i) galactic winds do not proceed according to the simple mechanism envisioned by \citet{Larson1974}, sudden 
ejection of all gas
from the whole galaxy, but rather a  local loss of hot gas over a certain time 
while  star formation in other regions of the same galaxy may still be  under way 
\citep[as indicated by NB-TSPH models of][]{Merlin06,Merlin07,Merlin2012}. Therefore in real galaxies all 
situations 
encompassed by the two paradigmatic cases  are possible.      
Despite the presence of galactic winds, 
galaxies may 
still retain some gas and continue to form stars. (ii) Since in models  with sudden galactic winds  the 
lines of constant age from 13 to 4 Gyr are  squeezed in a narrow range of red colors, they 
cannot account for the 
large dispersion of colors towards the blue side of  
in the CMD. Therefore there must be other concomitant causes at work, chief among them  is 
dispersion in the galaxy formation redshift $z_f$ 
soon after which the dominant star forming episode occurs. It this sense galactic winds  are a
 \textit{secondary process} with respect to the dispersion in $z_f$ and
rejuvenation of stellar populations. In alternative, one could also consider the effects of the
dynamical interaction of a galaxy with the environment on star formation and AGN fueling. Gas could be 
stripped 
from a galaxy by ram pressure, harassment, and mergers 
\citep[see][for a recent review on galaxy formation and evolution]{Silk_Mamon_2012}.

\subsubsection{More realistic CMDs for merging galaxies}

The last step of our analysis is to generate CMDs simulating mergers of galaxies with different ages 
and masses. 
The galaxy models incorporate a realistic description of the star formation and metal enrichment histories 
and also  the occurrence of galactic winds following the energy feedback by supernova explosions and 
stellar winds.  Ages and masses of model galaxies to be merged are chosen 
by means of the Monte-Carlo procedure, but the relative occurrence probability of each mass with respect to 
the whole permitted mass spectrum  is modulated by the HGF 
of DM halos at each redshift (that by construction contains the the physics expressed by the 
``funnel'' condition). 

The procedure is as follows:
we randomly choose a mass in $[M_L, M_U]$ and then  use its own photometric history in the rest-frame 
to place the object on the CMD;
in the space $n(M, z)$ we determine the halo number density of the HGF which upon normalization is used as a
probability of occurrence and replaces the factor $f_M$ in  eqn. (\ref{eqageseedmergfunneled}).

In other words, we are repeating the experiments of Section \ref{results} using the galaxy models of 
type ET$_{ms}$ (modulated star formation) and ET$_{gw}$ (sudden galactic wind) instead of the SSP-galaxies.
We remind the reader that the galaxy models are those already shown in the two panels of 
Fig.\ref{cmd42impro} (the red solid lines), whereas the SSP-galaxies are those already  shown in Fig. 
\ref{evo_burst} (black dotted lines).  We also make use of the LT$_{ms}$ models (for the sake of simplicity,
we do not  plot their paths in the CMD). 

In these simulations, the  HGF which is customarily used as number density of halos, it is now recast 
as the 
occurrence probability of a halo of given mass and formation redshift $z_f$. This by construction replaces 
the funnel condition with its physically grounded counterpart. 

Since all the three  cosmological HGFs  in usage
are very similar each other (see Fig. \ref{usedHGF}),  nearly identical  results are to be expected 
adopting one or another HGF. Therefore we limit ourselves to show here the results for the sole HGF of  
\citet{Warren_etal_2006}.

\begin{figure*}
	\centering{ \includegraphics[width=8.5cm]{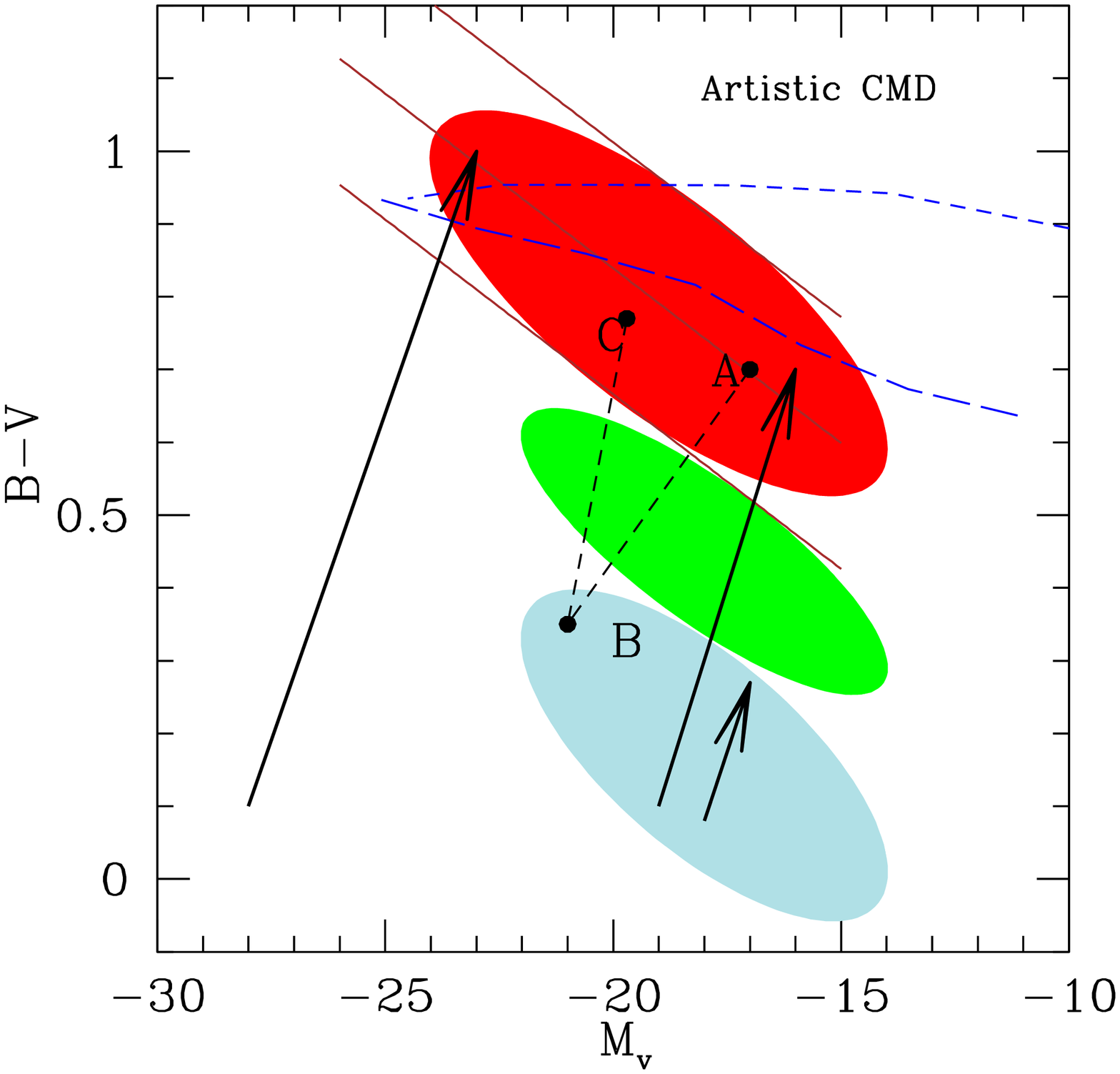}
		\includegraphics[width=8.5cm]{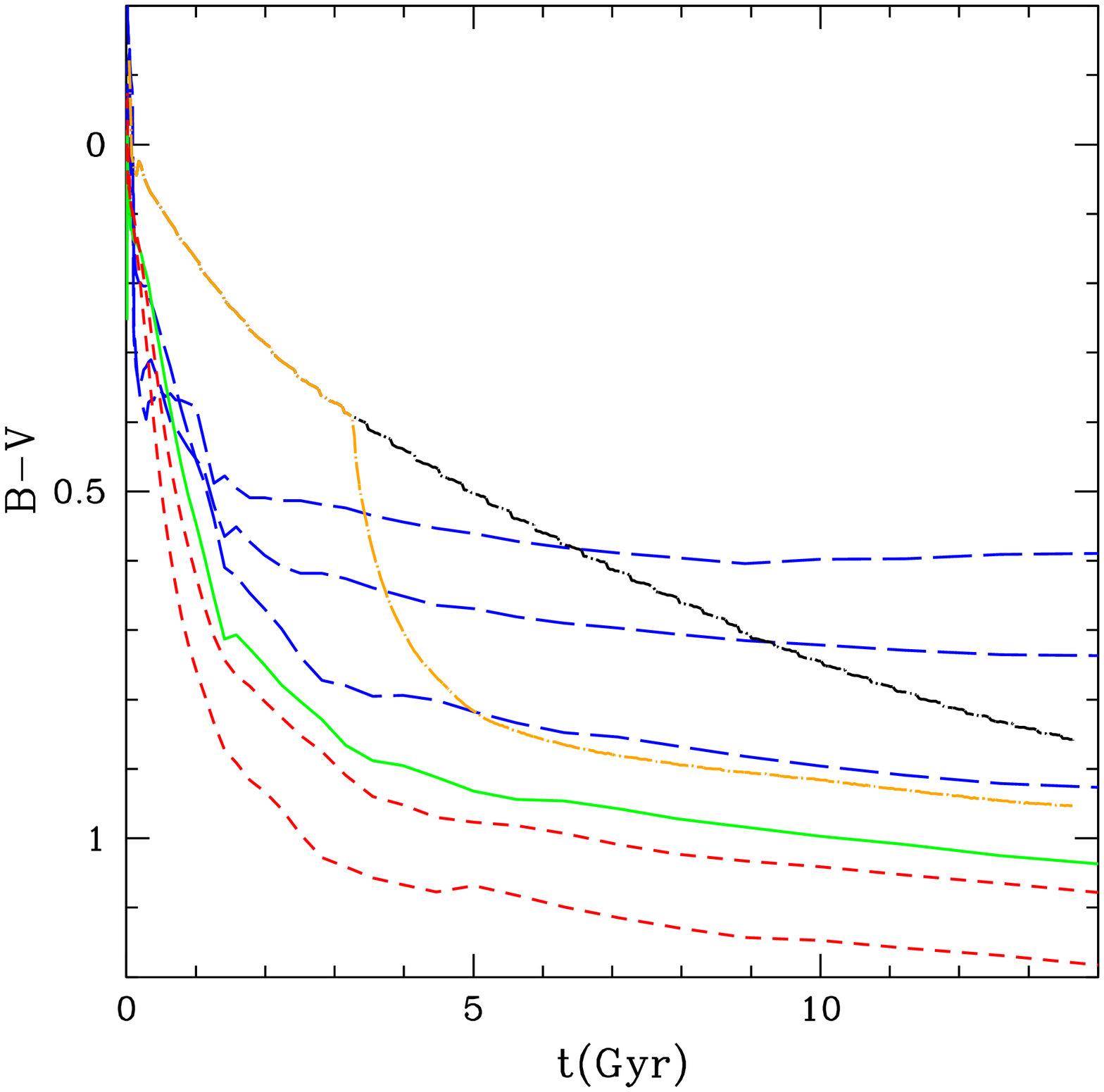}   }
	\caption{
		\textsf{Left Panel}: Artistic view of the Red Sequence, Green Valley and Blue Cloud on the CMD 
		of real galaxies. 
		Superposed are three schematic evolutionary sequences of model galaxies with 
                different mass 	(black arrows), the terminal stage at $z=0$ both in absence (long-dashed blue 
                line) and presence of galactic wind (short-dashed blue line). Schematically
		we also show the path of a red old galaxy (point A) undergoing a transient burst of star 
                formation either by
		internal or external (merger) causes (dotted line, point  B), and nearly recovering the original 
		position (point C).
		\textsf{Right Panel}: the evolutionary paths of SSPs and model galaxies  in the Color vs Age plane.
		SSPs are grouped according to their metallicity: long-dashed blue for $0.001\leq Z \leq 0.008$, solid green the solar 
		case ($Z=0.019$),  and short-dashed red for $0.40 \leq Z \leq  0.070$). The galaxy models are ET$_{ms}$ (dot short-dashed black)
		and ET$_{gw}$ (dot short-dashed orange) of $10^{11} M_\odot$.}
	\label{cmdschematic}
\end{figure*}

The results are plotted in Fig. \ref{gal_merger}, in which  the left panel displays mergers among 
galaxies randomly chosen from the ET$_{ms}$ and LT$_{ms}$ groups
and the right panel shows the same but now for groups ET$_{gw}$ and LT$_{ms}$.  
The simulated CMDs nicely reproduce the WINGS data of
\citet{Cariddi_etal_2018}, i.e. the existence of the three loci
in the CMD, i.e the Red Sequnce, the Green Valley, and the blue Group,   and  may also account for the  
extended overlapping of early and late-type galaxies pointed out 
in Fig. \ref{histo}. A remarkable crowding of points can be seen in the Blue Cloud, at variance 
with Fig. \ref{ccmjmergers} in
which only the reddest part was populated: this is essentially due to the broader temporal extension of the 
star forming activity (and hence bluer colors) in the model galaxies compared to the sharp assignment of ages to the bulk 
stellar populations of the merging SSP-galaxies.
Interestingly this feature is visible in both panels: we already know that the Blue Cloud is 
the locus occupied
by galaxies hosting very young stars, but this comparison tells us that the star formation activity is 
truly at the peak
with gas being converted into stars at maximum efficiency otherwise they would fall into the Green Valley. 
This latter being  less populated than  the Blue Cloud
lends additional support to  the notion that the Green Valley should correspond to  a rather short 
evolutionary phase 
for galaxies. Finally, in both panels we see a large number of galaxies crowding the Red Sequence, 
The two panels also highlight a point that 
could be relevant in relation to the nature and mode in which galactic winds should occur. As a matter 
of facts, 
ET$_{gw}$ and ET$_{ms}$ models predict a different slope for the red boundary of the $B-V$ distribution, 
nearly 
flat the former, and close to the observational  one (negative slope) the latter. The reddest colors for 
the two 
distributions agree with those of the observational one. The simple conclusion one might derive from this 
is that the
\citet{Larson1974} model of galactic winds  does not find correspondence in what implied by the observational
data. This suggestion is not new  because it has been pointed out several times 
\citep[see for instance][]{Chiosi_etal_1998, Chiosi_Carraro_2002, Merlin2012}, 
however it is also confirmed by the analysis of the CMDs derived from  the integral  $B$ and $V$ 
magnitudes of cluster 
galaxies. Insufficient correction for reddening by dust could be a reasonable way out and an 
alternative at the same time.
However, the issue requires further investigation before drawing any conclusion.

\begin{table}
\begin{center}
\caption{Age and redshift variation of the theoretical CMR $(B-V) = A\cdot M_V +B $  from models $ET_{ms}$. }
		\label{cmr_tab}
		\begin{tabular}{|c|c|c|c|c|}
\hline
$T_U$ (Gyr) &   $z$  &  A      & B         &  rms   \\
\hline
     13     &  0.06  & -0.023  &    0.359  & 0.011  \\
     12     &  0.14  & -0.023  &    0.329  & 0.011  \\
     11     &  0.23  & -0.023  &    0.301  & 0.012  \\
     10     &  0.34  & -0.022  &    0.282  & 0.012  \\
     9      &  0.46  & -0.021  &    0.266  & 0.014  \\
     8      &  0.60  & -0.019  &    0.263  & 0.012  \\
     7      &  0.78  & -0.017  &    0.259  & 0.013  \\
     6      &  0.99  & -0.014  &    0.262  & 0.011  \\
     5      &  1.27  & -0.011  &    0.263  & 0.008 \\
     4      &  1.66  & -0.009  &    0.261  & 0.006 \\
\hline
\end{tabular}
\end{center}
\end{table}

\subsubsection{The Red Sequence, Green Valley, and Blue Cloud }

The  evolutionary tracks  of model galaxies in Fig. \ref{cmd42impro} (red lines) and their evolutionary rate 
along these paths (the stages marked with different symbols according to the considered redshift interval)
clearly show that depending on the galaxy 
mass and mean metallicity, and the complicated physics of the galactic gas, various cases are possible. 

In view of the discussion below, in Fig. \ref{cmdschematic} we schematically show
the  Red Sequence, Green Valley, and Blue Cloud   together with the evolutionary 
paths of galaxies on the CMD. Specifically, we draw the Red Sequence with inclination and thickness 
using the values quoted in Section 2.      
Let us shortly present the different  cases  as a function of the galaxy mass: 
(i) Galaxies in the mass interval 
$10^{13} > M > 10^{12}\, M_\odot$ can be found on the Red Sequence only. This means that today no galaxy 
with this mass can be seen in  the Green Valley or Blue Cloud: ongoing star formation either in isolation 
or as consequence
of mergers does no longer occur. (ii) Galaxies in the mass interval $10^{12}  >  M > 10^{9}\, M_\odot$ can 
be in any of the three regions 
depending on their individual evolutionary history. If they evolve in isolation, at beginning when the 
star formation is very strong they populate  the Blue Cloud, cross the Green Valley towards the end of 
the star forming
activity they cross the Green Valley (fewer stars are formed and the stellar content gets older and redder),  
finally when the star formation activity extinguishes, they become old and red.
If they undergo mergers when they are on the Red Sequence, they first fall either on the Green Valley or Blue Cloud 
depending on the intensity 
of the burst of star formation. For small bursts engaging a small fraction of the total star mass, 
when the burst is over
the  CMD galaxy may recover the old position on the Red Sequence. In contrast, for  mergers among objects of 
comparable mass but different ages, the burst is very intense, and afterwards the composed object  can hardly recover 
the original position on the Red Sequence but remains bluer and brighter for very long times. 
(iii) Galaxies with masses in the interval $10^{9} > M > 10^{7} \, M_\odot$     
are not present in the WINGS data because of the limit magnitude of the survey but they exist in the ILLUSTRIS-1 
simulations, that in turn likely suffer  of incompleteness due to numerical resolution in the mass interval
$10^{8} > M > 10^{6} \, M_\odot$. Therefore the discussion of their evolution on the CMD is not of interest here.

The above considerations are supported and
strengthened by the variation of $B-V$ color as a function of the age for both SSPs and model galaxies, shown 
in the right panel of Fig. \ref{cmdschematic}. Here we display a few SSPs with different metallicity 
(blue for $0.001\leq Z \leq 0.008$, green for the solar case $Z=0.019$,  and red for $0.40 \leq Z \leq  0.070$). 
We also show two galaxy models of the same mass,  $10^{11} M_\odot$ but different evolutionary histories, namely  
ET$_{ms}$ (yellow)  and ET$_{gw}$ (black). Both SSPs and galaxy models behave as expected: the SSPs are very blue
when young massive stars are present; they get red as these latter disappear and when AGB first and RGB  
plus AGB stars are present; they get even redder on rather short timescales (fast evolution on the CMD plane)
because the RGB  phase dominates the color evolution, and remain red at nearly constant color afterwards. The 
two model galaxies have a similar appearance at the older ages (namely older than 12 Gyr):
they remain very blue as long as star formation is active, because  massive young stars are present, and 
plunge to red colors after star formation ceases, with a rate depending on the strength of winds inside the
galaxy itself.

\begin{figure}
	\centering{
		\includegraphics[width=8.5cm]{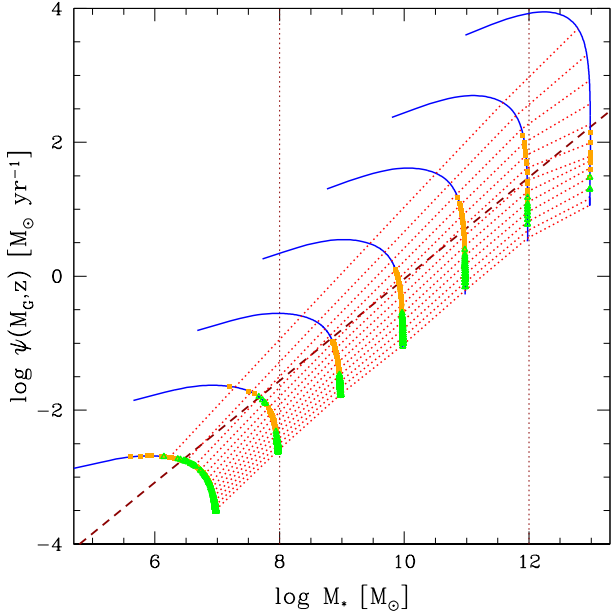}}
	\caption{SFR versus Stellar Mass: the solid squares are model galaxies observed at
		$0.1<z\leq0.7$ (orange solid squares) and $z\leq0.1$ (green open triangles);
		the blue solid lines are the evolutionary paths (from left to right) of each model galaxy; the
		dark-red dashed line represents the \citet{RenziniPeng2015} average relation
		for star-forming galaxies, which range of validity is
		identified by the two black dashed vertical lines; the 
		red dotted lines connect points of same 
		age for each
		mass, in time steps of 1 Gyr  from 1 to 13 Gyr from top to bottom.}
	\label{RenziniPeng}
\end{figure}

\subsection{Physical meaning and evolution of the CMR}
The observational data show two distinct groupings of red galaxies: a first one made by those crowding 
along the Red Sequence 
with mean slope $A=-0.048$, and a second  less numerous group whose  (B-V) colors exceed the upper limit 
of the CMR itself and  run nearly constant with the absolute magnitude (mass). Furthermore, among the brightest 
galaxies of both groups the colors become nearly independent of the luminosity (mass): the CMR flattens out. 

The theoretical models also
show  a similar behavior that seems to  
depend on the interplay between star formation and galactic winds. The models ($ET_{ms})$ in which 
the energy  feedback instead of energizing the galactic wind goes into modulating (decreasing) the SFR, 
their present-day (B-V) colors run along a sequence whose slope (see the entries of Table \ref{cmr_tab}) 
is comparable to the observational one of the first group above. In contrast, models ($ET_{gw})$ whose 
energy feedback  
straightly goes into the galactic wind, so that they suddenly extinguish star formation and evolve passively 
ever since, reach present-day (B-V) colors close to the limit values of the second group and remain nearly 
constant with the absolute magnitudes. Finally, for  the brightest galaxies the colors seem to become constant  at 
the mean value (B-V)=0.95  (see both panels of Fig.\ref{gal_merger}), the typical value for luminous early type objects.

Furthermore, it is implicit that when galactic winds occur for whatever reason
 a galaxy  switches from the first to the second group. 

Finally, the constant (B-V) color of the most luminous galaxies could be attributed to a constant metallicity: the 
mass-metallicity sequence of the 
CMR breaks down for this type of galaxies, in agreement with recent observational data and theoretical NB-TSPH simulations
\citep[e.g.][and references]{Tremonti_etal_2004, Mannucci_etal_2010,Mannucci_Cresci_2012, Cresci_etal_2012, 
Bothwell_etal_2013, Torrey_etal_2014, Torrey_etal_2017, Matsuoka_etal_2018} 
   and theoretical NB-TSPH simulations \citep{Chiosi_Carraro_2002,Merlin06,Merlin07,Merlin2012}.

The last interesting issue to address here is the time dependence of the CMR as function of the age. 
In both 
panels of Fig.\ref{cmd42impro} the red dashed line visually show the CMR at different ages going from 13 Gyr 
(top)  to 1 Gyr (bottom).  While the CMR gradually varies from 1 Gyr to the present and the slope gently gets 
steeper with  increasing age for  galaxy models with no galactic winds, much more drastic changes occur in 
the 
presence of sudden galactic winds that usually set in  early in the evolutionary history ($T_G \lesssim 4$ 
Gyr)  of a galaxy. Past 4 Gyr the CMR is much more regular, uniform and nearly flat. This unlikely 
behavior of the CMR strongly suggest that  the \citet{Larson1974} mode of galactic wind is too a crude 
description of this phenomenon: therefore in the following we limit ourselves to the case of $ET_{ms}$ models 
to discuss the time evolution of the CMR(t). In Table \ref{cmr_tab} for a few select value of the 
rest-frame galaxy ages, and galaxy masses from $10^7$ to $10^{12} \, M_\odot$, and using a linear 
least-square fit of the data ($B-V = A\cdot M_V + B$) we derive the slope $A$, the zero-point $B$ and the 
r.m.s.  The slope remains nearly the same from 13 Gyr down to 8 Gyr (z=0.6), whereas in the age interval 8 to 
4 (z=1.66) it goes down to nearly zero. To discuss the time dependence of the CMR with the observational data 
to our disposal is not possible. Because of the very small redshift coverage \citep[e.g. 
z=0.04  to z=0.07 in ][]{Cariddi_etal_2018} and also because theoretical magnitudes and colors in other 
photometric systems and pass-bands would be required. Therefore we leave the subject to a future study.
In any case, we would like to mention here that starting from the rest-frame age of 6 Gyr, i.e. z=0.99 
$\simeq$ 1 the slope of the CMR is significantly different from zero and stable in agreement with observational data on high 
redshift objects \citep{Gladders_etal_1998, Tran_etal_2007, Mei_etal_2009}.  

The conclusion one would draw from the above picture is that the inclination of the crowded Red 
Sequence implies that  a minor 
residual stellar activity is at work or ceased in a very recent past. The upper boundary may be sensitive to 
the underlying mechanism of energy feed-back in driving galactic winds and regulating the efficiency of star 
formation and  chemical enrichment. In other words the CMR in addition 
to being an indicator of mass-metallicity sequence is also the signature of the past  star formation 
activity. 
Finally,  looking at the Red Sequence sub-populations one could infer about the kind of star formation 
history of galaxies.

Before closing this section, we like to shortly comment on the potential of the CMR for cosmological 
studies. A great  deal of observational data suggest that early-type galaxies are nearly coeval, had their 
last episode of intense star formation at high redshift ($z \geq 2$) and evolved nearly passively since then 
\citep[see e.g.][]{Bower_etal_1992a, Bower_etal_1992b,  Aragon-Salamanca_etal_1993, 
Pahre_etal_1996, vanDokkum_etal_1998, Stanford_etal_1998, Gladders_etal_1998, 
Chiosi_2002, Chiosi_Carraro_2002, 
Peebles_James_2002, Chiosi_2007,Chiosi_etal_2014}. Consequently, the CMR can be in place at high redshift and 
thus be a great potential tool for cosmological studies \citep{Sandage_1972}. In particular, any change in 
its slope could signal changes in the star formation
history (and/or metallicity). In this context, starting from previous work by 
\citet{Visvanathan_Sandage_1977}, \citet{Sandage_Visvanathan_1978a, Sandage_Visvanathan_1978b}, and 
\citet{Bower_etal_1992a,Bower_etal_1992b},
 \citet{Lopez_Cruz_etal_2004} addressed the question of the universality of the CMR in cluster early-type  
galaxies. Analyzing the CMR of 57 galaxy clusters of the Abell sample over the redshift interval $0.02 < z< 
0.2$,  they found that the CMR is 
linear over  about eight magnitudes with no breaks in the slope, and that changes in the slope can be 
explained by metallicity and passive evolution. Furthermore, they pointed out that the CMR could be an 
accurate indicator of membership and redshift. 
We would like to note that the results presented in Figs. \ref{cmd42impro}, \ref{gal_merger}, and \ref{cmd42iall} 
could be used for this kind of studies. However, this is beyond the aims of this study and is left to 
future investigation.

\subsection{SFR vs Stellar Mass Relationship}
We conclude the analysis of model galaxies by addressing the issue of the relation
between SFR and stellar mass, vividly debated in literature
\citep[see e.g. ][and references therein]{RenziniPeng2015}. This gives also the possibility
to show the underlying physics at work for model galaxies exploited until now in terms of photometry.
Since one needs galaxies with ongoing star formation at 
the present time ($z=0$) we consider only galaxy models with no galactic winds, i.e. with residual
star-forming activity. The theoretical SFR versus stellar mass relation  
is shown in Fig. \ref{RenziniPeng}. 
We schematically build it up as Fig. \ref{cmd42impro}, because randomness of
$z_f$ plays a key role in shaping also this relation.  The solid blue lines show the relations for galaxy models of 
increasing mass as indicated, whereas the dotted lines indicate loci of different ages in the rest-frame: 
we refer to these latter as the age grid. Finally, the 
dark red dashed line is the empirical mean relation by  \citet{RenziniPeng2015}
for star-forming galaxies in the SDSS DR7 sample  spanning the  
redshift range  $0.02<z<0.08$.   
To better compare the observational data for the SFR vs stellar mass relationship with our randomly-born 
model galaxies, 
for these latter we also plot the stages in the redshift interval  $0 \leq z\leq 0.1$  (the green triangles). 

An overall agreement with observations
is evident: the slope of our synthetic dataset is nearly identical to that of the observational data or 
their best-fit relationship.  However,
the model galaxies at the ridge line for  $z\leq0.1$  are on average 0.5 dex below  the \citet{RenziniPeng2015}
relation. This means that either the model galaxies have a SFR 0.5 dex lower than observed  or a stellar mass 0.5 dex 
larger
than estimated for the observed galaxies or a suitable combination of the two effects. As far as the SFR is concerned, 
the one of the model galaxies in the redshift interval $0 \leq z\leq 0.1$ is  already in the long tail following the 
peak value  that occurred in the past. To evaluate the effect of it,  we also plot the model galaxies
as observed  in the redshift interval $0.1<z\leq0.7$ (orange solid squares),
which broadly corresponds to include galaxies at a look-back time of
$7 \lesssim t_{lb}(\text{Gyr}) \lesssim 12.5$: by including these younger and gas-richer galaxies,
the \citet{RenziniPeng2015} relation is fully recovered. We remark that the case of model galaxies with the 
\citet{Larson1974} scheme for 
galactic wind  cannot be plotted in this diagram, because their present-day SFR would be zero. However 
keeping mind that in real galaxies the galactic winds seem to occur any time a fraction of gas may be hot enough to 
acquire enough kinetic energy to escape the galaxy, whereas the remaining cold part may still form stars 
\citep[see the NB-TSPH models of ][for more details]{Merlin06,Merlin07,Merlin2012} so that some 
residual star formation may always be present, the analysis could also be extended to include this case. 
However, doing so would not add very much to the present discussion.    

The comparison between the \citet{RenziniPeng2015} relation and our age grid
sheds light on two important features. First, the mean SFR vs mass relationship
(red dashed line) crosses the rest-frame tracks of the models at values of the ages that vary with the mass: this
can be seen by looking, for each mass, at the point of the age grid nearest to the dark red dashed line.
On this basis, our simple model shows that objects belonging to the group  of ``star-forming galaxies" are not coeval.
In addition, the dark red dashed line crosses each rest-frame curve
at an age which increases with increasing mass: shortly speaking, on average greater stellar masses are associated 
with older galaxies.
This argument, in addition to the need of bursts and mergers to explain colors of cluster galaxies, leads to 
enforce the idea
that galaxies as stellar assemblies are further older systems.
Secondly,  the slopes of the theoretical SFR vs stellar-mass relationship 
are practically identical to that of the \citet{RenziniPeng2015} relation.
This is a reminiscence of the result found in \citet{Chiosi_etal_2017} (their Fig. 11,
top panel) with same model galaxies but a slightly different cosmological approach.  All this, 
confirms that, in spite of its complexity and caveats,
our model does  a good job in placing galaxies inside a physically sounded cosmological 
context. 

In conclusion, we can briefly say that both the distribution of galaxies on the CMD and the SFR vs stellar mass 
relationship are not
straightforward results because they actually stem from  the contributions of several
different physical properties of galaxy formation and evolution.

\section{Final remarks and perspectives for future studies} \label{conclusions}

We have investigated the CMD of cluster galaxies by exploring the phenomena
that could affect  the intensity and color  of the light they emit: these can be bursts and/or mergers 
inducing  rejuvenation of the stellar component and chemical enrichment,  or galactic winds of internal origin 
or gas removal by external causes both  stripping a galaxy of
its gas and initiating passive evolution of this latter. All these phenomena are more likely to occur in galaxy clusters 
than in the field.
We have analyzed the impact of these two different physical processes on the evolution by means of two
approaches, i.e.  galaxy-sized SSPs and  
realistic chemo-photometric models of galaxies.
Among the main results of our study, we recall the following ones:

\noindent 1) Simulations of bursts of star formation based on SSP-galaxies show the effect of a younger stellar component 
of arbitrary age, metallicity and mass
on the light emitted by an assembly of  coeval stars with different  age, metallicity and mass. On the $M_V-(B-V)$  plane, 
adding a younger component  leads the host  galaxy to shift leftwards and downwards, 
i.e. towards higher $V$ luminosities and bluer $B-V$ colors:
the shift has an amplitude increasing with the mass of the stars engaged in the burst  with respect 
to the mass of the host galaxy and the difference between the mean ages of the two components. Masses and ages are the 
dominant parameters, the metallicity plays a secondary role.
The net effect of random bursts is to broaden the Red Sequence at $z=0$ making it to stretch down to 
 the reddest part of the Green Valley. Roughly speaking,  bursts younger than 3 Gyr already account for more than 
60$\%$ of the width of the Green Valley.

\noindent 2) To further test the dispersion inside CMDs engendered by
age differences among the underlying stellar populations, we present the simulations of single
mergers, which at a first glance provide a much more effective scatter 
than similarly massive bursts. This is due to the fact that mergers involve a broader mix
of ages, i.e. the more massive galaxy in the merger episode may not necessarily be the
older one, together with a broader mix of metallicities. The merger simulations present a rich Red Sequence, 
which slope nicely matches the observational one, and a well crowded
Green Valley. Finally, they also partially populate the Blue Cloud, thus suggesting that
fractions of young stellar
populations  larger  than allowed by typical bursts are needed.

\noindent 
3) Basing on the CMDs for mergers of SSP-galaxies, we heuristically demonstrated that 
mergers  between galaxies  cannot occur independently of the age and mass of the interacting galaxies. 
Were this the case, mergers would populate regions of the CMD that are seen void of objects. The region in 
question is the triangle with basis $-20 > M_V > -25$ and height $0.2 < B-V < 0.9$ 
(see the left panel of Fig. \ref{ccmjmergers}), which we named \textit{zone of avoidance}). 
In other words, not all values of age and mass for the merging 
galaxies are permitted. The avoidance zone is simply telling the following: \textit{the probability that 
a galaxy can merge with another of similar mass decreases at increasing the  galaxy mass and hence, in the 
hierarchical scheme,  the age}.  As matter of facts, only a merger (burst) involving less that about ten percent 
of the total star mass may leave a post-merger (burst) descendant similar to the progenitor object.  The reddest and
brightest objects of the Red Sequence  crowd in a rather compact group and have no blue counterparts of the same 
luminosity. This means that they have not acquired a significant amount of mass in a recent 
time and by virtue of their own color they have remained the same since long time. Therefore, the red bright objects are 
preferentially 
made of old stars. The goal has been achieved by introducing a filter (named funnel) for masses and ages of 
the merging galaxies.  The same filter allows us to account for the galaxies  crowding the Green Valley in a 
coherent fashion.  Later the filter has been identified with the HGF of the cosmological building up of 
galaxies. 

\noindent 4) Spurred by the results obtained with SSPs, we turned to galaxy models
nicely simulating the collapse of BM into the gravitational potential well of DM, 
including star formation and chemical
enrichment,  energy feedback from various sources, and the effect of this in modulating the SFR and triggering
galactic winds. We coupled them with a set of HGFs, in turn
obtained from their respective HMFs \citep[i.e.][]{Warren_etal_2006,Angulo_etal_2012,Behroozi_etal_2013},
in order to assign random formation redshifts.
In this context, we  showed that a great deal of the information carried by the precision cosmology
is lost when analyzing the integrated properties of galaxies, so that any of the three
HMFs used in this work provides qualitatively the same results. We exploited
the HGFs to build up our samples of synthetic galaxies, the  demography of which  has been
validated by comparison with the most massive galaxy cluster in the ILLUSTRIS-1 simulation.

\noindent 5) We have set up  three groups of galaxy models, namely the early-type galaxies with
sudden global galactic wind (ET$_{gw}$) and those with star formation modulated by the feedback energy 
injection
and continuous galactic winds that remove part of the gas without stopping star formation (ET$_{ms}$), and 
late-type galaxies (LT$_{ms}$). With aid of these models we have  repeated
the merger simulations by applying the HGF as the cosmological funnel to assign ages and masses to 
galaxies.
The results of this experiments mainly recover  those with SSP-galaxies, however the major difference that
a more prominent and adequate occupation of the three loci is at work. Again a simply tuned
approach.  

\noindent 6) The simulated samples have proven to be solid in explaining CMDs of cluster galaxies,
together with the SFR vs stellar mass relationship  which is also related to the light emitted by stars.
Although with some caveats regarding the crudeness of our  approach (e.g. only one merger per galaxy 
is allowed), we
can conclude that our model provides consistent insights on phenomena at work during
the formation and evolution histories of galaxies. 

\noindent 
7) Finally, we would like to note  that despite the many drawbacks and limitations
our simple  tool for describing the population of galaxies in a cluster is able to get the essence of a 
number of physical phenomena shaping  the CMDs of cluster galaxies. 
The present approach is not meant to replace full cosmological simulations but simply to  
to play an ancillary role in testing many physical assumptions and ingredients before they are 
incorporated in heavy numerical simulations. 

\section*{Acknowledgments}
We would like to thank the anonymous referee for the many constructive remarks and suggestions that 
grealty helpd us to improve the content and layout of the paper.
M.S. thanks Giulia Despali, Chiara Mancini and Alessia Moretti for helpful discussions and hints.
C.C. would like to thank Rosaria Tantalo and Emanuela Chiosi for valuable collaboration, and the
Department of Physics and Astronomy  of the Padua University for the 
friendly hospitality and computational support. The tool HMFcalc and the public data from the ILLUSTRIS-project have
been amply used in this study.

\bibliographystyle{mn2e}

\bibliography{CCMD_BIBLIO}


\appendix

\section{Magnitudes and colors of SSPs}\label{SSP_IMF_COL}

Table \ref{table_mag_col}  provides a short summary of magnitudes and colors in the Johnson-Bessell-Brett 
photometric system for SSPs with solar composition (metallicity $Z=0.019$), a few selected values of the age,
and three popular IMFs, namely \citet{Chabrier_2014}, \citet{Kroupa2008}, and  \citet{Salpeter1955}, and 
references therein. The SSPs are from the library calculated by \citet{Tantalo05}. 
It is evident that while 
the magnitudes depend on the 
adopted IMF, the colors are much less affected, so that  passing from one IMF to another could be detected 
only with very accurate photometry. Ages are in years.

\begin{table*}[h]
\begin{center}
\caption{ The first column is the logarithm of the age in years. }
\begin{tabular}{|c c c c c c c c c c c c c c|}
\hline
  log(Age)  &    V   &  BOL   &    BC  &  U-B    & B-V    &  V-R   &  R-I  & V-J   & V-H   &  V-K  & V-L   & V-M   & V-N  \\ 
\hline
\multicolumn{14}{|c|}{ CHABRIER }\\
 \hline
 7.00 &  4.935 &  3.228 & -1.707 & -0.632  & 0.216  & 0.269  & 0.326 & 1.222 & 1.811 & 1.888 & 1.973 & 1.797 & 1.993 \\
 8.00 &  6.658 &  5.987 & -0.671 & -0.267  & 0.133  & 0.172  & 0.224 & 0.861 & 1.358 & 1.422 & 1.485 & 1.347 & 1.494 \\ 
 9.00 &  8.476 &  8.074 & -0.402 &  0.194  & 0.546  & 0.363  & 0.419 & 1.687 & 2.393 & 2.536 & 2.620 & 2.515 & 2.613 \\ 
 9.50 &  9.686 &  9.049 & -0.638 &  0.407  & 0.867  & 0.539  & 0.567 & 2.104 & 2.866 & 2.996 & 3.086 & 2.953 & 3.084 \\ 
10.00 & 10.851 & 10.061 & -0.789 &  0.625  & 0.991  & 0.595  & 0.604 & 2.236 & 3.010 & 3.143 & 3.234 & 3.103 & 3.232 \\ 
10.15 & 11.181 & 10.282 & -0.899 &  0.702  & 1.030  & 0.616  & 0.623 & 2.308 & 3.092 & 3.229 & 3.322 & 3.193 & 3.320 \\
\hline
\multicolumn{14}{|c|}{ KROUPA }\\
\hline            
 7.00 &  0.261 & -1.425 & -1.686 & -0.620  & 0.200  & 0.256  & 0.314 & 1.181 & 1.761 & 1.837 & 1.922 & 1.748 & 1.942 \\ 
 8.00 &  1.735 &  1.091 & -0.644 & -0.255  & 0.128  & 0.163  & 0.212 & 0.816 & 1.294 & 1.357 & 1.418 & 1.286 & 1.426 \\
 9.00 &  3.238 &  2.869 & -0.368 &  0.183  & 0.537  & 0.356  & 0.407 & 1.627 & 2.312 & 2.450 & 2.533 & 2.430 & 2.527 \\ 
 9.50 &  4.251 &  3.638 & -0.613 &  0.393  & 0.856  & 0.533  & 0.558 & 2.067 & 2.820 & 2.948 & 3.037 & 2.906 & 3.035 \\ 
10.00 &  5.231 &  4.452 & -0.779 &  0.624  & 0.989  & 0.595  & 0.603 & 2.226 & 2.995 & 3.127 & 3.218 & 3.089 & 3.217 \\ 
10.15 &  5.527 &  4.633 & -0.893 &  0.705  & 1.031  & 0.618  & 0.623 & 2.304 & 3.086 & 3.222 & 3.315 & 3.188 & 3.314 \\ 
\hline
\multicolumn{14}{|c|}{ SALPETER }\\
\hline   
 7.00 & -0.705 & -2.388 & -1.683 & -0.619  & 0.253  & 0.291  & 0.346 & 1.285 & 1.881 & 1.959 & 2.045 & 1.869 & 2.066 \\   
 8.00 &  1.238 &  0.571 & -0.668 & -0.264  & 0.137  & 0.175  & 0.227 & 0.872 & 1.370 & 1.435 & 1.499 & 1.362 & 1.506 \\  
 9.00 &  3.010 &  2.614 & -0.396 &  0.192  & 0.546  & 0.364  & 0.419 & 1.676 & 2.375 & 2.517 & 2.601 & 2.498 & 2.594 \\ 
 9.50 &  4.138 &  3.506 & -0.633 &  0.406  & 0.866  & 0.541  & 0.568 & 2.097 & 2.853 & 2.983 & 3.074 & 2.944 & 3.070 \\   
10.00 &  5.189 &  4.398 & -0.791 &  0.634  & 0.997  & 0.602  & 0.613 & 2.245 & 3.010 & 3.145 & 3.238 & 3.114 & 3.232 \\  
10.15 &  5.479 &  4.581 & -0.899 &  0.714  & 1.038  & 0.626  & 0.633 & 2.319 & 3.095 & 3.233 & 3.328 & 3.208 & 3.322 \\ 
\hline\end{tabular}
\end{center}
\label{table_mag_col}
\end{table*}

\section{HGFs and CMD simulations for cluster galaxies}\label{app_galnum_analysis}

\begin{table*}
	\begin{center}
		\caption{ Coefficients of the polynomial interpolation of \citet{Warren_etal_2006} HGF.}
		\label{coef_w06}
		\begin{tabular}{|c|c|c|c|c|c|}
			\hline
			$\log M\,(M_{\odot}h^{-1})$ & A$_4$      &    A$_3$       &    A$_2$      &    A$_1$       &    A$_0$ \\
			\hline
			7  &  2.00391e-05 & -0.00069 & -0.00172 & 0.06186 &  2.75945 \\
			8  &  1.96696e-05 & -0.00064 & -0.00597 & 0.07659 &  2.50057 \\
			9  &  1.68349e-05 & -0.00048 & -0.01349 & 0.09530 &  1.57895 \\
			10 &  1.05238e-05 & -0.00019 & -0.02622 & 0.11542 &  0.67160 \\
			11 & -5.80424e-07 &  0.00031 & -0.04848 & 0.13077 & -0.21784 \\
			12 & -1.90090e-05 &  0.00113 & -0.09057 & 0.12322 & -1.09097 \\
			13 & -4.46934e-05 &  0.00218 & -0.17571 & 0.04452 & -2.21666 \\
			\hline
		\end{tabular}
	\end{center}
\end{table*}

\begin{table*}
	\begin{center}
		\caption{ Coefficients of the polynomial interpolation of \citet{Angulo_etal_2012} HGF.}
		\label{coef_a12}
		\begin{tabular}{|c|c|c|c|c|c|}
			\hline
			$\log M\,(M_{\odot}h^{-1})$ & A$_4$      &    A$_3$       &    A$_2$      &    A$_1$       &    A$_0$ \\
			\hline
			7  &  2.32293e-05 & -0.00090 &  0.00198 & 0.02171 &  2.87802 \\
			8  &  2.77379e-05 & -0.00103 & -0.00076 & 0.03137 &  2.60891 \\
			9  &  3.11068e-05 & -0.00110 & -0.00692 & 0.04606 &  1.67251 \\
			10 &  3.03550e-05 & -0.00098 & -0.01955 & 0.06583 &  0.74492 \\
			11 &  2.19534e-05 & -0.00054 & -0.04451 & 0.08647 & -0.17097 \\
			12 & -8.06725e-07 &  0.00049 & -0.09497 & 0.08979 & -1.07614 \\
			13 & -3.75088e-05 &  0.00197 & -0.19783 & 0.01701 & -2.23637 \\
			\hline
		\end{tabular}
	\end{center}
\end{table*}

\begin{table*}
	\begin{center}
		\caption{ Coefficients of the polynomial interpolation of \citet{Behroozi_etal_2013} HGF.}
		\label{coef_b13}
		\begin{tabular}{|c|c|c|c|c|c|}
			\hline
			$\log M\,(M_{\odot}h^{-1})$ & A$_4$      &    A$_3$       &    A$_2$      &    A$_1$       &    A$_0$ \\
			\hline
			7  &  2.16291e-05 & -0.00070 & -0.00221 &  0.04123 &  2.78376 \\
			8  &  2.09428e-05 & -0.00064 & -0.00643 &  0.05307 &  2.52830 \\
			9  &  1.93874e-05 & -0.00054 & -0.01316 &  0.06578 &  1.61067 \\
			10 &  1.68678e-05 & -0.00039 & -0.02392 &  0.07592 &  0.70708 \\
			11 &  1.34064e-05 & -0.00019 & -0.04253 &  0.07624 & -0.18019 \\
			12 &  7.44482e-06 &  0.00016 & -0.07894 &  0.04842 & -1.05531 \\
			13 & -2.09921e-06 &  0.00059 & -0.15666 & -0.05354 & -2.19023 \\
			\hline
		\end{tabular}
	\end{center}
\end{table*}

\begin{figure*}[h]
	\centering{
		\includegraphics[width=8.0cm]{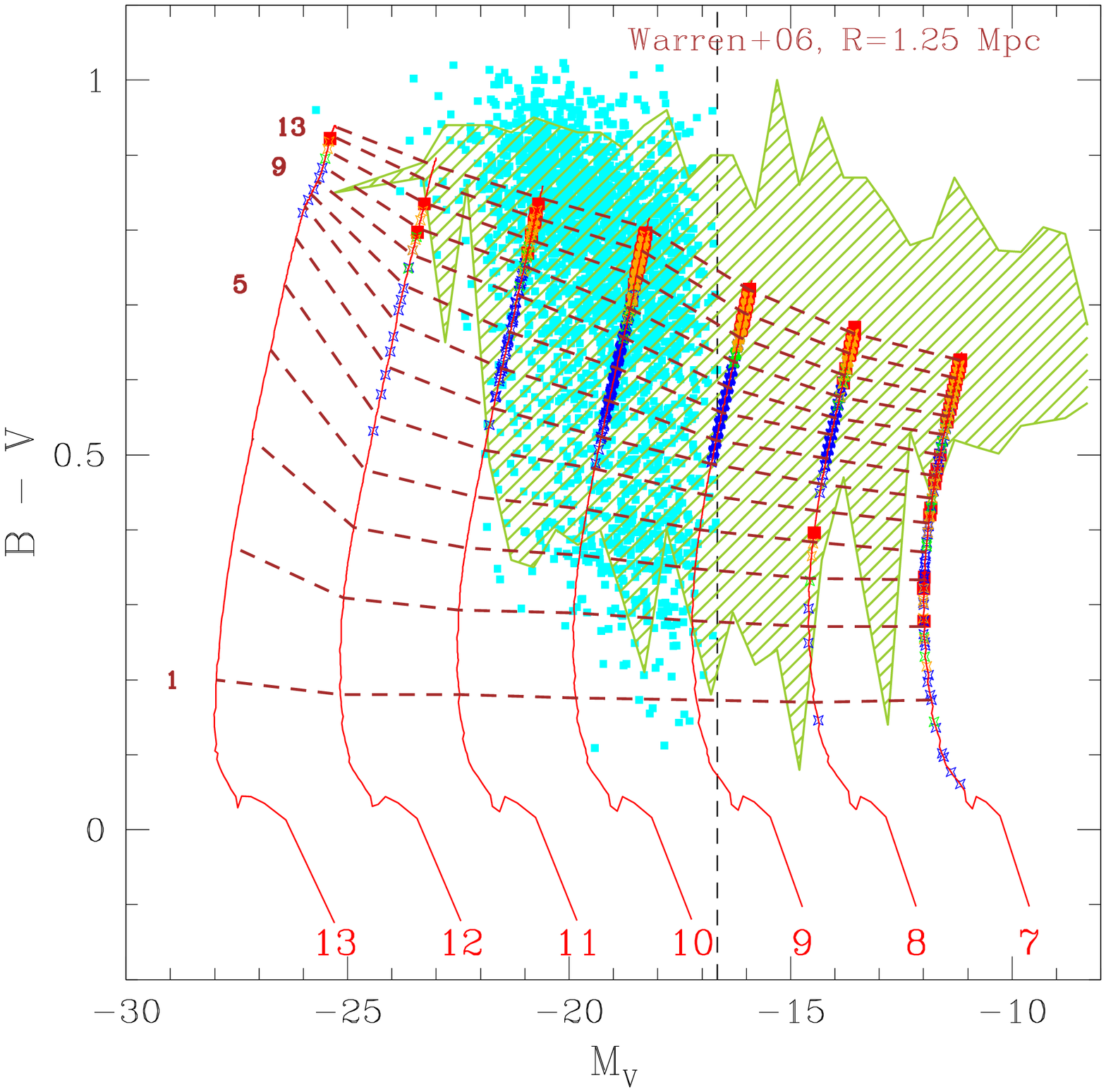}
		\includegraphics[width=8.0cm]{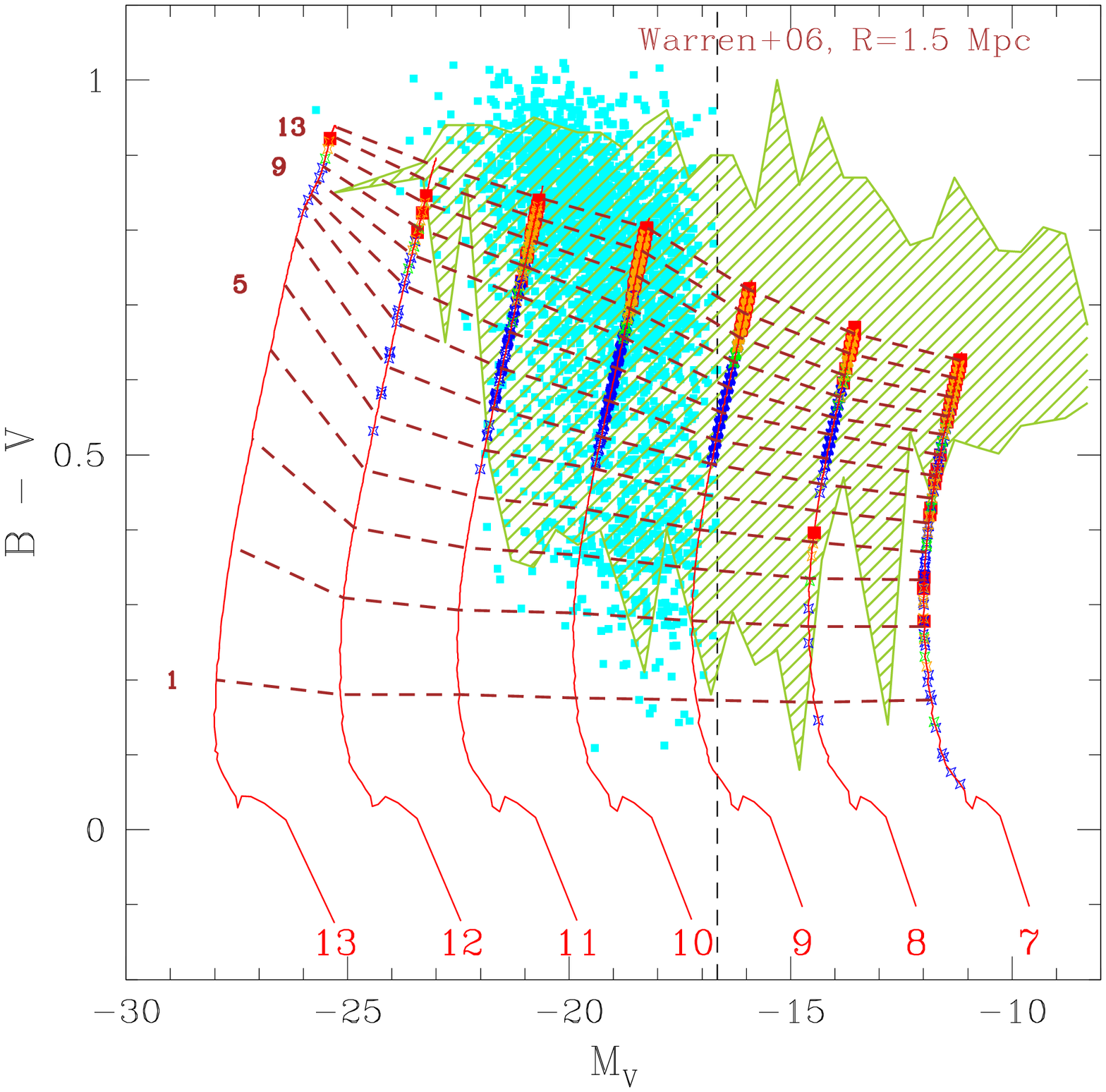}
		\includegraphics[width=8.0cm]{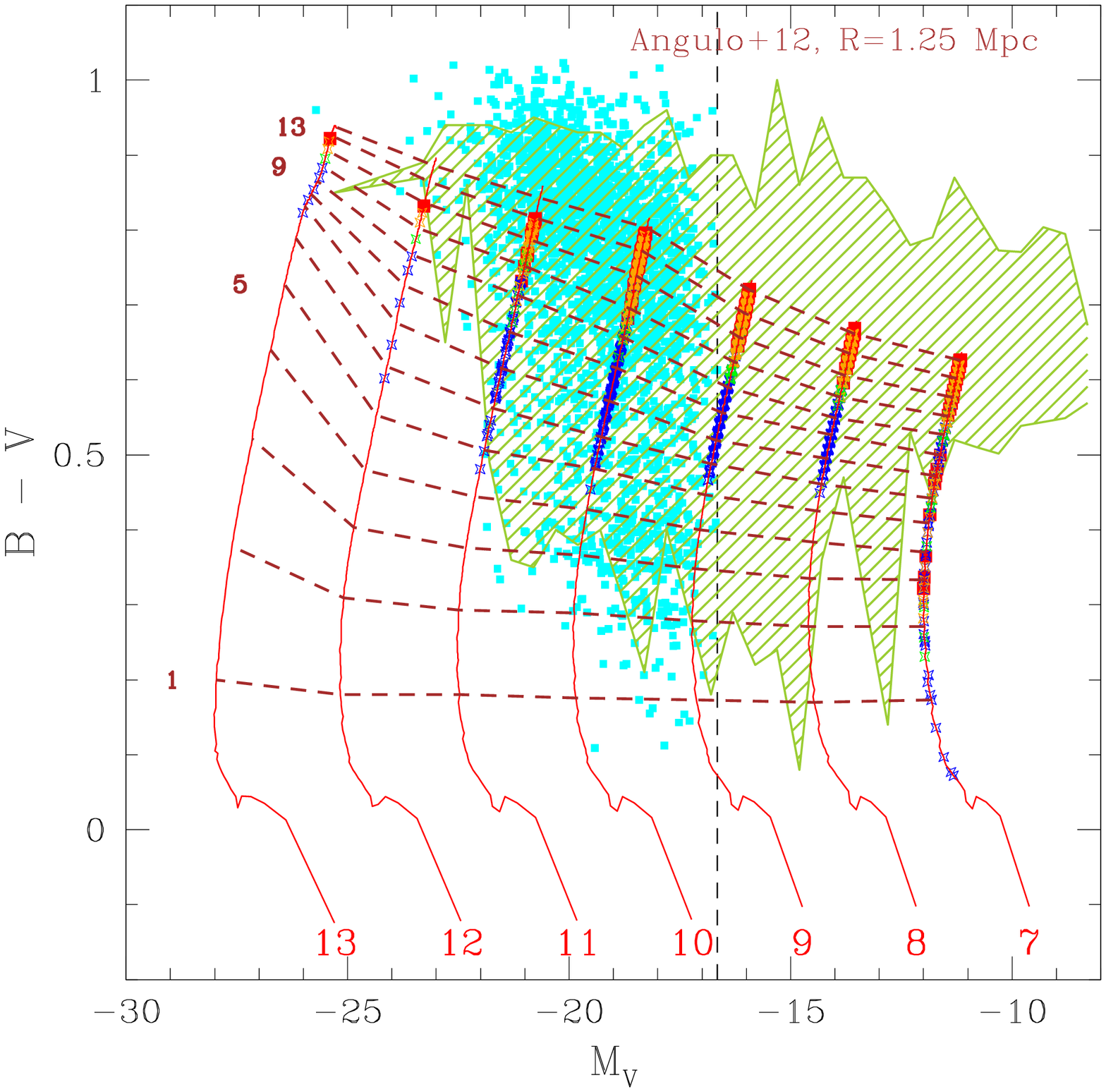}
		\includegraphics[width=8.0cm]{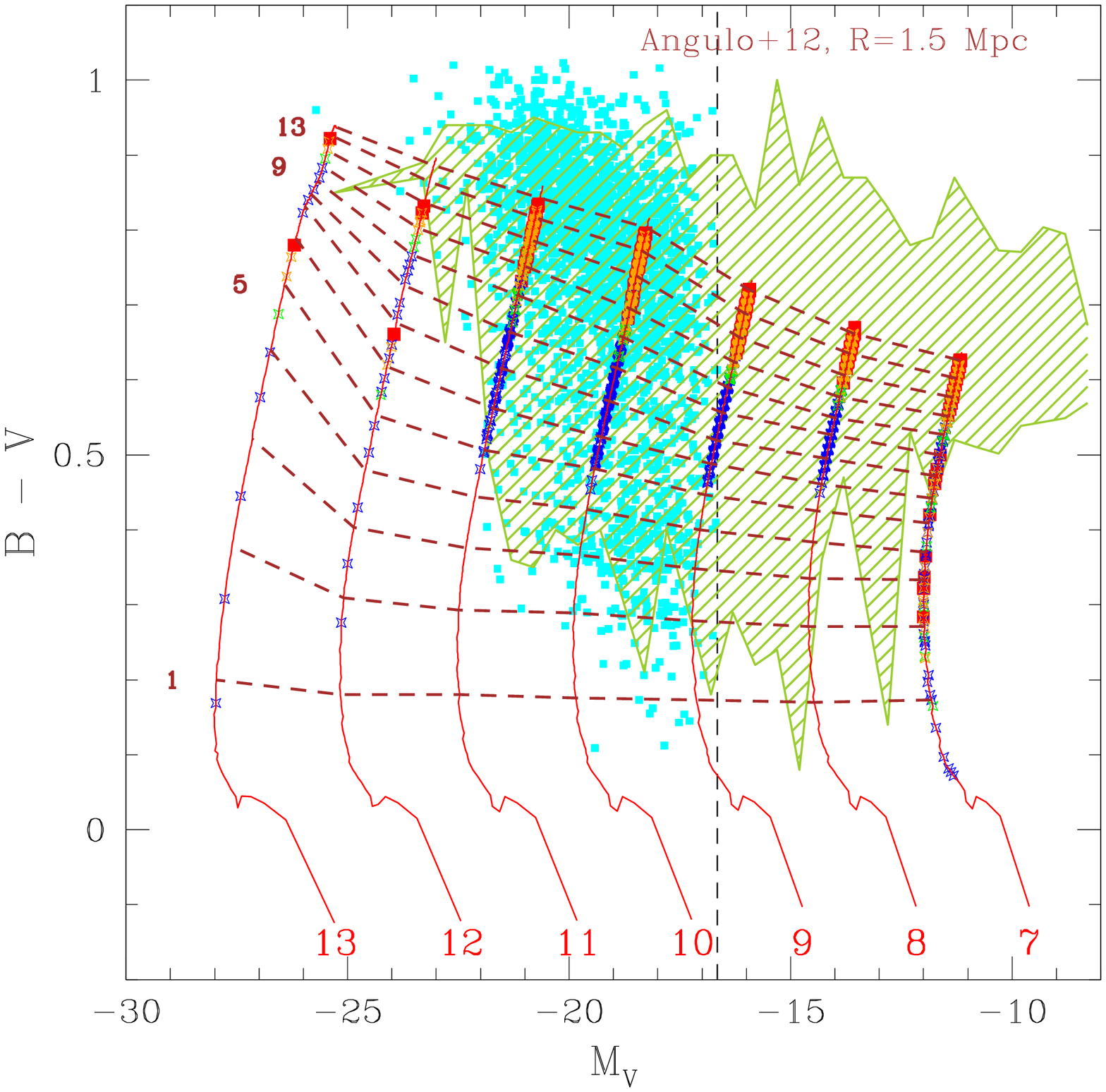}
		\includegraphics[width=8.0cm]{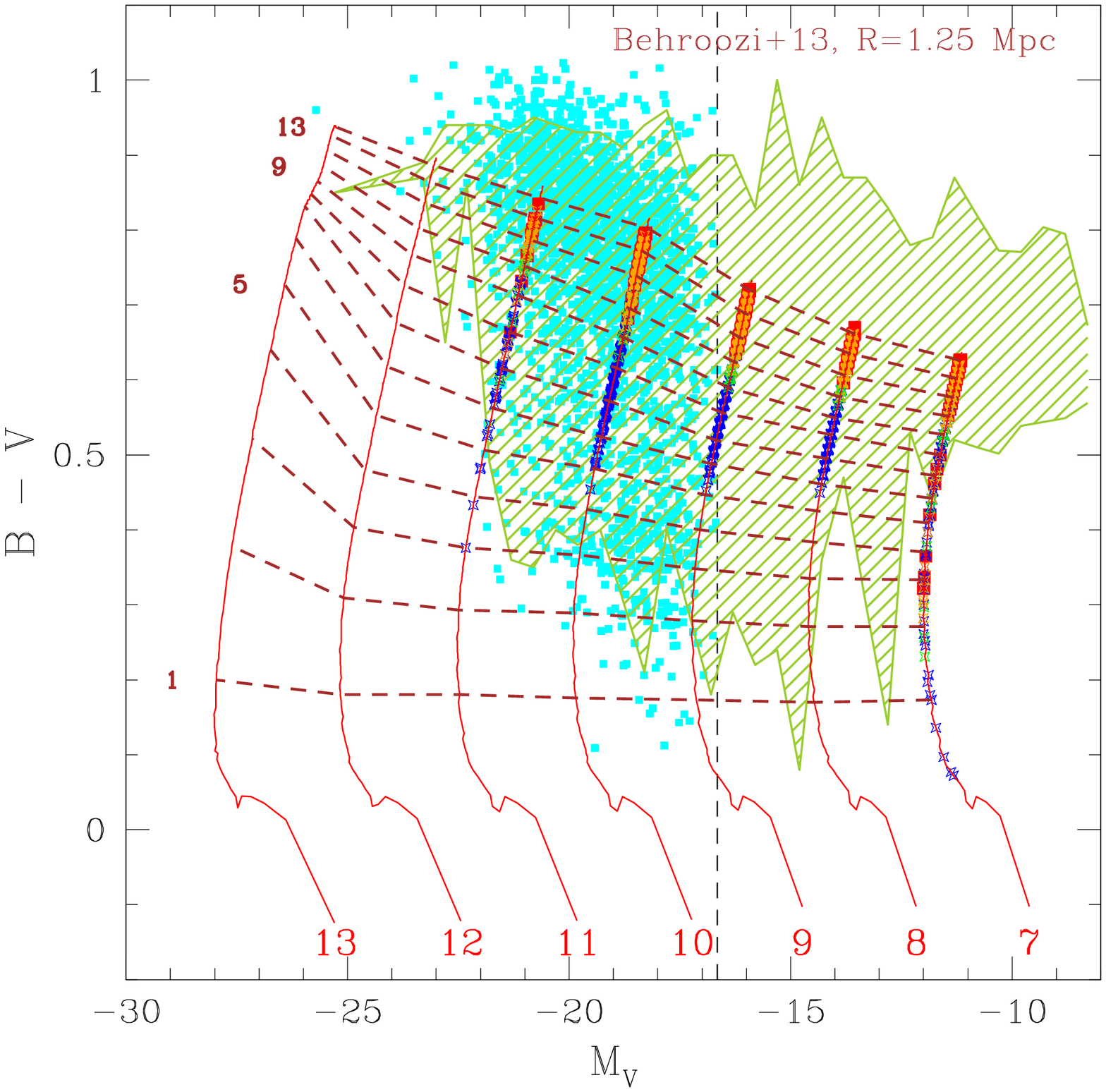}
		\includegraphics[width=8.0cm]{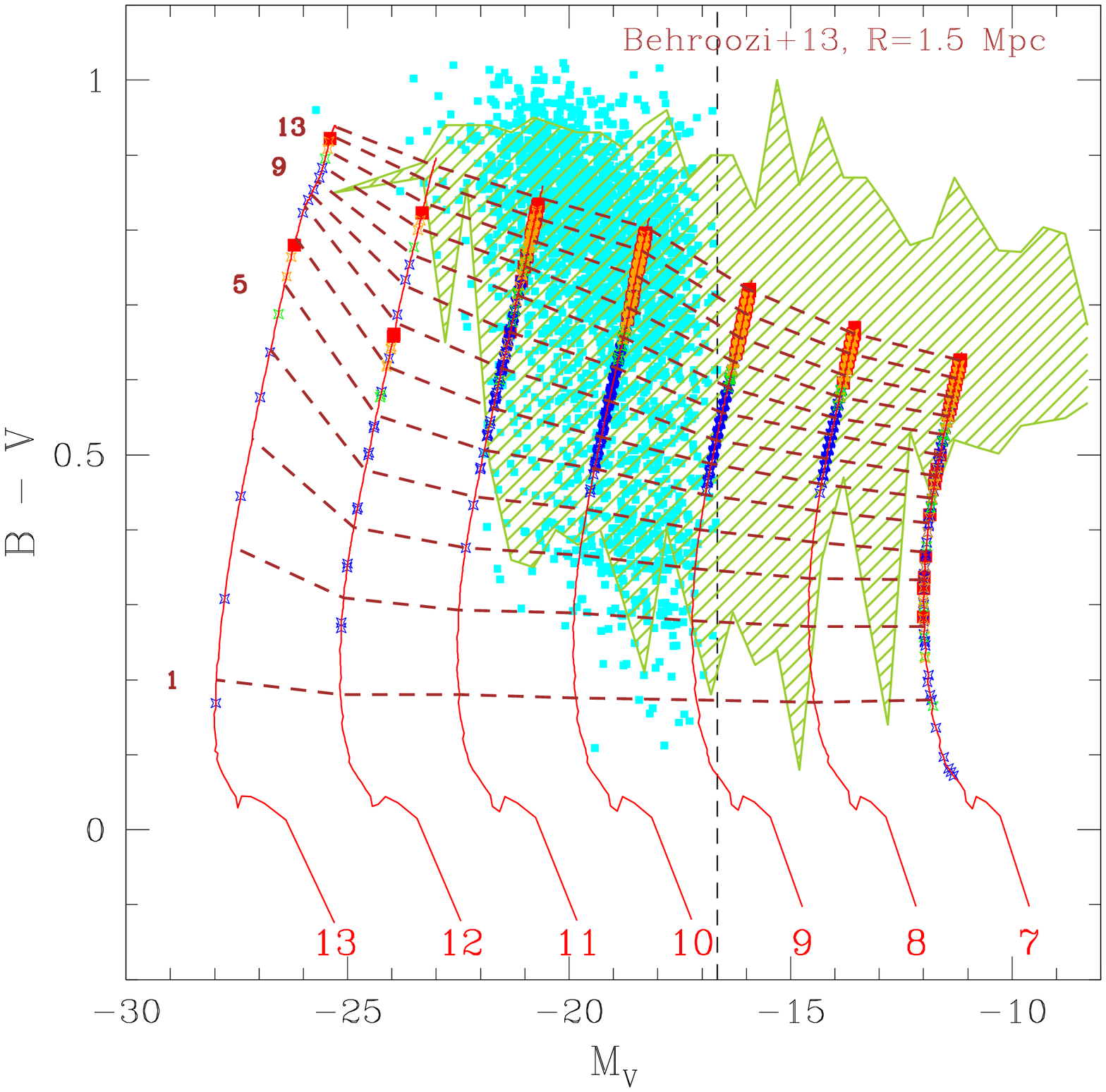}}
	\caption{CMDs of all simulated clusters: left panels are for radius 1.25 Mpc, right
	panels 1.50 Mpc; from top to bottom, \citet{Warren_etal_2006}, \citet{Angulo_etal_2012}, 
        \citet{Behroozi_etal_2013}.
	Color code and symbols  as in Figure \ref{cmd42impro}. }
\label{cmd42iall}
\end{figure*}

In this section first we provide the coefficients of the  4-th order interpolating polynomials of eqn. (23) that are
used to derive the HGFs 
as a function of the redshift, with the resulting galaxy populations expected at redshift  $z=0$ in clusters of typical size visible in Table
\ref{galdemo_sicofi}. The
coefficients  are listed in Tables from \ref{coef_w06} to \ref{coef_b13}. The clusters radii under consideration and 
total number of galaxies per cluster are 
given in Table \ref{galdemo_sicofi}, finally the associated CMDs are shown in  Fig. \ref{cmd42iall}. For the sake of
simplicity, the CMDs are for  galaxy 
models with no galactic winds. See the main text for more details on the subject.

\label{lastpage}

\end{document}